\documentclass[aps,floatfix]{revtex4-1}

\draft 
\usepackage{graphicx}

\usepackage{bm}
\usepackage{dcolumn}
\usepackage{ascmac}
\usepackage{amsmath,amssymb}

\newcommand{\mathbd}[1]{\mbox{\boldmath ${#1}$}}
\newcommand{\pd}[2]{{\frac{\partial {#1}}{\partial {#2}}}}

\newcommand{\pdpd}[3]{{\frac{\partial^2 {#1}}{\partial {#2}\partial {#3}}}}

\begin{document}


\title{Improving accuracy of turbulence models by neural network} 



\author{Satoshi Miyazaki}
\affiliation{Graduate School of Information Sciences, Tohoku University, Sendai 980-8579, Japan}
\author{Yuji Hattori}
\email[]{hattori@ifs.tohoku.ac.jp}
\thanks{corresponding author}
\affiliation{Institute of Fluid Sciences, Tohoku University, Sendai 980-8577, Japan}


\date{\today}

\begin{abstract}
Neural networks of simple structures are used to construct a turbulence model 
for large-eddy simulation (LES). 
Data obtained by direct numerical simulation (DNS) of homogeneous isotropic turbulence 
are used to train neural networks. 
It is shown that two methods are effective for improvement of accuracy of the model: 
weighting data for training 
and addition of the second-order derivatives of velocity to the input variables.  
As a result, 
high correlation between the exact sub-grid scale stress 
and the prediction by the neural network is obtained 
for large filter width; 
the correlation coefficient is about $0.9$ and $0.8$ 
for filter widths $48.8\eta$ and $97.4\eta$, respectively,  
where $\eta$ is the Kolmogorov scale. 
The models established by neural networks are close to but not identical 
with the gradient models. 
LES with the neural network model is performed for 
the homogeneous isotropic turbulence 
and the initial-value problem of the Taylor-Green vortices. 
The results obtained with the neural network model 
are in reasonable agreement with those of the filtered DNS.  
However, symmetry in the latter problem is broken 
since the neural network model does not possess rigorous symmetry 
under orthogonal transformations. 
\end{abstract}


\maketitle 

\section{Introduction}

Machine learning has been extensively applied to a wide range of 
problems in fluid dynamics, 
now forming an active area in data-driven fluid dynamics \cite{BEF-2019, BNK-2020}. 
Turbulence modelling is one of the important topics in this area 
since more accurate and robust turbulence models are wanted 
to predict and control turbulent flows 
and to optimize and design devices affected by turbulence. 
Many efforts have been devoted not only to RANS (Reynolds-Averaged Navier-Stokes) modeling \cite{LKT-2016, PD-2016, WWXL-2017, SMD-2017, WWX-2017, MD-2019, ZZKL-2019, DIX-2019} 
but also to LES (Large-Eddy Simulation) modeling \cite{GH-2017, MS-2017, ZHWJ-2019, BFM-2019, XWE-2020} 
seeking better models beyond the human knowledge. 

In the LES modeling, 
several works have shown that machine learning is a promising tool 
for improving the sub-grid scale (SGS) model beyond the existing models like 
the Smagorinsky model \cite{S-1963}, the similarity model \cite{BFR-1984}, the gradient model \cite{CFR-1979}, 
and their variants \cite{G-1992, VGK-1996, ZSK-1993, H-1997}.  
Gamahara and Hattori \cite{GH-2017} 
used a simple neural network (NN) to find a new model 
of the subgrid-scale (SGS) stress; 
the relation between the velocity gradient tensor as the input variables 
and the SGS stress as the output variable was established by neural networks. 
Data required for training and test of the neural network were
provided by direct numerical simulation (DNS) of a turbulent channel flow. 
They showed that neural networks can establish models similar to the gradient model. 
{\textit{A posteriori}} test using the NN model was performed to show that 
neural networks are a promising
tool for establishing a new subgrid model, 
although further improvement is required.
Zhou et al. \cite{ZHWJ-2019} took a similar approach 
with DNS data of homogeneous isotropic turbulence. 
Beck et al. \cite{BFM-2019} used a deep neural network 
with DNS data of weakly compressible isotropic turbulence. 
Xie et al. \cite{XWE-2020} also took an approach similar to Gamahara and Hattori \cite{GH-2017};  
they showed that addition of the velocity gradient tensor at points 
in the neighborhood to the input variables 
significantly improves the results including the correlation between 
the correct SGS stress and the prediction by the neural network.

There are two goals of different levels in this line of approach.  
One is to establish a turbulence model by machine learning as a black box;  
in this approach one does not pay attention to what is happening in e.g. neural networks  
pursuing just better prediction for a particular flow. 
Nowadays it can be done easily for a particular flow chosen for training 
with powerful tools of machine learning. 
However, whether the model gives accurate results for other flows is unknown 
since machine learning may have used some characteristics specific to the flow. 
The other goal is to establish an accurate and robust turbulence model 
which has an explicit expression as a function of resolved-scale variables 
and physical interpretation; 
this should be the ultimate goal.

Although recent works including those mentioned above 
showed some promising results, 
several important problems should be yet overcome to achieve our ultimate goal. 
They can be classified into three: 
(i) {\textit{reliability}}  
which implies that the SGS model obtained by machine learning 
should give accurate results within a reasonable range of errors; 
it should be numerically stable in actual LES calculations;   
(ii) {\textit{universality}} 
which implies that 
the obtained SGS model should be applicable 
not only to a wide range of the Reynolds numbers 
but also to flows of which types and/or geometries are different from 
the flow used in training process;  
and (iii) {\textit{usability}}  
which implies that the model can be implemented without difficulties 
and heavy numerical costs. 
Although the recent works mentioned above showed promising results, 
there has been no model that satisfies the three properties.  
Some NN models lack numerical stability in LES calculations.  
Applicability to flows other than the flow used in training is shown in 
some works but still limited. 
Moreover, 
no explicit expression of the NN model has been cultivated;  
in this sense no NN model has a firm physical basis, 
although our previous work pointed out similarity to the gradient model \cite{GH-2017}.

In order to solve the above problems, 
we should proceed step by step.  
Although a number of excellent tools of machine learning are available in these days, 
more knowledges and experiences should be explored to use them efficiently;  
we should clarify what are most useful in development of turbulence models.  
In particular, 
more should be investigated with simple tools like shallow neural networks 
which have a smaller number of parameters and options 
than deep neural networks 
and can allow us to infer an explicit expression of the turbulence model.

In this paper, 
we pursue methods for improving prediction accuracy of 
neural networks following the above line. 
Our objectives are 
to show that weighting training data 
and choice of the input variables can improve the prediction accuracy of neural networks
and to check the accuracy in the actual LES calculations using the turbulence model 
constructed by neural networks.

This paper is organized as follows.  
Numerical methods are described in Sec.~\ref{sec-method}.  
The turbulence models constructed by neural networks 
are evaluated by {\textit{a priori}} test in Sec.~\ref{sec-apriori};  
it is shown that the prediction accuracy of the NN model is 
improved by weighting training data and choice of the input variables.  
The NN models are used in actual LES calculations in Sec.~\ref{sec-aposteriori}; 
they are applied to two problems: the homogeneous isotropic turbulence 
and the initial value problem of the Taylor-Green vortices;  
the results are compared to LES using the existing models. 
Summary and future works are given in Sec.~\ref{sec-summary}.

\section{Numerical Methods}
\label{sec-method}

\subsection{Outline}
\label{sec-outline}
The numerical procedure is similar to that of Gamahara and Hattori \cite{GH-2017}. 
Our task consists of three steps: 
(i) development of a SGS model by a neural network, 
(ii) {\textit{a priori}} test in which the SGS stress 
is compared between the developed model and the true values, 
and (iii) {\textit{a posteriori}} test 
in which LES using the developed model 
is performed and evaluated. 

In LES small fluctuations of a flow variable $f$ are filtered out 
and we are concerned with the resolved-scale or grid-scale (GS) flow field  
$\overline{f} = \int G(\pmb{x}') f (\pmb{x}-\pmb{x}') d\pmb{x}'$, 
where $G$ is a filter function.  
The filtered Navier-Stokes equations for the GS flow field read 
\begin{eqnarray}
\pd{\overline{u_i}}{t} + \pd{}{x_j} \left(\overline{u_i} \ \overline{u_j}\right) 
&=& -\pd{\overline{p}}{x_i} + \frac{1}{Re} \pdpd{\overline{u_i}}{x_k}{x_k} - \pd{\tau_{ij}}{x_j}, 
\label{LESeq-1} \\
\pd{\overline{u_j}}{x_j} &=& 0,   \label{LESeq-2} 
\end{eqnarray}
where the residual or SGS stress tensor  
\begin{eqnarray}
\tau_{ij} = \overline{u_i u_j} -\overline{u_i} \ \overline{u_j} 
\end{eqnarray}
depends not only on the GS flow field but also on the fluctuations.  
The Gaussian filter is chosen as a filter function 
except that the top-hat filter is also used for comparison in Sec.~\ref{sec-input-res}. 
The filter width $\overline{\Delta}$ is the same in the $x$, $y$, and $z$ directions.

Our first step is to express the SGS stress tensor using the GS flow field 
\begin{eqnarray}
\tau_{ij} = {\cal{F}} \left[\overline{\mathbd{u}}\right].  
\end{eqnarray}
In the expression above, 
$\tau_{ij}(\mathbd{x})$ can depend on the GS velocity components $\{\overline{u_i}\}$ 
and their derivatives $\{ \overline{u_i}, \partial_j\overline{u_i}, \partial_j\partial_k\overline{u_i}, \cdots\}$ 
at any point in general.  
In the present study, however, we seek a pointwise relation 
\begin{eqnarray}
\tau_{ij} (\mathbd{x}) = {\cal{F}} \left(\{\partial_j\overline{u_i}(\mathbd{x}), \partial_j\partial_k\overline{u_i}(\mathbd{x}), \cdots\}\right), 
\end{eqnarray}
since it can be widely applied to flows in complex geometry 
and physical interpretation will be easier. 
The actual choice of the independent (or input) variables 
will be explained in Sec.~\ref{sec-input}. 
In the present study we use neural networks to 
establish a functional relation between 
the GS flow field and the SGS stress tensor.

\subsection{Training data and direct numerical simulation}
\label{sec-dns}
In Gamahara and Hattori \cite{GH-2017}, 
the training data for neural networks were provided by DNS of a turbulent channel flow. 
It turned out that the near-wall region requires careful treatment 
both in {\textit{a priori}} and {\textit{a posteriori}} tests. 
Thus, in the present study,  
we use the data obtained by DNS of homogeneous isotropic turbulence, 
which is not affected by walls, 
as the training data. 
DNS was performed in the same way as in Ishihara et al.~\cite{Ietal-2007} 
and Hattori and Ishihara \cite{HI-2012}. 
The three-dimensional incompressible Navier-Stokes equations 
were solved by the Fourier spectral method. 
Forcing was introduced at low wavenumbers to keep the kinetic energy constant. 
See Ishihara et al.~\cite{Ietal-2007} and Hattori and Ishihara \cite{HI-2012} for the details. 
The number of the Fourier modes was $N^3=512^3$ or $1024^3$. 
The values of simulation parameters and the turbulence characteristics 
are listed in Table \ref{tab:DNS_CASE}, 
where $R_{\lambda}$, $k_{\rm{max}}$, $\Delta t$, $\nu$, $\varepsilon$, $\eta$, and $T$ 
denote the Reynolds number based on the Taylor microscale, 
the effective maximum wavenumber, 
the size of the time step, 
the kinematic viscosity, 
the averaged rate of energy dissipation, 
the Kolmogorov scale, 
and the eddy turnover time. 

\begin{table}[h]
\begin{center}
  \vspace{5mm}
  \caption{DNS parameters and turbulence characteristics at final time. }
  \begin{tabular}{ccrrrrrrr}
\hline\hline
	Case & $N^3$ & $R_{\lambda}$ & $k_{\rm{max}}$ & $10^3 \Delta t$ & $10^4\nu$ & $\varepsilon$ & $10^3\eta$ & $T$ \\ \hline
	Case 1 &$512^3$ & 173 & 241 & 1.0 & 7.0 & 0.0795 & 8.10 & 2.10 \\
	Case 2 &$1024^3$& 268 & 483 & 0.625 & 2.8 & 0.0829 & 4.03 & 1.94 \\
\hline\hline
  \end{tabular}
  \label{tab:DNS_CASE}
\end{center}
\end{table}

The training data were calculated and extracted from a snapshot of the DNS data. 
The number of data $n_d$ was chosen from the range $5000 \le n_d \le 50000$. 
How to choose the data is described in section \ref{sec-sampling}. 
The trained neural networks were tested using data extracted from snapshots 
different from those used for training. 

\subsection{Neural network}
\label{sec-nn}
As discussed in the introduction, 
we employ a shallow feed-forward neural network for training 
to establish a functional relation 
between the GS flow field and the SGS stress tensor. 
Our neural network consists of three layers:  
the input, hidden, and output layers. 
A single neuron of the $l$-th layer receives a set of inputs $\left\{X_j^{(l-1)}\right\}$ 
and then outputs $X_i^{(l)}$ which is calculated as 
\begin{eqnarray}
X_i^{(l)} &=& {\cal{B}}\left(s_i^{(l)}+b_i^{(l)}\right), \\
s_i^{(l)} &=& \sum_j W_{ij}^{(l)} X_j^{(l-1)},  
\end{eqnarray}
where ${\cal{B}}(z)= 1/(1+e^{-\alpha z})$ is the activation function, 
$b_i^{(l)}$ is the bias parameter, and $W_{ij}^{(l)}$ is the weight. 
The bias parameters and the weights are corrected iteratively 
so that the final output $X^{(L)}$ approximates well the given SGS stress. 
The data of the first layer $\left\{X_j^{(1)}\right\}$ are given by the GS flow field. 
The back propagation is used as a method for training to 
minimize the difference between the output and the given SGS stress 
$\sum |X^{(3)}-\tau_{ij}|^2$. 

The neural network is trained for one diagonal component $\tau_{11}$ and 
one off-diagonal component $\tau_{12}$ separately. 
The other components can be predicted by permutation of the indices: e.g. 
$\tau_{23}$ can be predicted by the network trained for $\tau_{12}$ by 
permutation of indices $(1,2,3) \to (2,3,1)$.

\subsection{Sampling training data}
\label{sec-sampling}
\begin{figure}
  \centering
  \includegraphics[width=72mm]{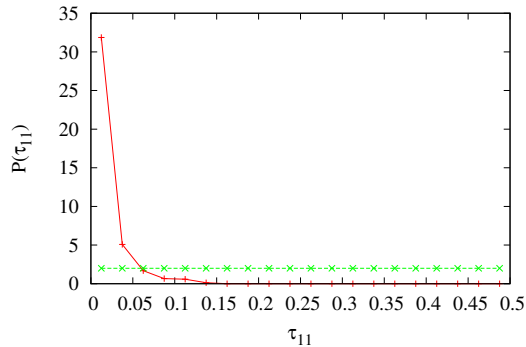}
  \caption{P.d.f. of $\tau_{11}$. Red line: continuous sampling, green line: uniform sampling. }
  \label{fig:weight-pdf}
\end{figure}

In this study a sufficient amount of training data are available 
since the number of data is $N^3$ times the number of the instantaneous field data. 
The training data are normally chosen without any preference from the available data; 
they are chosen randomly or on particular parts (lines, sections etc.).  
However, these methods of data sampling can be disadvantageous 
since large values of the SGS stress, which are important in the filtered equation, 
are rarely encountered and training can be insufficient (Fig.~\ref{fig:weight-pdf}). 
Therefore, we consider two methods of data sampling:  
One is the continuous sampling, in which the data on several lines are chosen;  
and the other is the uniform sampling, in which the data are chosen 
so that the probability density function (p.d.f.) of the sampled SGS stress becomes a uniform distribution 
within a range containing large values.  
In other words, the sampled data are weighted in the uniform sampling.  
We expect that the uniform sampling improves the efficiency of training. 

Figure \ref{fig:weight-pdf} shows the probability density function 
of $\tau_{11}$ obtained by two different methods of sampling.  
The total number of data is $5000$. 
The data obtained by the continuous sampling are concentrated at small values. 
On the other hand, the probability density function of the uniform sampling is constant by definition.  

\subsection{Input variables}
\label{sec-input}
The choice of the input variables is one of the most important points 
for successful learning. 
The base set of the input variables is the velocity gradient tensor $\{\partial_j{\overline{u_i}}\}$ 
as in Gamahara and Hattori \cite{GH-2017}, 
while the distance from the wall is absent for the homogeneous isotropic turbulence.  
In the present study, we add the second-order derivative of velocity 
to the input variables and see how the prediction is improved.   
There are variants of the above two sets in which some components are excluded. 
The sets of the input variables are listed below:    
\begin{itemize}
\item Set S: (the Smagorinsky type) the components of the velocity gradient tensor 
sufficient to reproduce the Smagorinsky model  
\begin{eqnarray}
\tau_{11} = {\cal{F}}^{({\rm{S}})} \left(\left\{ \overline{S_{ij}} \right\}\right),   
\end{eqnarray}
where $\overline{S_{ij}} = \frac{1}{2} \left(\pd{\overline{u_i}}{x_j}+\pd{\overline{u_j}}{x_i}\right)$. 
\item Set D1: the first-order derivative of velocity 
\begin{eqnarray}
\tau_{11} = {\cal{F}}^{({\rm{D1}})} \left(\left\{\pd{\overline{u_i}}{x_j}\right\}\right).  
\end{eqnarray}
\item Set G1: the first-order gradient of the velocity components whose indices appear in the 
corresponding SGS stress component 
\begin{eqnarray}
\tau_{11} &=& {\cal{F}}^{({\rm{G1}})} \left(\left\{\pd{\overline{u_1}}{x_j}\right\}\right) \\  
\tau_{12} &=& {\cal{F}}^{({\rm{G1}})} \left(\left\{\pd{\overline{u_1}}{x_j}, \pd{\overline{u_2}}{x_j}\right\}\right) 
\end{eqnarray}
\item Set D2: the first-order and the second-order derivatives of velocity 
\begin{eqnarray}
\tau_{11} = {\cal{F}}^{({\rm{D2}})} \left(\left\{\pd{\overline{u_i}}{x_j}, 
\pdpd{\overline{u_i}}{x_j}{x_k}\right\}\right).  
\end{eqnarray}
\item Set G2: the first-order and the second-order derivatives of the velocity components 
whose indices appear in the corresponding SGS stress component 
\begin{eqnarray}
\tau_{11} &=& {\cal{F}}^{({\rm{G2}})} \left(\left\{\pd{\overline{u_1}}{x_j}, 
\pdpd{\overline{u_1}}{x_j}{x_k}\right\}\right), \\  
\tau_{12} &=& {\cal{F}}^{({\rm{G2}})} \left(\left\{\pd{\overline{u_1}}{x_j}, \pd{\overline{u_2}}{x_j}, 
\pdpd{\overline{u_1}}{x_j}{x_k}, \pdpd{\overline{u_2}}{x_j}{x_k}\right\}\right).  
\end{eqnarray}
\end{itemize}
The numbers of the input variables in the sets above are listed in Table \ref{ta:input}. 

\subsection{Large-Eddy Simulation}
\label{sec-LES}
In {\textit{a posteriori}} test 
LES was performed using the trained neural networks. 
The equations (\ref{LESeq-1}) and (\ref{LESeq-2}) were solved 
numerically by the same method as DNS, 
while the SGS stress $\tau_{ij}$ was calculated by the trained neural networks. 
LES with existing models such as the Smagorinsky model 
\begin{eqnarray}
	\tau_{ij}^{\rm{(SM)}}-\frac{1}{3}\delta_{ij}\tau_{kk}^{\rm{(SM)}}
= -2(C_s\overline{\Delta})^2 \left(2\overline{S}_{ij} \, \overline{S}_{ij}\right)^{1/2} \overline{S}_{ij}, \ \ \ 
\label{eq:eddy_visc}
\end{eqnarray}
where $C_S$ is the Smagorinsky coefficient, 
and the Bardina model 
\begin{eqnarray}
	\tau_{ij}^{\rm{(B)}}
= \overline{\overline{u_i}\, \overline{u_j}} - \overline{\overline{u_i}}\, \overline{\overline{u_j}} 
+ \tau_{ij}^{\rm{(SM)}}, 
\label{eq:eddy_bardina}
\end{eqnarray}
where the value of $C_S$ can be different from that of the Smagorinsky model,  
was also performed for comparison. 
The gradient model 
\begin{eqnarray}
	\tau_{ij}^{\rm{(GM)}}
= \frac{{\overline{\Delta}}^2}{12}\pd{\overline{u_i}}{x_k}\pd{\overline{u_j}}{x_k}, 
\label{eq:GM}
\end{eqnarray}
which can be derived by the Taylor expansion, 
is compared to neural networks in {\textit{a priori}} test.  
The next-order terms can be derived by the Taylor expansion 
and can be included in the above model \cite{Yeo-1987, YB-1988, Leonard-1997}
\begin{eqnarray}
	\tau_{ij}^{\rm{(EGM)}}
= \frac{{\overline{\Delta}}^2}{12}\pd{\overline{u_i}}{x_k}\pd{\overline{u_j}}{x_k} 
+ \frac{{\overline{\Delta}}^4}{288}\pdpd{\overline{u_i}}{x_k}{x_l}\pdpd{\overline{u_j}}{x_k}{x_l}.  
\label{eq:EGM}
\end{eqnarray}
It is called the extended gradient model in this paper; 
this is also used in LES and in {\textit{a priori}} test.   

\section{Results of {\textit{a priori}} test}
\label{sec-apriori}

\subsection{Effects of data sampling method}
\label{sec-sampling}

Table \ref{ta:train_defCC} shows 
the correlation coefficient between the exact SGS stress obtained by filtering DNS data of Case 1 
with filter width $\overline{\Delta} = 8\Delta_{\rm{DNS}}=12.2\eta$ 
and the prediction by a trained neural network with the input variables of Set D1, 
which is hereafter called NN-D1 (similar abbreviation is used for the other Sets). 
The two methods of data sampling are compared. 
The uniform sampling gives better correlation than the continuous sampling as expected. 
The improvement of correlation is more pronounced for the off-diagonal components. 
Table \ref{ta:train_defL2} shows the error evaluated by 
the $L^2-$norm between the exact SGS stress 
and the prediction by the neural networks obtained by the two methods of data sampling. 
For the off-diagonal components 
the uniform sampling gives smaller errors than the continuous sampling, 
while the errors in the diagonal components 
are slightly larger for the uniform sampling than for the continuous sampling; 
the latter result is due to that the uniform sampling slightly overpredicts 
the magnitude of $\tau_{11}$ as shown later in Fig.~\ref{fig:SGSjPDF512-8d}.  
This over-prediction, however, disappears when Sets D2 and G2 are used as the input variables. 
Thus, we use the uniform sampling in the following.  
Tables \ref{ta:train_defCC} and \ref{ta:train_defL2} also confirm 
that the neural networks trained with $\tau_{11}$ and $\tau_{12}$
give similar correlation for the other diagonal components ($\tau_{22}$ and $\tau_{33}$) 
and off-diagonal components ($\tau_{23}$ and $\tau_{31}$), respectively. 
Therefore, we show the results for $\tau_{11}$ and $\tau_{12}$ in the rest of this section. 

\begin{table}[h]
\begin{center}
  \vspace{5mm}
  \caption{Correlation coefficient between exact SGS stress and prediction by neural network. 
Dependence on the method of data sampling. 
Case 1, $\overline{\Delta} = 8\Delta_{\rm{DNS}}=12.2\eta$. 
The input variables of Set D1 are used in the neural network (NN-D1). }
  \begin{tabular}{c||rrr|r||rrr|r}
\hline \hline 
    sampling & $\tau_{11}$ & $\tau_{22}$ & $\tau_{33}$ & mean & $\tau_{12}$ & $\tau_{23}$ & $\tau_{31}$ & mean \\ \hline
continuous &0.745&0.755&0.754&0.751&0.575&0.592&0.578&0.582 \\
uniform &0.821&0.818&0.825&0.822&0.872&0.878&0.877&0.876 \\ 
\hline\hline 
  \end{tabular}
  \label{ta:train_defCC}
\end{center}
\end{table}

\begin{table}[h]
\begin{center}
  \vspace{5mm}
  \caption{Error evaluated by $L^2-$norm between exact SGS stress 
and prediction by neural network. Dependence on the two methods of data sampling. 
Same as in Table \ref{ta:train_defCC}. }
  \begin{tabular}{c||rrr|r||rrr|r}
\hline\hline 
sampling & $\tau_{11}$ & $\tau_{22}$ & $\tau_{33}$ & mean & $\tau_{12}$ & $\tau_{23}$ & $\tau_{31}$ & mean \\ \hline
continuous &0.534&0.525&0.526&0.528&0.888&0.868&0.887&0.881\\
uniform &0.664&0.655&0.642&0.654&0.556&0.541&0.543&0.547\\ 
\hline\hline 
  \end{tabular}
  \label{ta:train_defL2}
\end{center}
\end{table}

\subsection{Dependence on input variables}
\label{sec-input-res}

Next we seek a set of the input variables which maximizes the correlation 
between the exact SGS stress and the prediction by neural networks. 
In particular, the effects of adding the second-order derivatives of velocity 
on the prediction accuracy are investigated. 
The results of prediction by the neural networks with the sets of the input variables introduced 
in Sec.~\ref{sec-input} are also compared to 
the gradient model (GM) and the extended gradient model (EGM).  
The DNS data and the filter width are the same as in Sec.~\ref{sec-sampling}: 
Case 1 and $\overline{\Delta} = 8\Delta_{\rm{DNS}}=12.2\eta$. 

Table \ref{ta:input} shows correlation between the exact SGS stress 
and the prediction by the neural network for each set of the input variables.  
In addition to the Gaussian filter, the top-hat filter is also used as filter function 
for a limited number of sets. 
The table shows that correlation is low for the Smagorinsky type (Set S), 
while high correlation is obtained for the other sets which use derivatives of velocity 
in accordance with Gamahara and Hattori \cite{GH-2017}.  
NN-G1 gives better correlation than NN-D1, 
although the number of the input variables is smaller. 
It implies that the derivatives of $u_i$ and $u_j$ are important for estimate of $\tau_{ij}$;  
it also implies that it is not always advantageous to increase the number of the input variables. 

The most important point in Table \ref{ta:input} 
is that addition of the second-order derivatives increases the correlation coefficients 
for the Gaussian filter significantly: 
e.g. $0.814$ for $\tau_{11}$ with NN-D1 increases to $0.946$ with NN-D2, 
while $0.901$ for $\tau_{12}$ with NN-G1 increases to $0.975$ with NN-G2. 
This is not the case for the top-hat filter as 
$0.746$ for $\tau_{11}$ with NN-D1 increases only a little to $0.783$ with NN-D2.  
This difference in the improvement of correlation is closely related with the fact that 
the extended gradient model (\ref{eq:EGM}) is derived for the Gaussian filter, 
while it cannot be derived for 
the top-hat filter which is the product of the filters in the three directions. 

The table also shows correlation for the gradient and the extended gradient models.  
The correlation is higher than the prediction by the neural networks;  
it is not striking as high correlation between the exact SGS stress 
and the gradient-type models has been reported in previous works \cite{BO-1998}.

\begin{figure}
  \begin{center}
   \includegraphics[width=70mm]{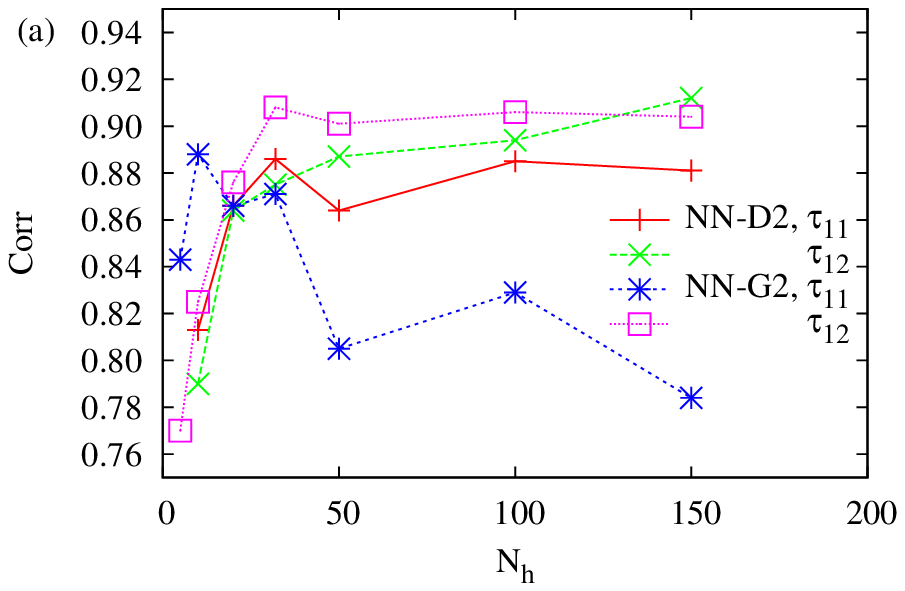}
   \includegraphics[width=70mm]{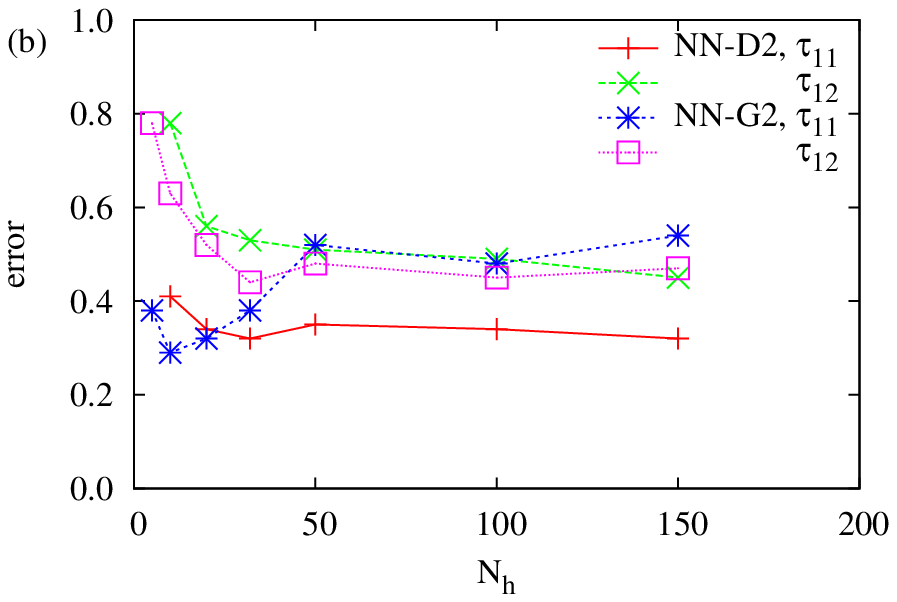}
  \end{center}
 \caption{(a) Correlation between exact SGS stress 
and prediction by NN-D2 and NN-G2 and (b) error. 
Dependence on the number $N_h$ of neurons in the hidden layer is shown. }
 \label{fig:neuronN}
\end{figure}

Table \ref{ta:input} also shows the number $N_h$ of the neurons in the hidden layer 
used for each set of the input variables. 
These values were optimized by investigating the performance of the neural networks 
as shown in Fig.~\ref{fig:neuronN} for NN-D2 and NN-G2;  
the filter size is set to $\overline{\Delta}=32\Delta_{\rm{DNS}}=48.7\eta$, 
which is four times the value used for obtaining correlation, 
to elucidate the dependence on $N_h$. 
Compelling effects of changing $N_h$ are observed in Fig.~\ref{fig:neuronN}: 
the prediction accuracy increases with $N_h$ for small values of $N_h$, 
while overfitting can lower the regression ability of the neural network with Set G2 for large $N_h$. 
The dependence on $N_h$ is weak for $N_h \gtrsim 50$. 
Since these effects depend on the set of the input variables, 
the optimized value of $N_h$ is different between the cases in general. 

\begin{table}[h]
\begin{center}
  \vspace{5mm}
  \caption{Correlation coefficient between exact SGS stress and prediction by neural network. 
Dependence on the set of the input variables is shown. 
The input variables in parentheses are used only for $\tau_{12}$. 
Correspondingly, the values of $N_i$ and $N_h$ in parentheses are those for $\tau_{12}$ 
which are different from $\tau_{11}$. 
Case1, $\overline{\Delta}=8\Delta_{\rm{DNS}}=12.2\eta$. 
}
  \begin{tabular}{c|rr|rr|ll|cc|cc}
\hline \hline 
    & & & & & & & \multicolumn{2}{c|}{Corr (Gaussian)} & \multicolumn{2}{c}{Corr (top-hat)} \\
    Set & \multicolumn{2}{c|}{$N_i$} & \multicolumn{2}{c|}{$N_h$} & \multicolumn{2}{c|}{input variables} & $\tau_{11}$ & $\tau_{12}$ &  $\tau_{11}$ &  $\tau_{12}$ \\ \hline
	S& 6 & & 20 & &$\overline{S}_{ij}$& & 0.227&0.180&--&--\\
	D1& 9 & & 20 & &$\partial_j\overline{u}_i$ & &0.814&0.878&0.746&0.811\\
	G1& 3 & (6)& 10 & &$\partial_j\overline{u}_1$&$(\partial_j\overline{u}_2)$&0.922&0.901&0.894&0.893\\
	D2& 27 & & 32 & &$\partial_j\overline{u}_i,\,\partial_j\partial_k\overline{u}_i$& &0.946&0.972&0.783&0.838\\
	G2&9 & (18) &20 & (32)&$\partial_j\overline{u}_1,\,\partial_j\partial_k\overline{u}_1$ & $(\partial_j\overline{u}_2,\,\partial_j\partial_k\overline{u}_2)$&
	0.981&0.975&--&--\\ \hline 
	GM&3 & (6)&-- & &$\partial_j\overline{u}_1$&$(\partial_j\overline{u}_2)$&0.962&0.964&--&--\\
	EGM&9 & (18)& --& & $\partial_j\overline{u}_1,\,\partial_j\partial_k\overline{u}_1$&$(\partial_j\overline{u}_2,\,\partial_j\partial_k\overline{u}_2)$&
	0.993&0.994&--&--\\
\hline \hline 
  \end{tabular}
  \label{ta:input}
\end{center}
\end{table}

\begin{figure}
  \begin{center}
   \includegraphics[width=55mm]{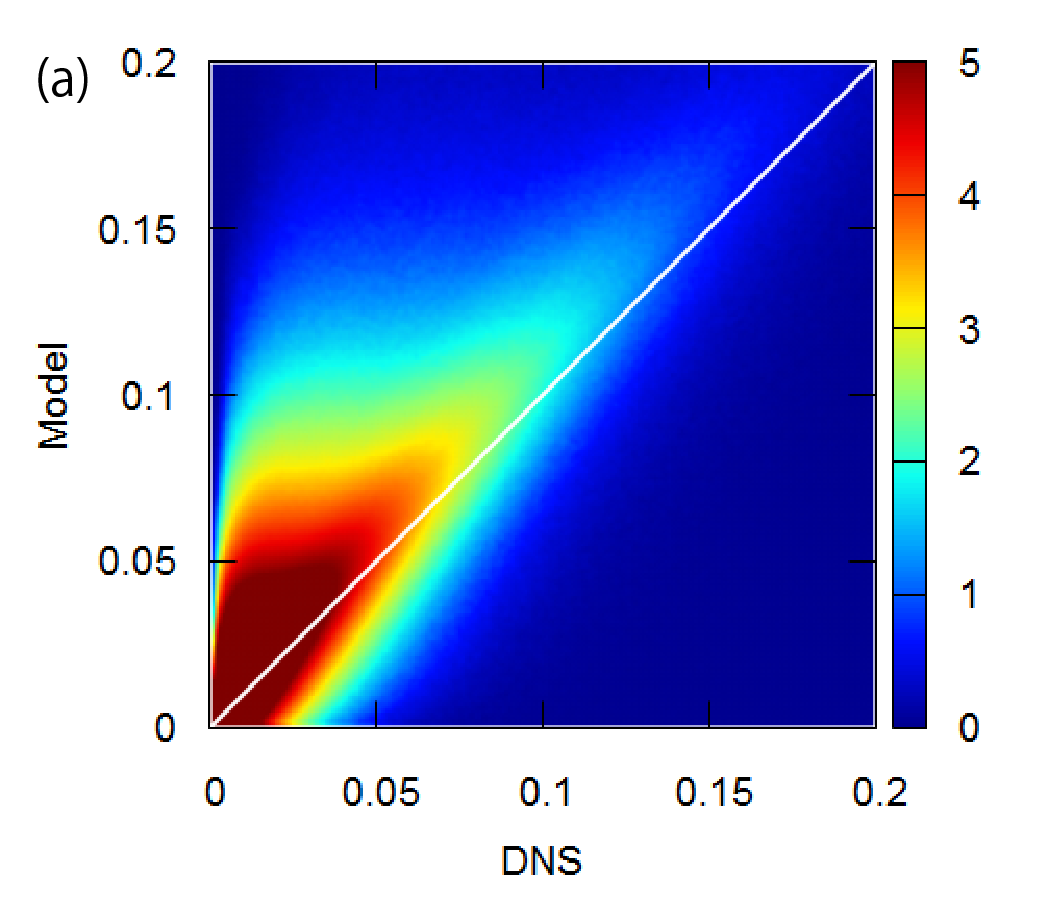}
   \includegraphics[width=55mm]{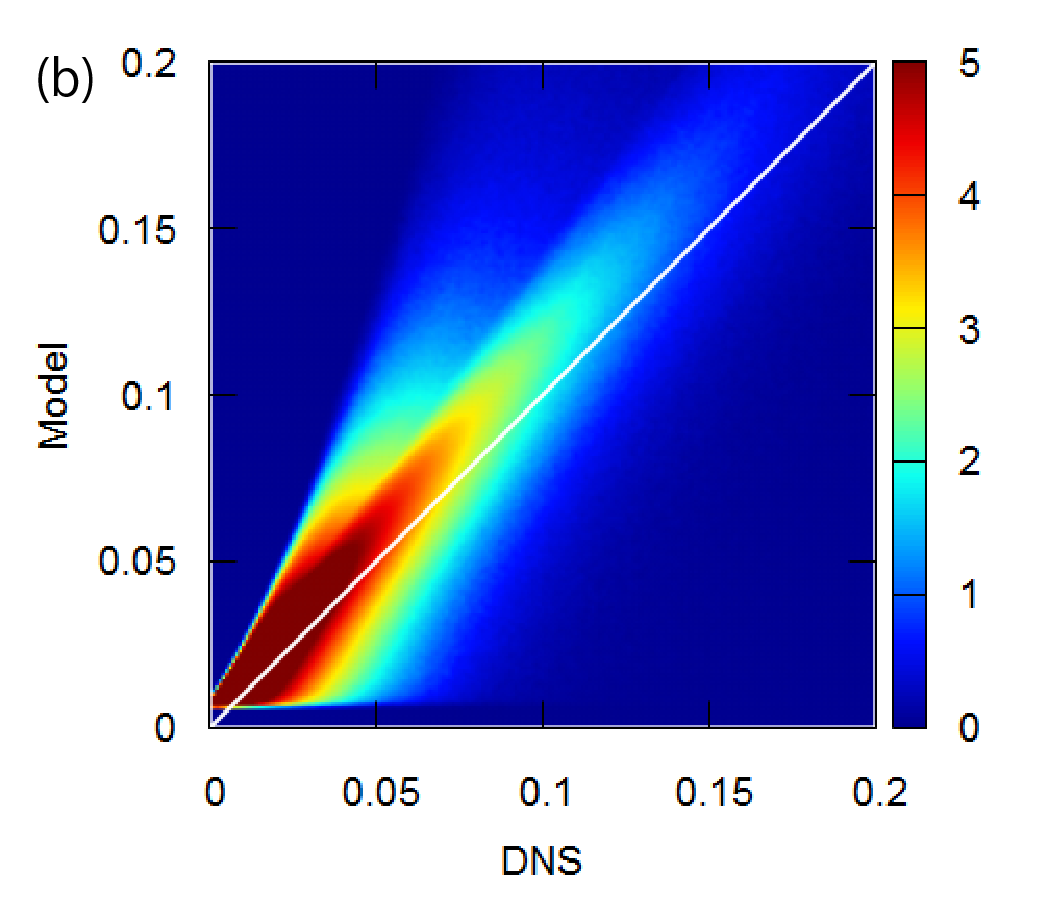}
   \includegraphics[width=55mm]{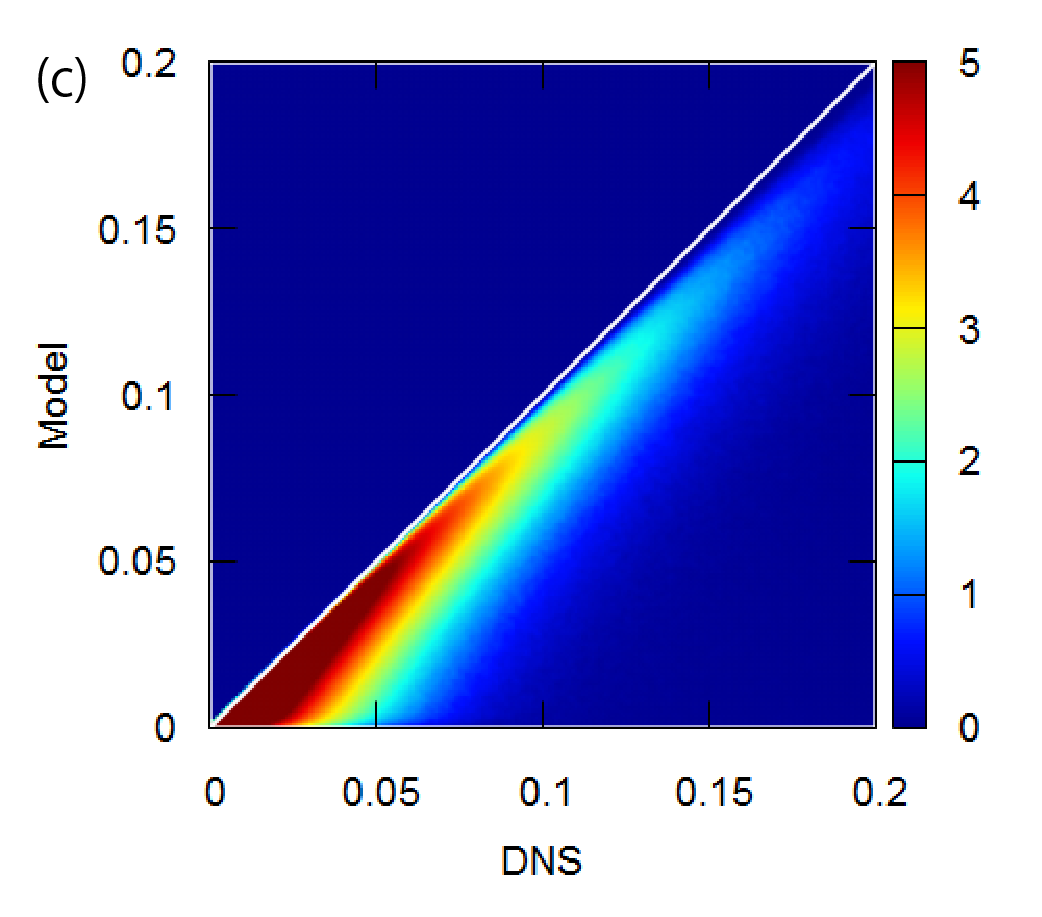}

   \includegraphics[width=55mm]{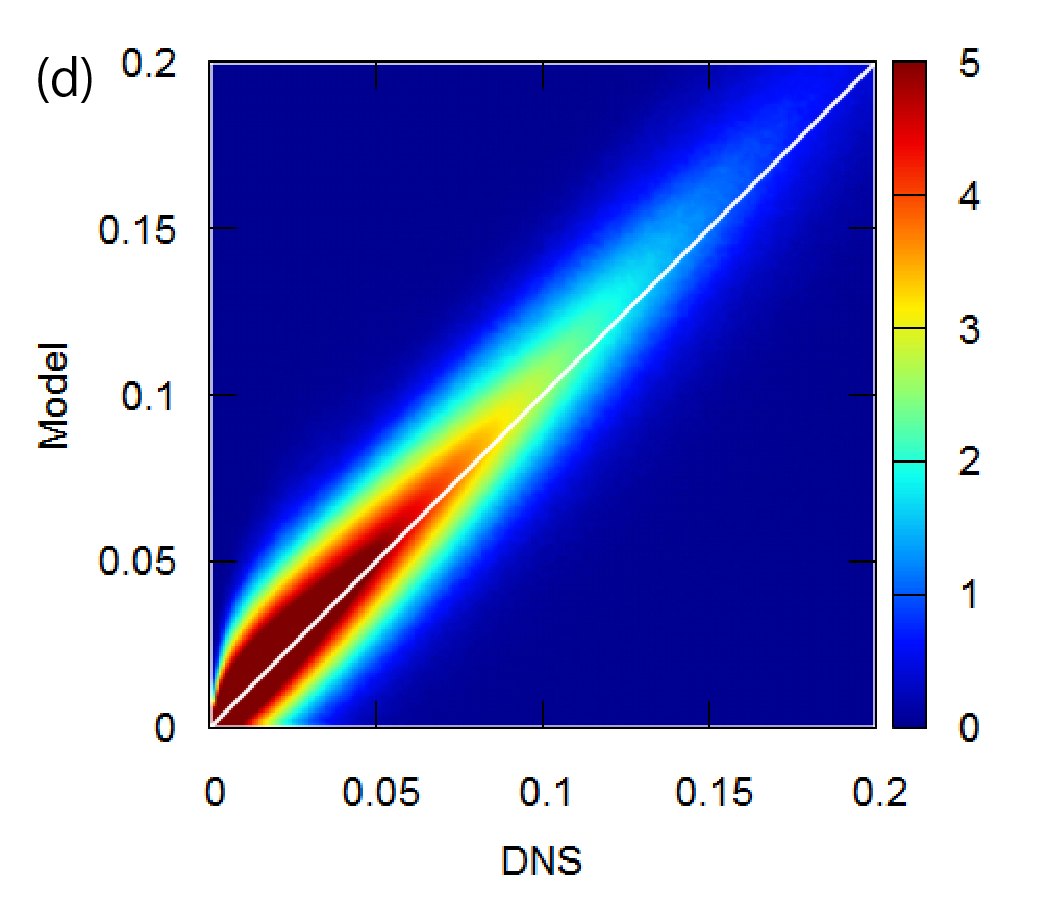}
   \includegraphics[width=55mm]{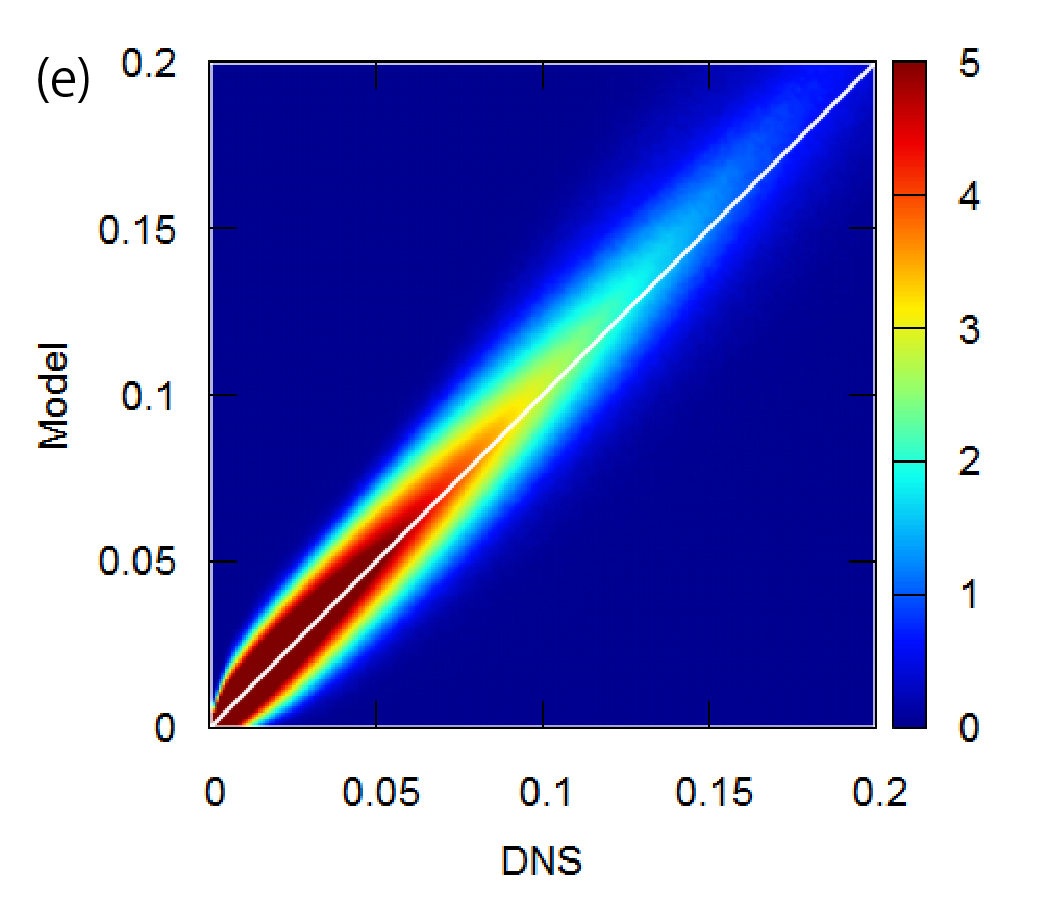}
   \includegraphics[width=55mm]{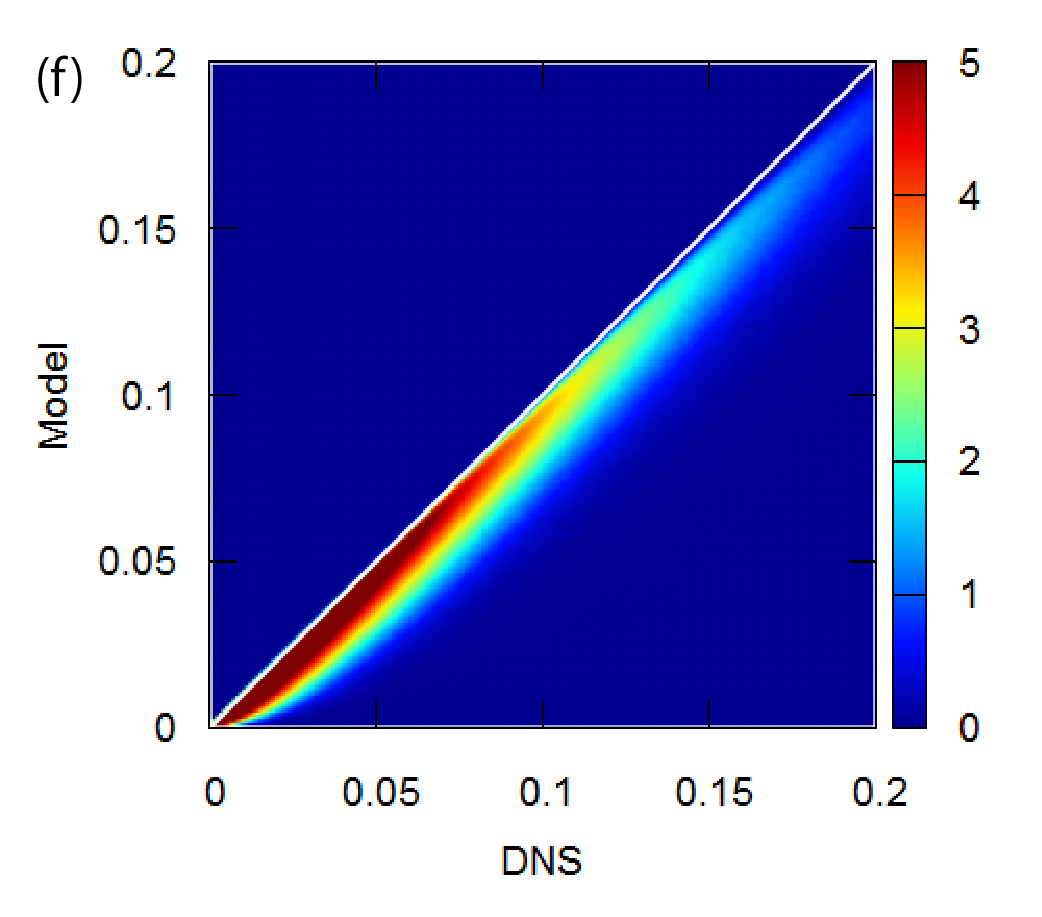}
  \end{center}
 
 \caption{Joint p.d.f. of SGS stress $\tau_{11}$. \,Case 1,\,$\overline{\Delta}=12.2\eta$. 
The horizontal axis is the exact SGS stress obtained by filtering the DNS data. 
The vertical axes are predictions by (a) NN-D1, 
(b) NN-G1, (c) GM (gradient model), (d) NN-D2, (e) NN-G2, (f) EGM (extended gradient model).  }
 \label{fig:SGSjPDF512-8d}
\end{figure}

\begin{figure}
  \begin{center}
   \includegraphics[width=55mm]{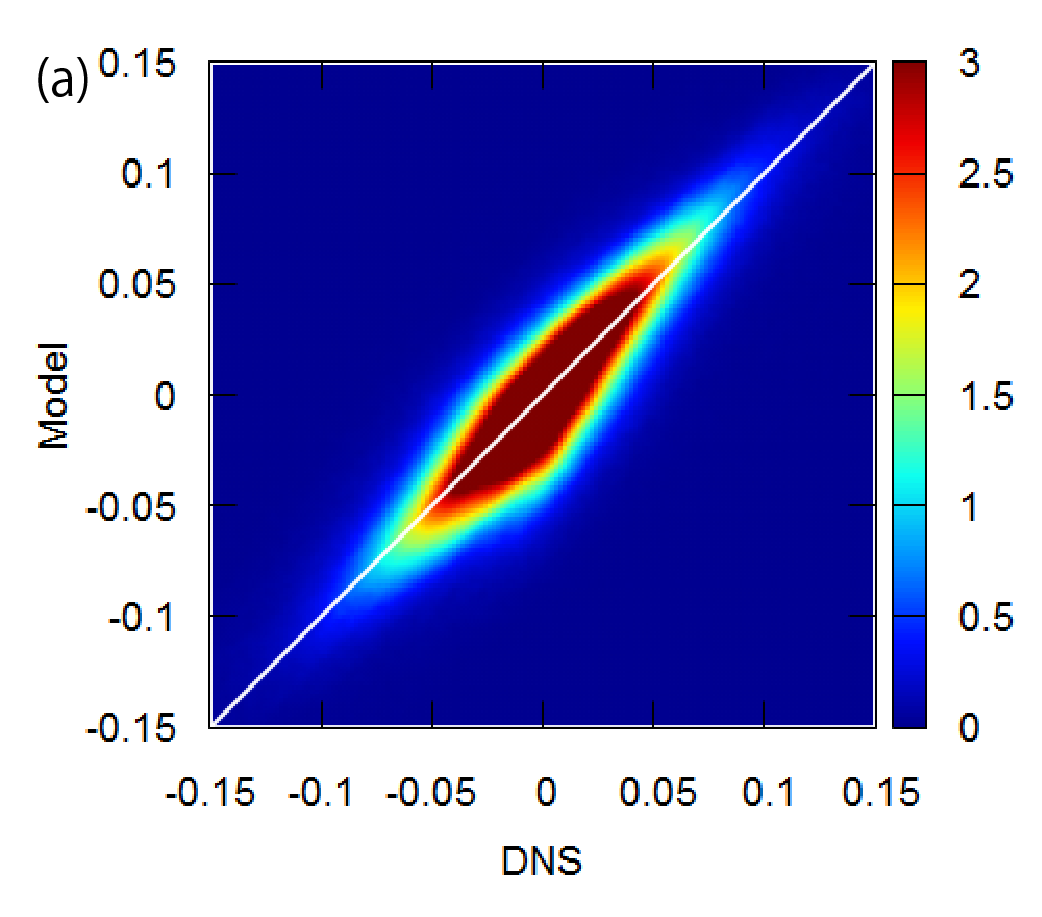}
   \includegraphics[width=55mm]{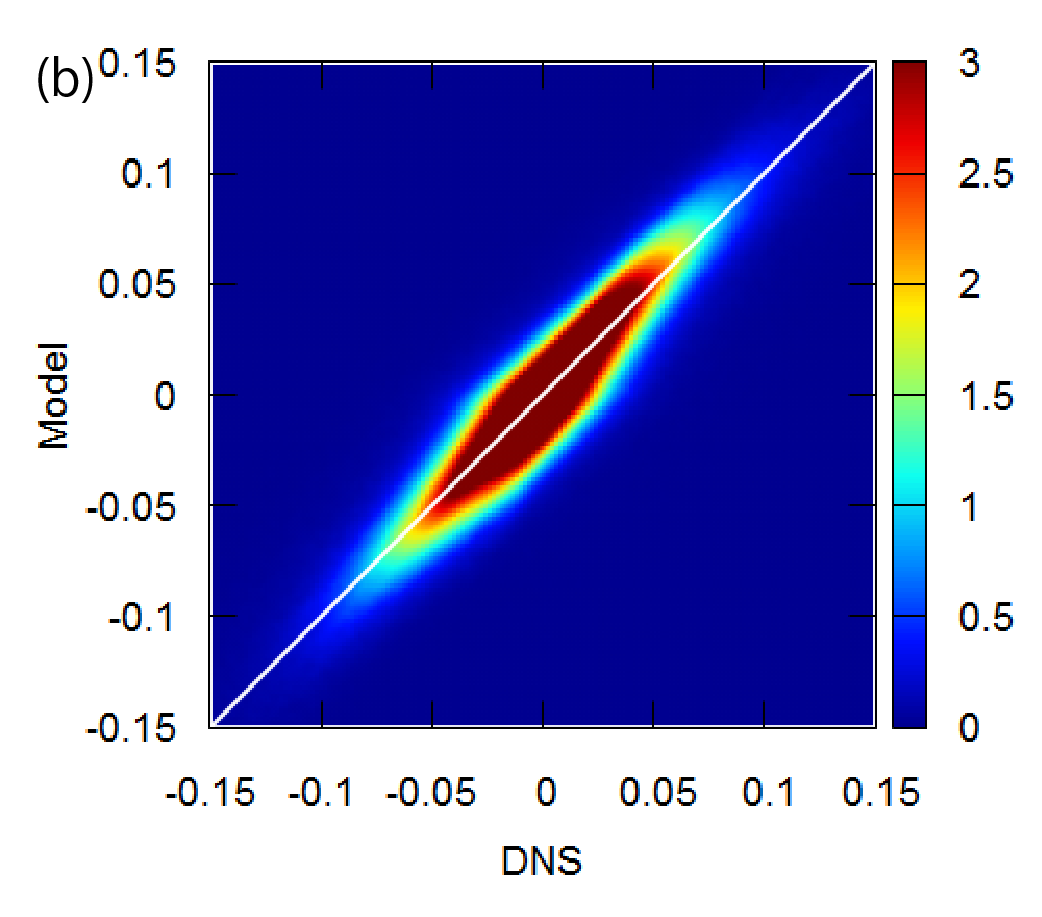}
   \includegraphics[width=55mm]{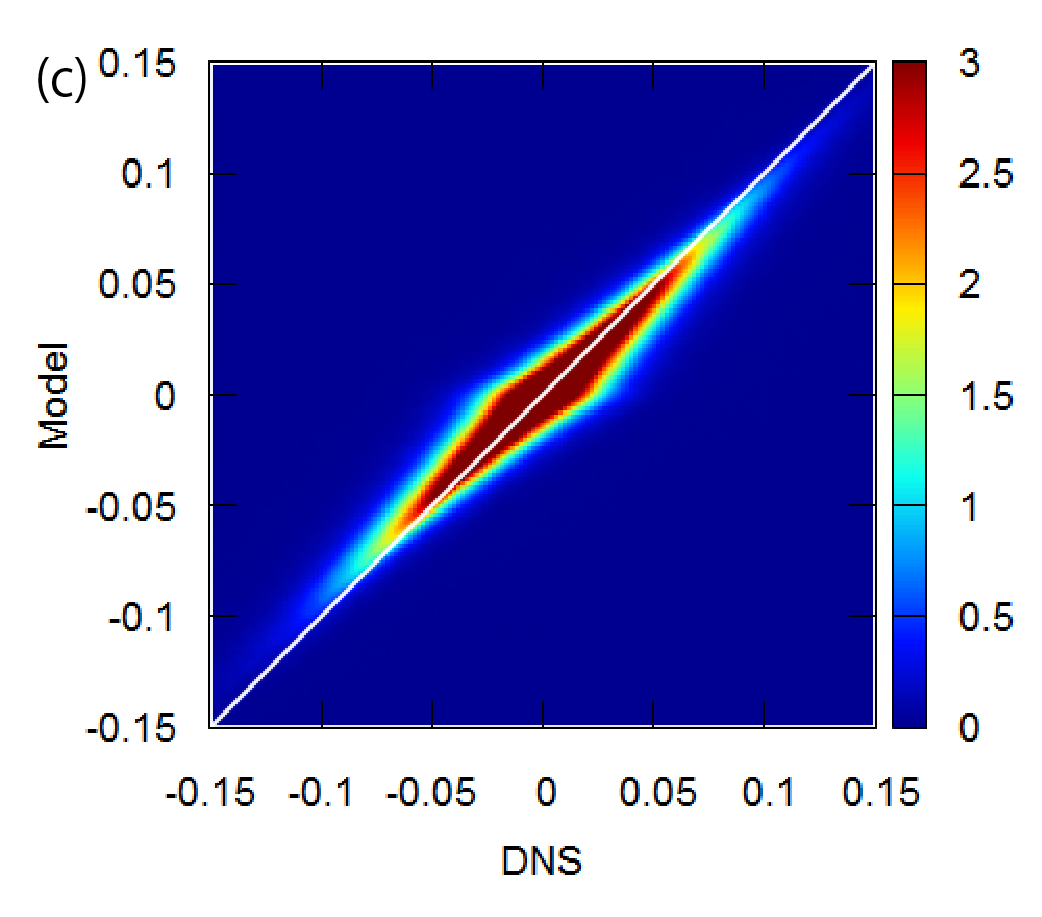}

   \includegraphics[width=55mm]{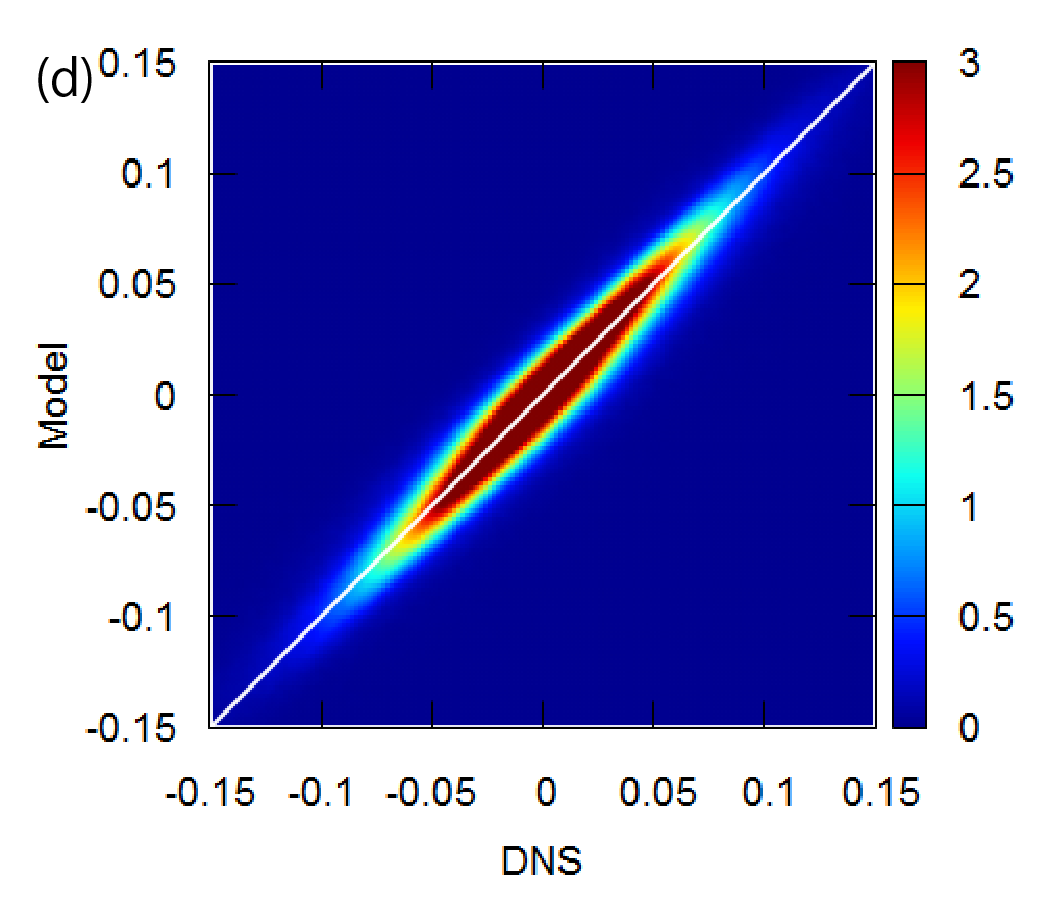}
   \includegraphics[width=55mm]{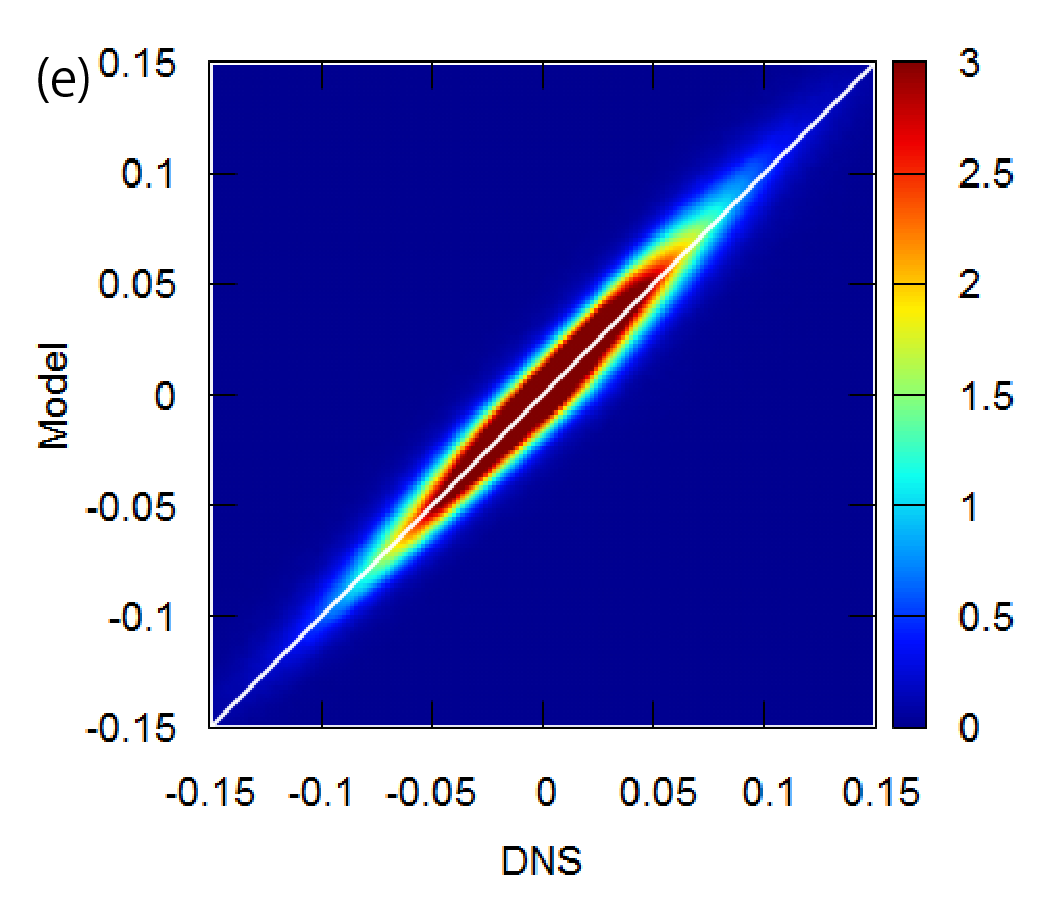}
   \includegraphics[width=55mm]{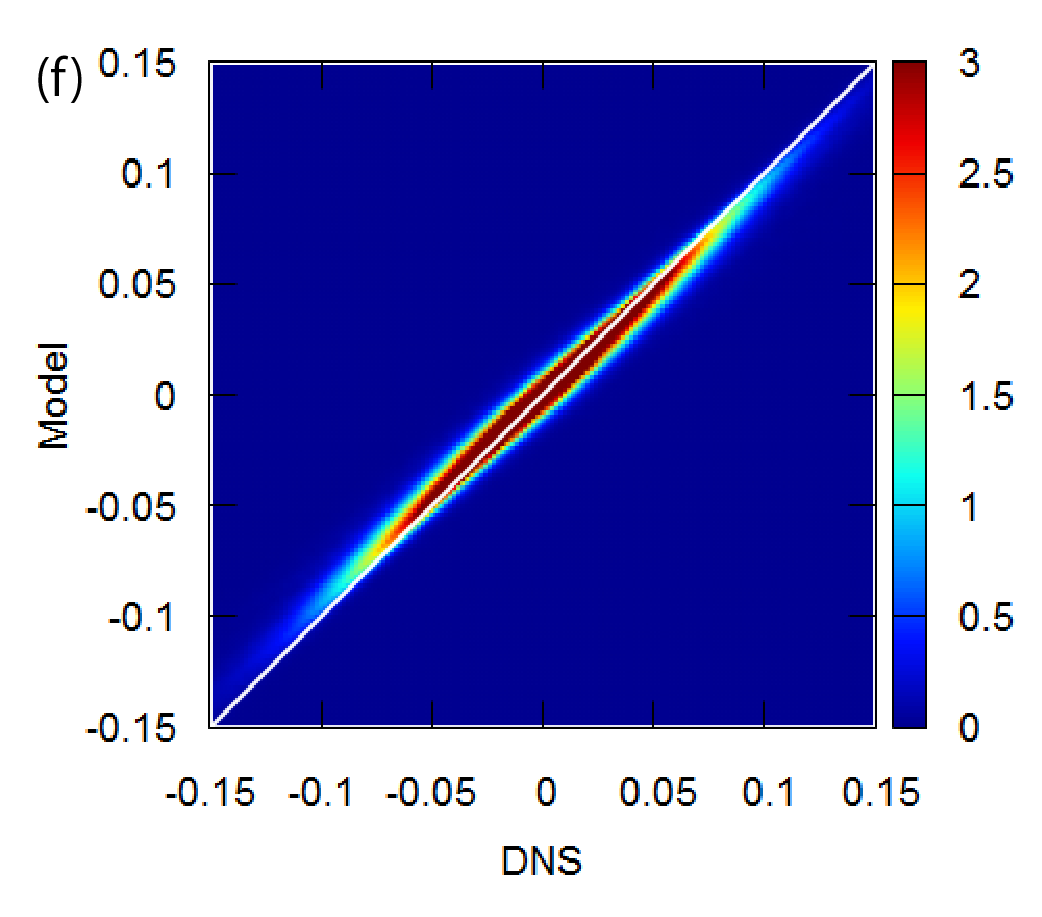}
  \end{center}

 \caption{Joint p.d.f. of SGS stress $\tau_{12}$,\,Case 1,\,$\overline{\Delta}=12.2\eta$. 
The horizontal axis is the exact SGS stress obtained by filtering the DNS data. 
The vertical axes are predictions by (a) NN-D1, 
(b) NN-G1, (c) GM, (d) NN-D2, (e) NN-G2, (f) EGM.
}
 \label{fig:SGSjPDF512-8n}
\end{figure}

Figures \ref{fig:SGSjPDF512-8d} and \ref{fig:SGSjPDF512-8n} compare 
joint probability density functions (joint p.d.f.s) 
of the exact SGS stress and the prediction by neural networks 
for different sets of the input variables. 
The joint p.d.f.s of the exact SGS stress and the gradient and extended gradient models 
are also shown.  
The figures show that 
addition of the second-order derivatives makes the distributions more concentrated 
near the diagonal lines of perfect match. 
The joint p.d.f.s for NN-G1 and NN-G2 are more concentrated 
than those for NN-D1 and NN-D2, respectively, 
in accordance with the results on correlation.  
The distributions of $\tau_{11}$ for the gradient and extended gradient models 
are below the diagonal lines 
showing that these models underestimate $\tau_{11}$; 
the similar but weaker trend is observed for the magnitude of $\tau_{12}$.  
This underestimate is attributed to absence of the higher-order terms which are positive for the 
diagonal components $\tau_{ii}$. 
On the other hand, the distributions for the neural networks 
are nearly symmetric with respect to the diagonal lines.

\begin{figure}
  \begin{center}
   \includegraphics[width=55mm]{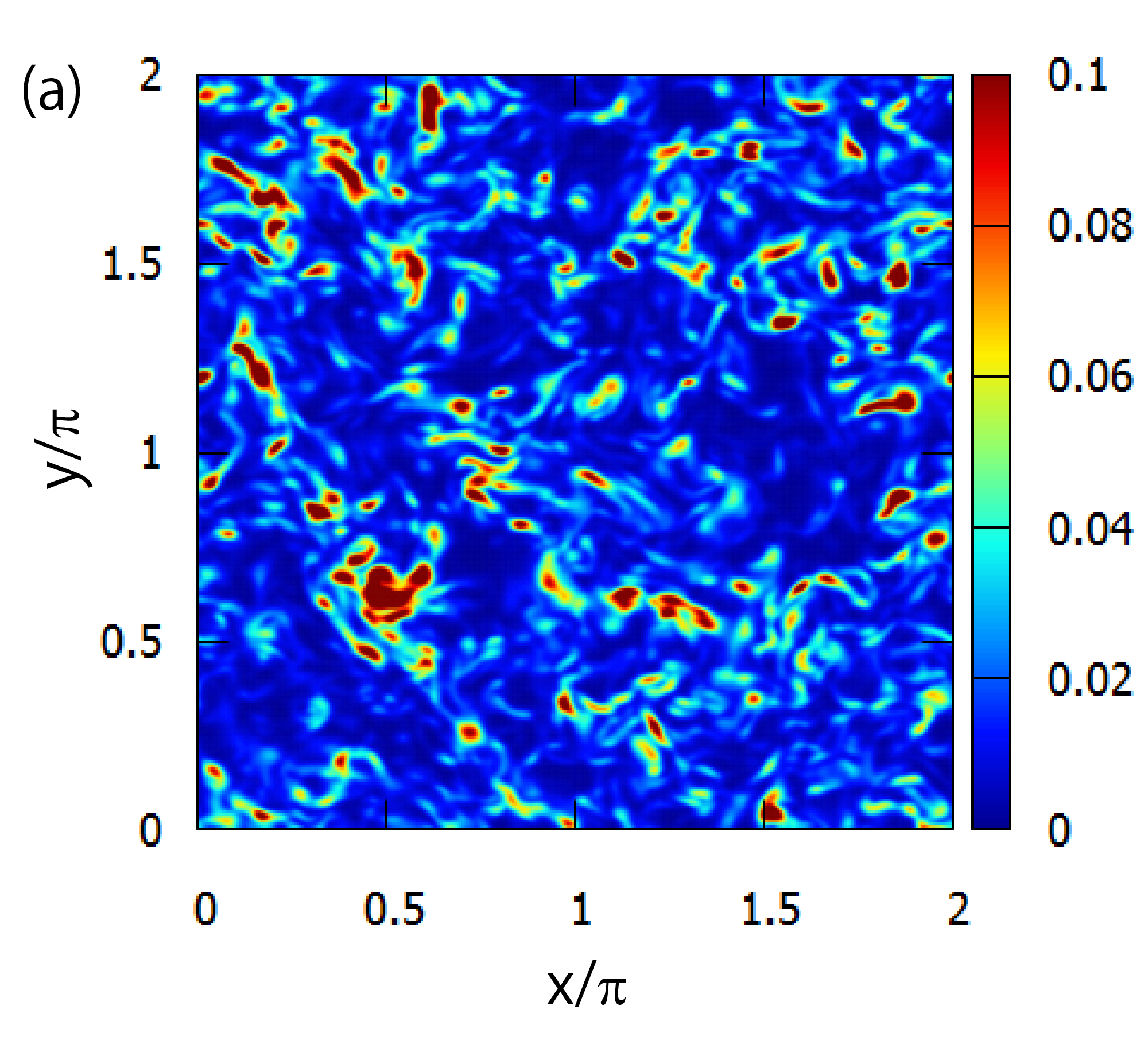}

   \includegraphics[width=55mm]{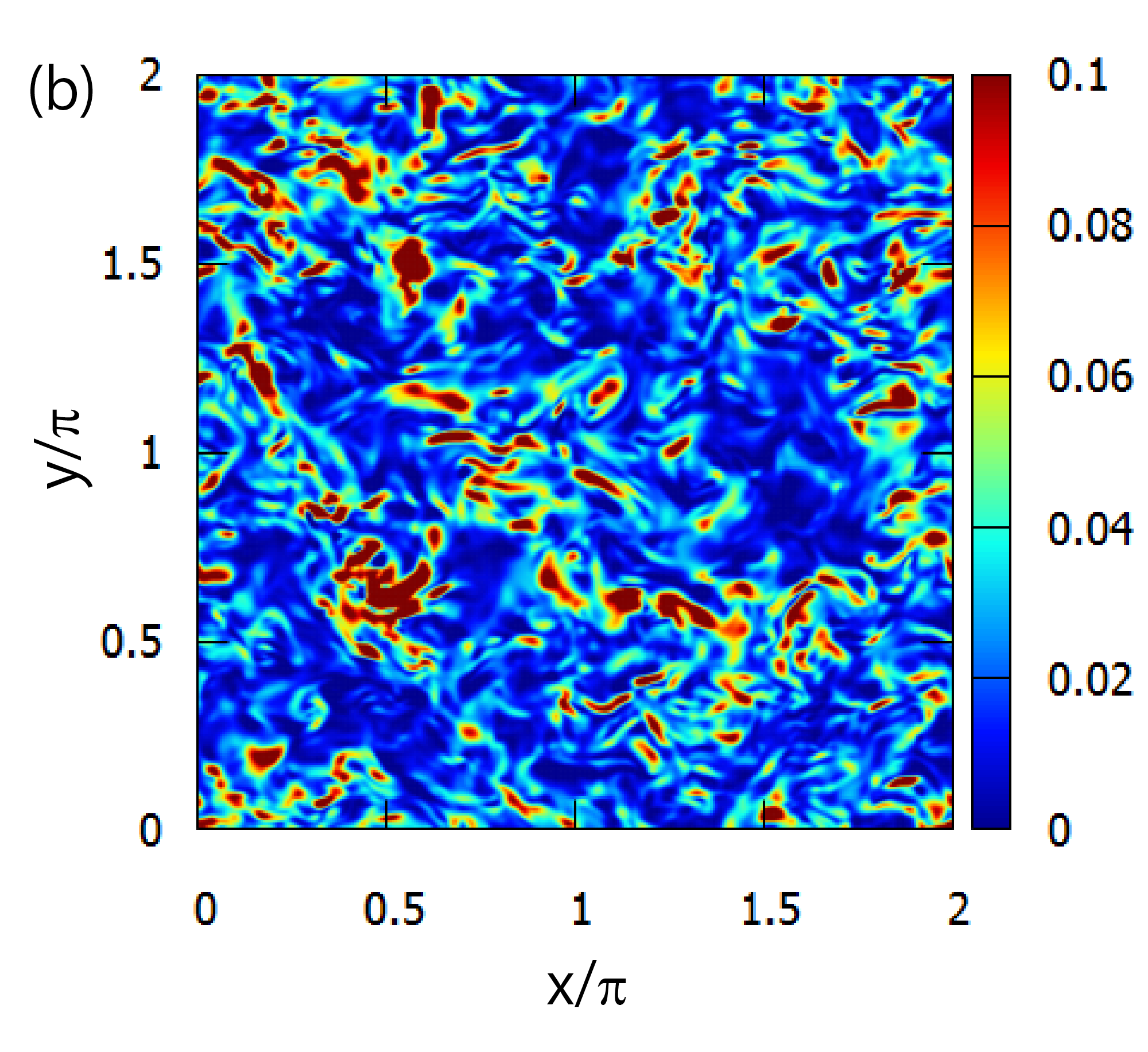}
   \includegraphics[width=55mm]{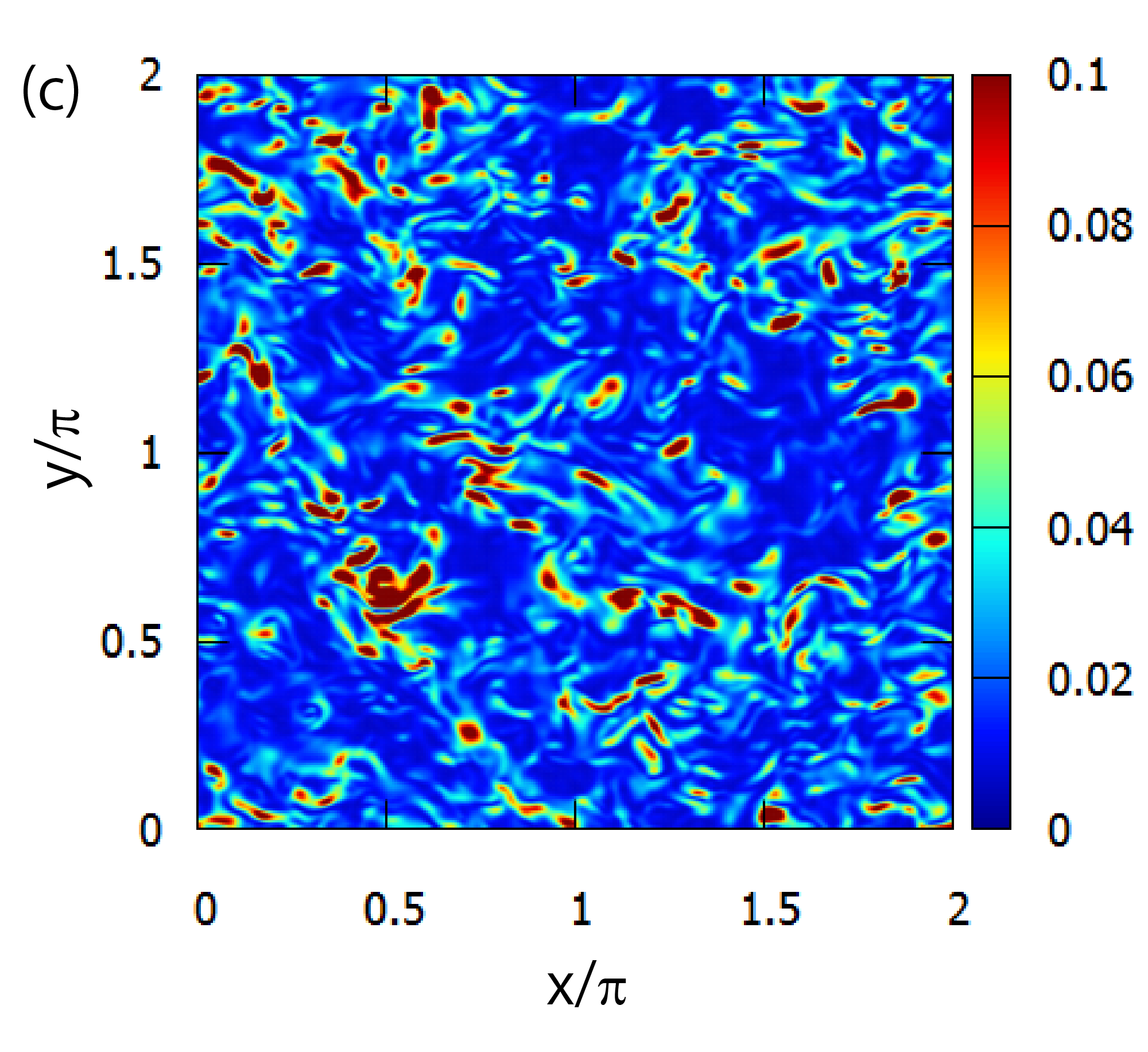}
   \includegraphics[width=55mm]{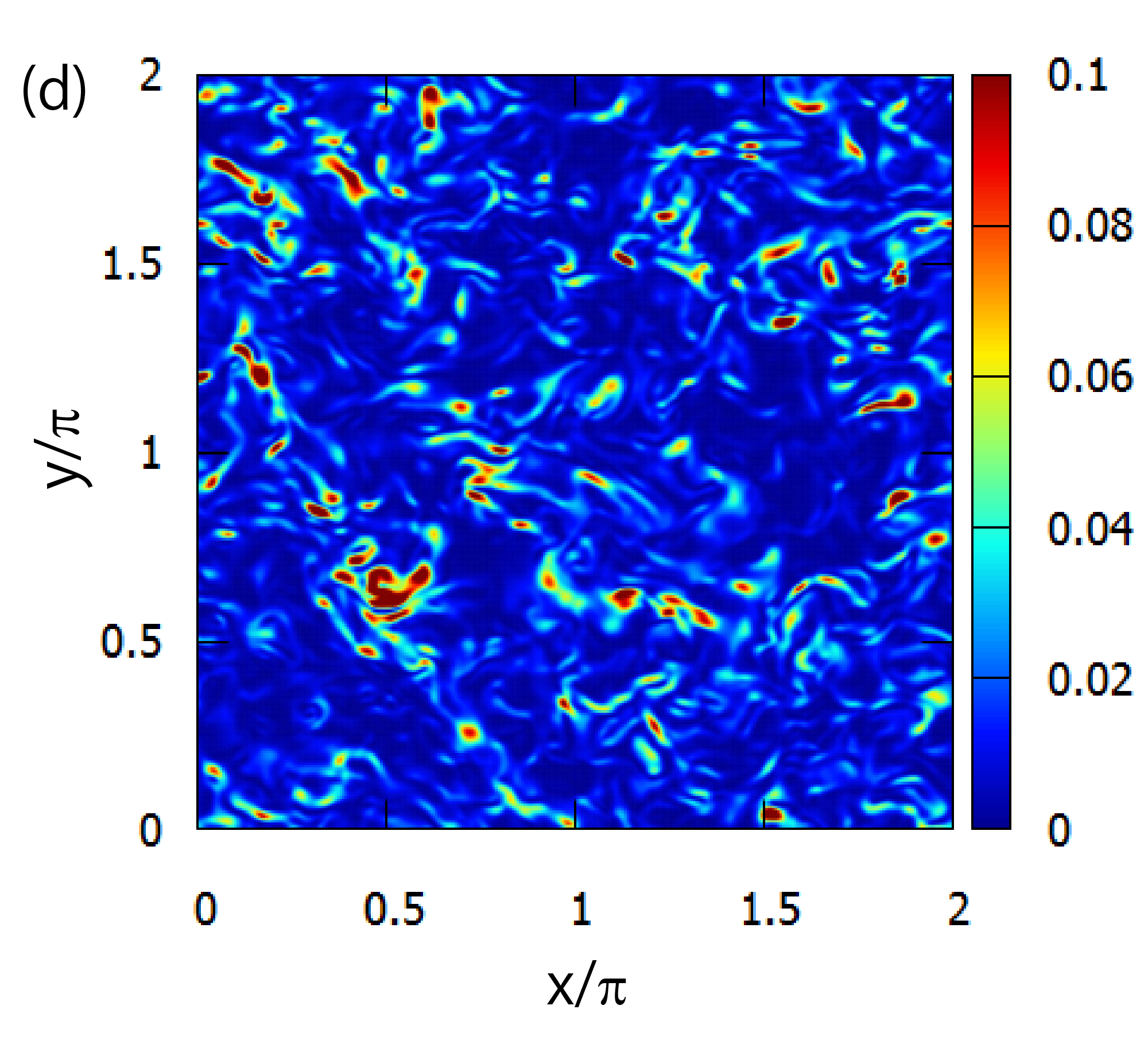}

   \includegraphics[width=55mm]{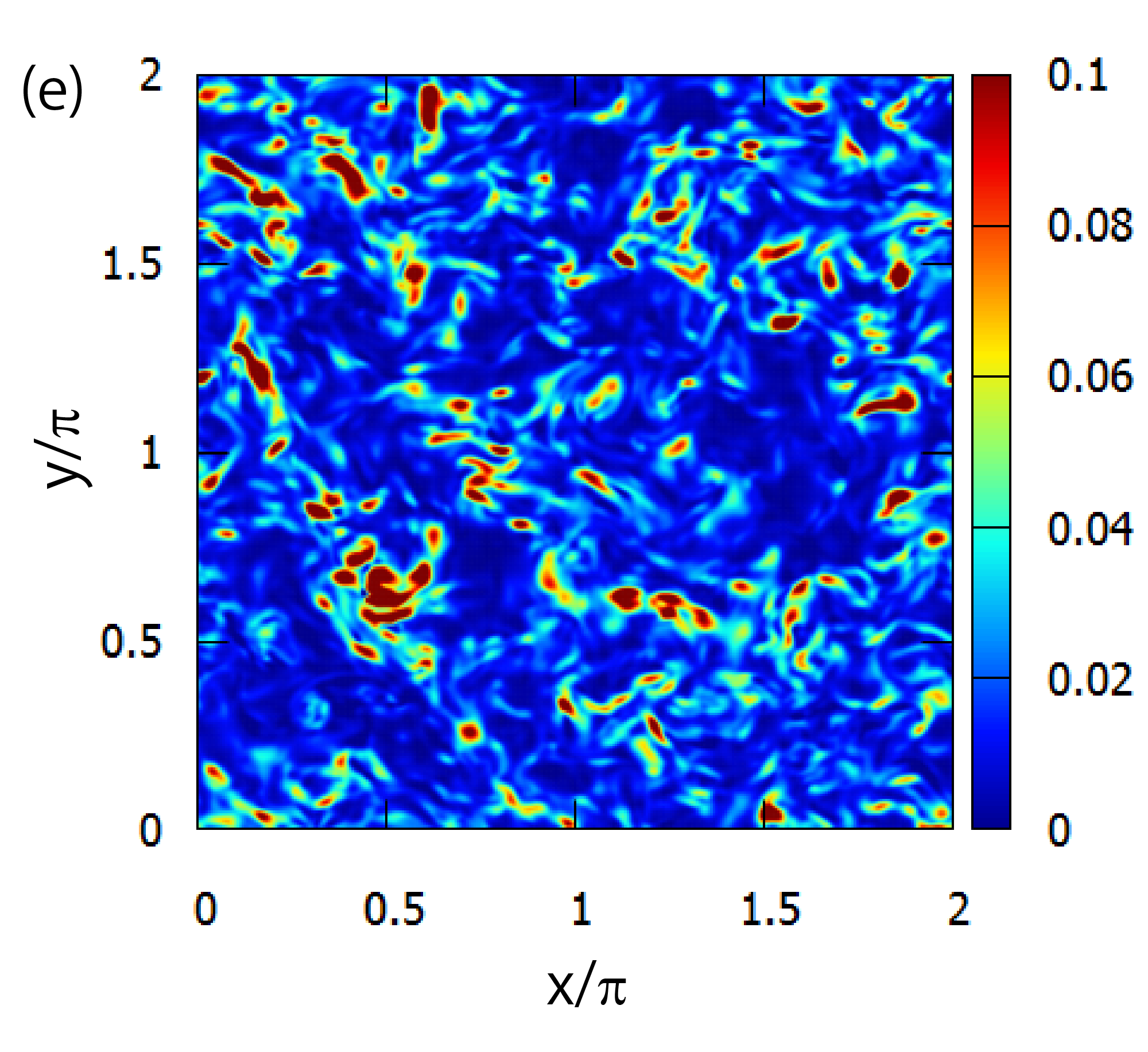}
   \includegraphics[width=55mm]{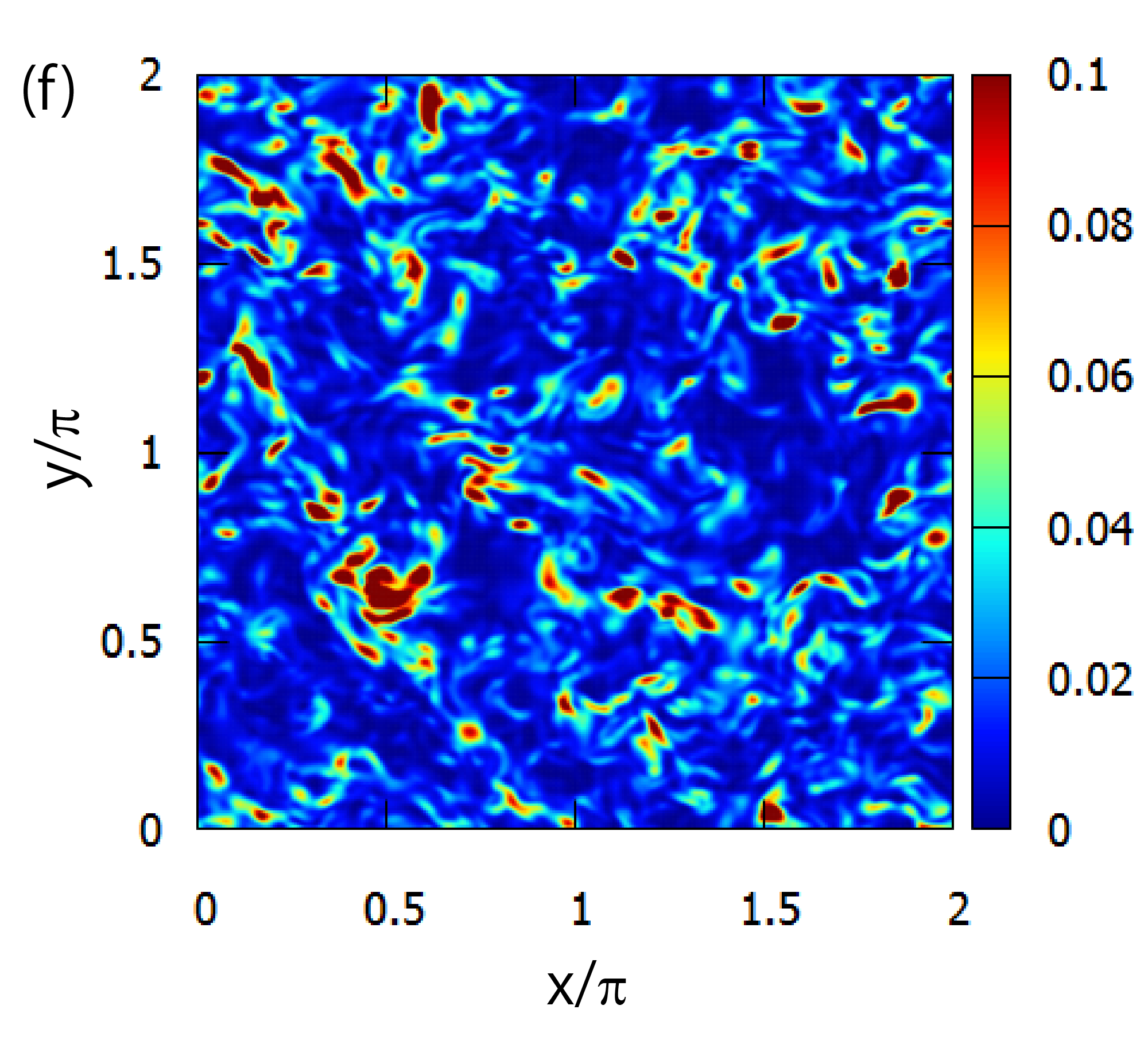}
   \includegraphics[width=55mm]{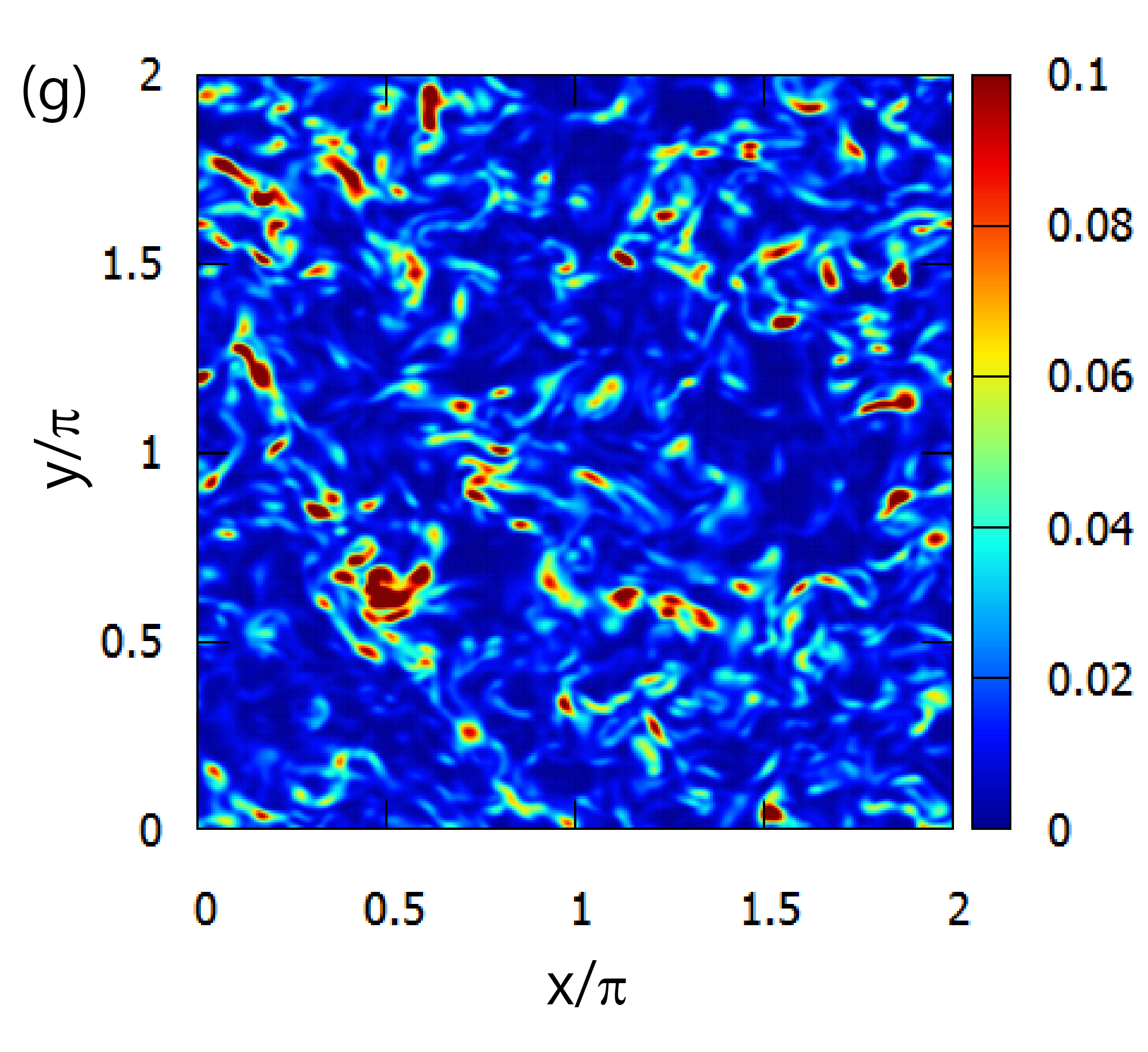}
  \end{center}

 \caption{Spatial distribution of SGS stress $\tau_{11}$ on $z=0$.  Case 1,\,$\overline{\Delta}=12.2\eta$. 
(a) (Filtered) DNS, (b) NN-D1, 
(c) NN-G1, (d) GM, (e) NN-D2, (f) NN-G2, (g) EGM.}
 \label{fig:apriori512-8d}
\end{figure}

\begin{figure}
  \begin{center}
   \includegraphics[width=55mm]{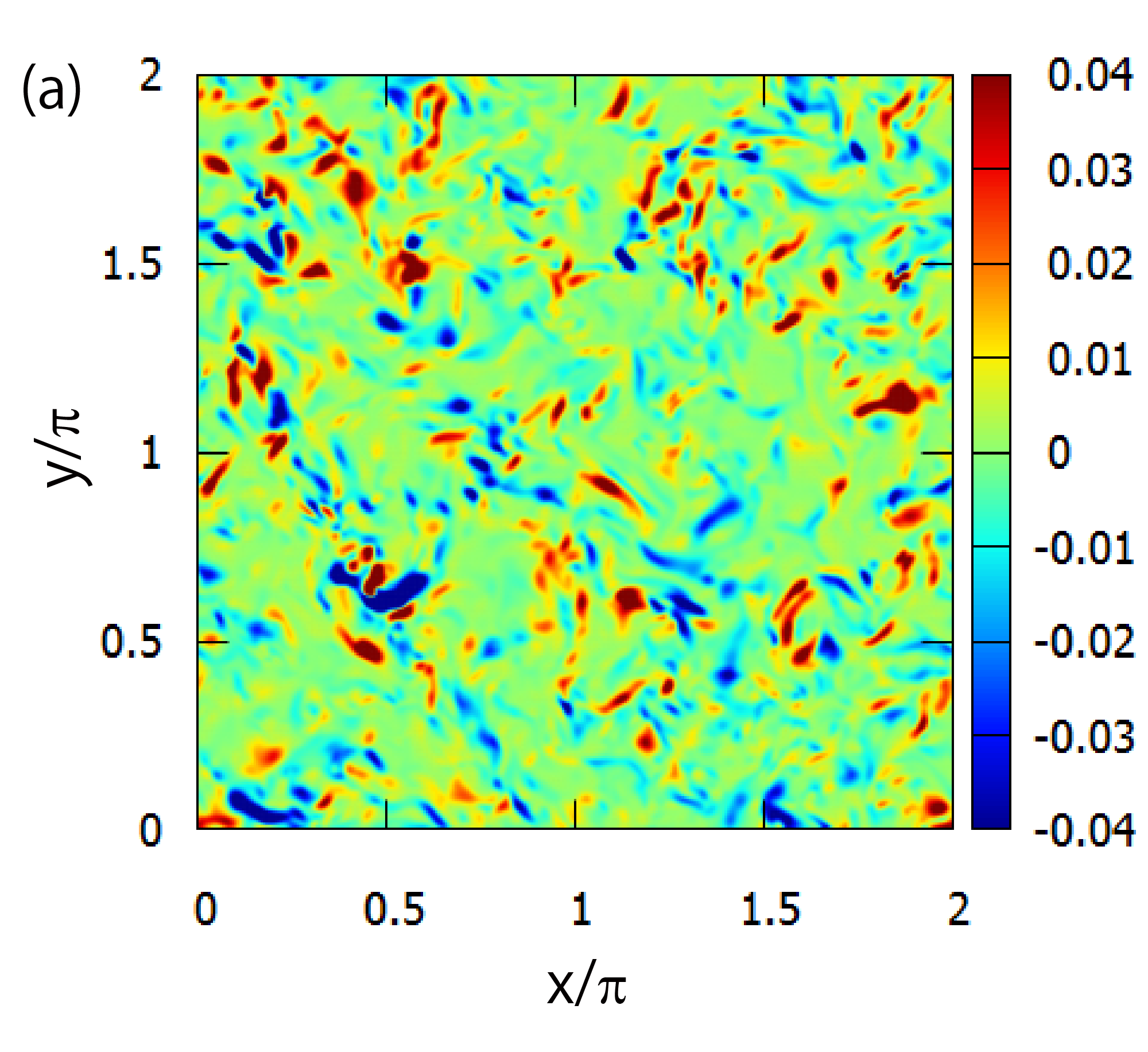}

   \includegraphics[width=55mm]{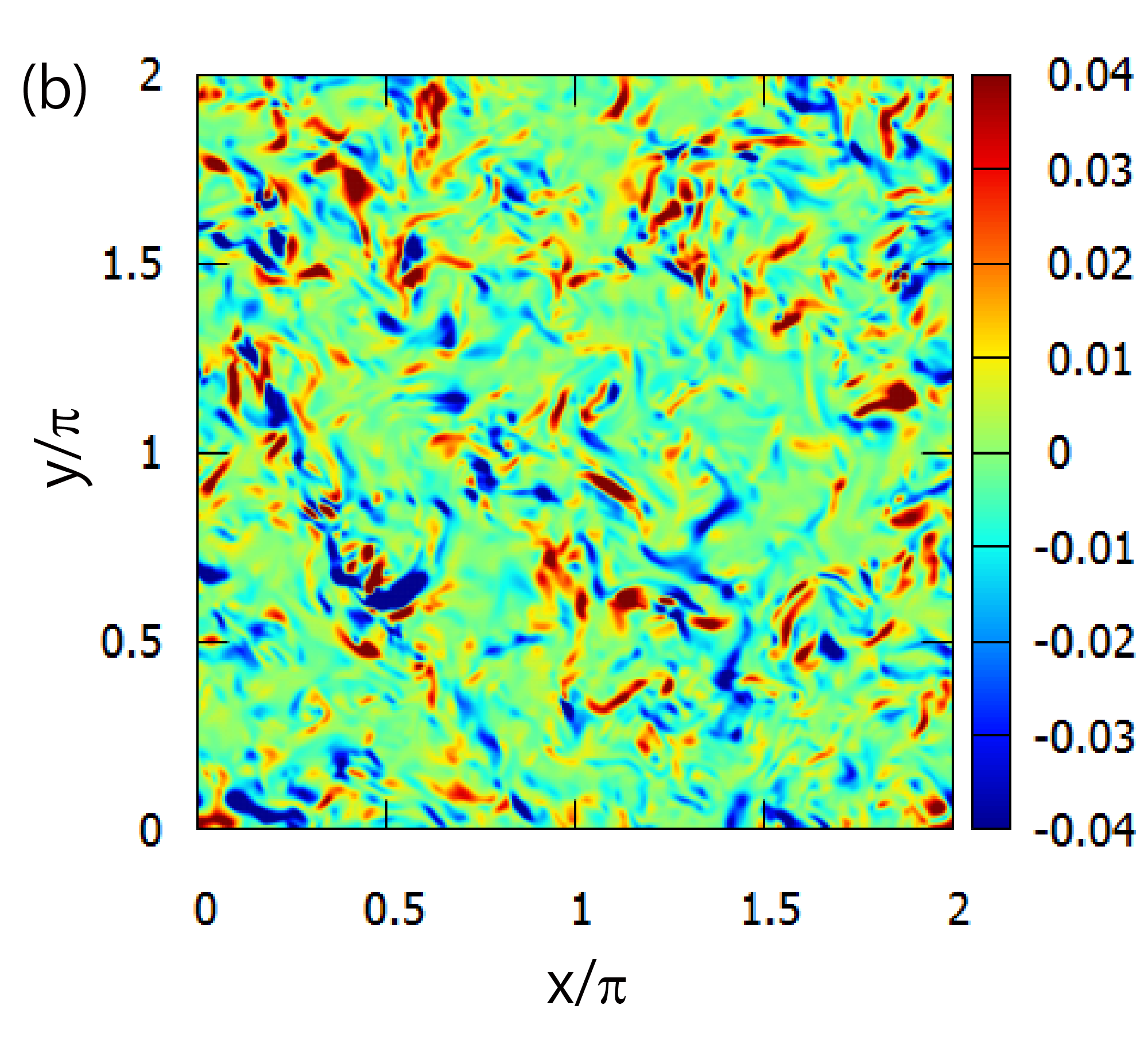}
   \includegraphics[width=55mm]{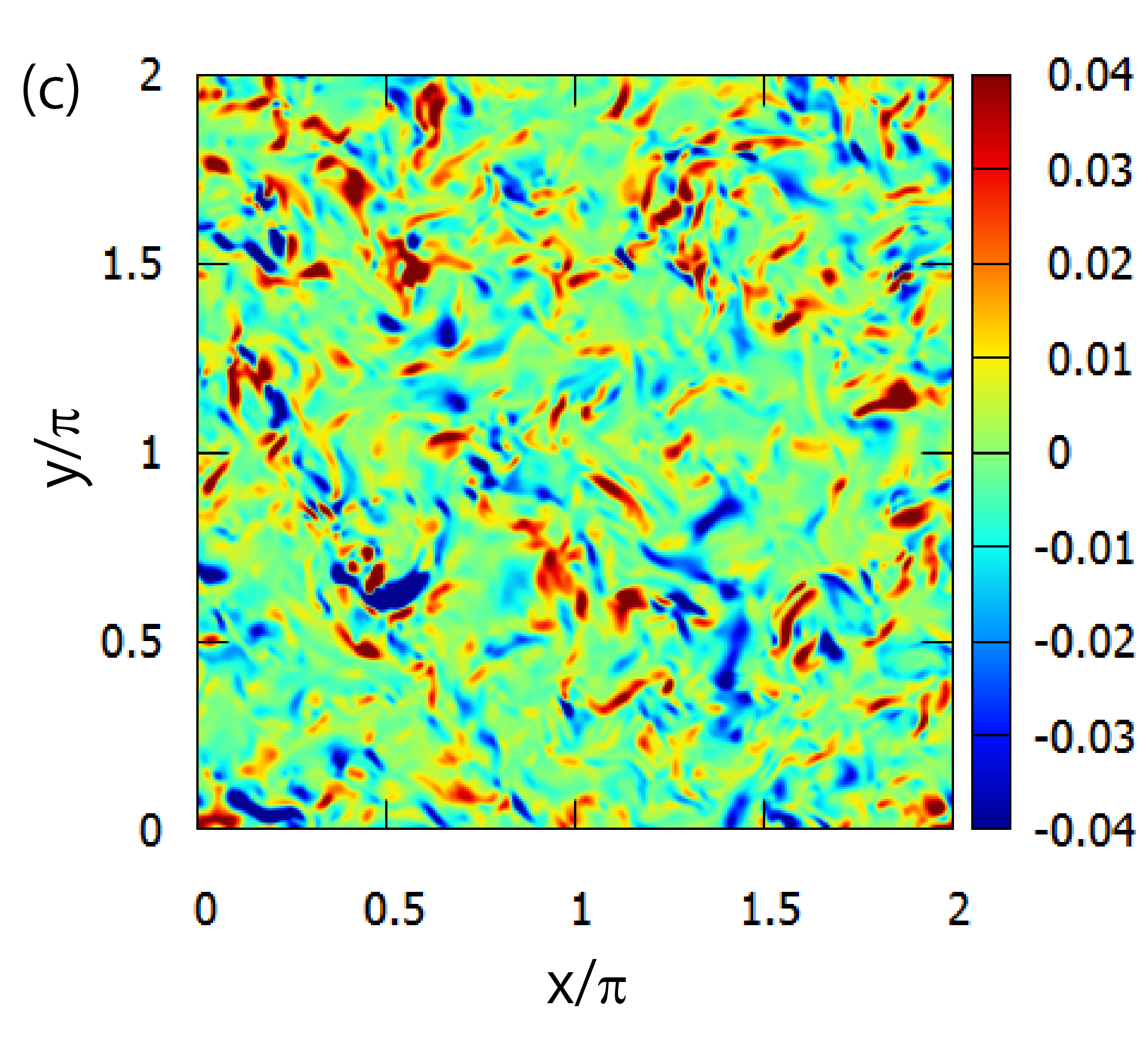}
   \includegraphics[width=55mm]{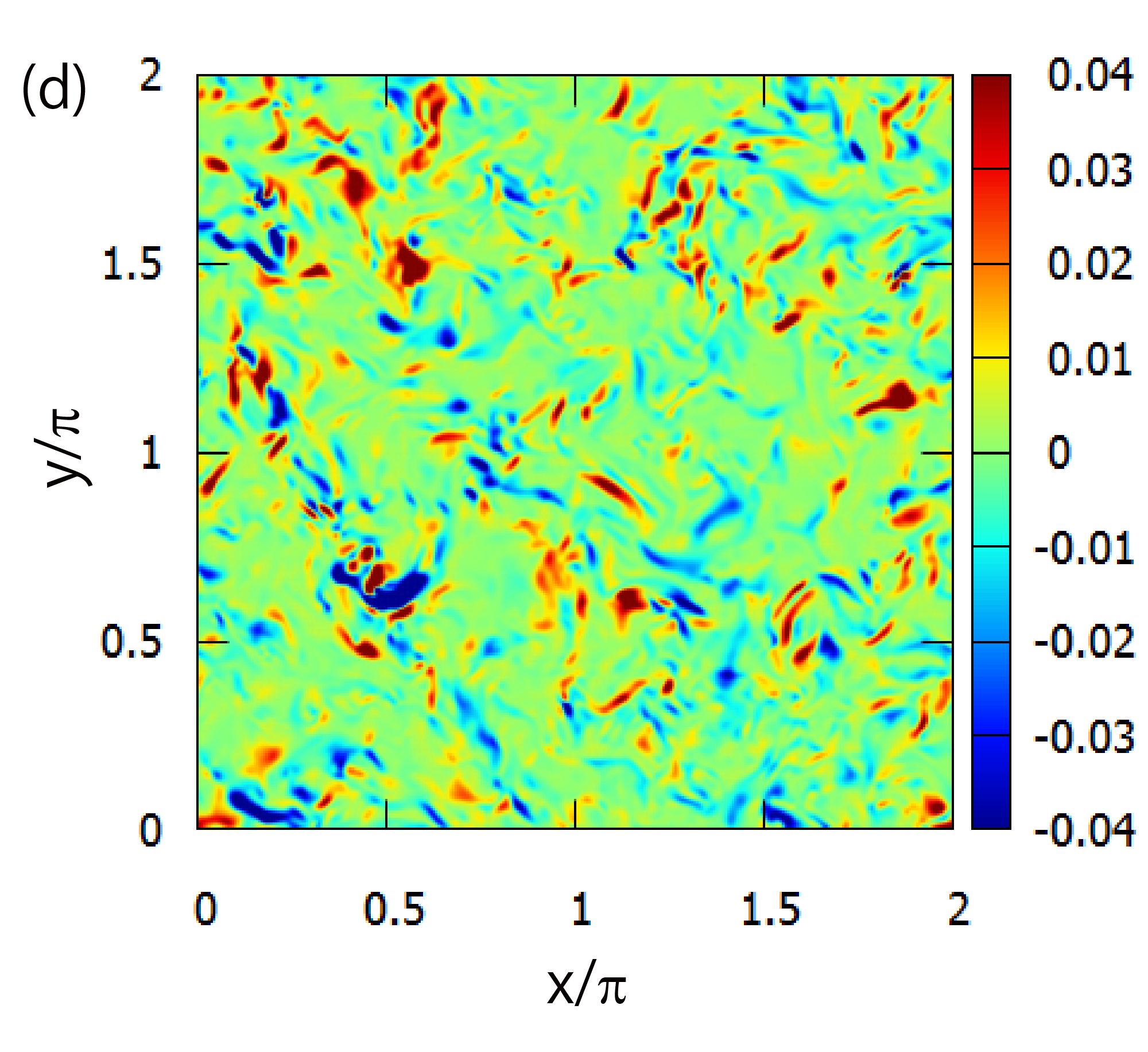}

   \includegraphics[width=55mm]{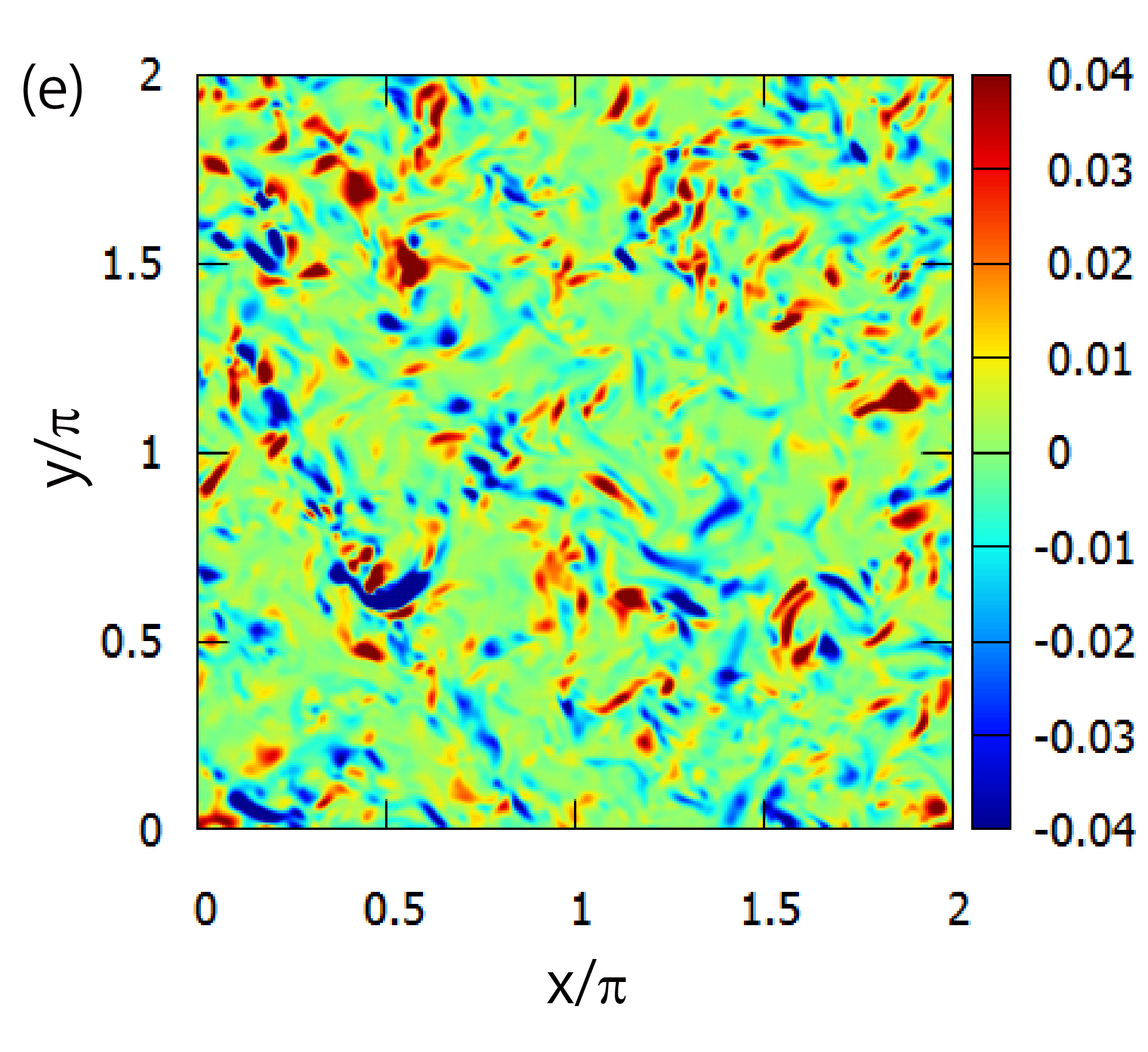}
   \includegraphics[width=55mm]{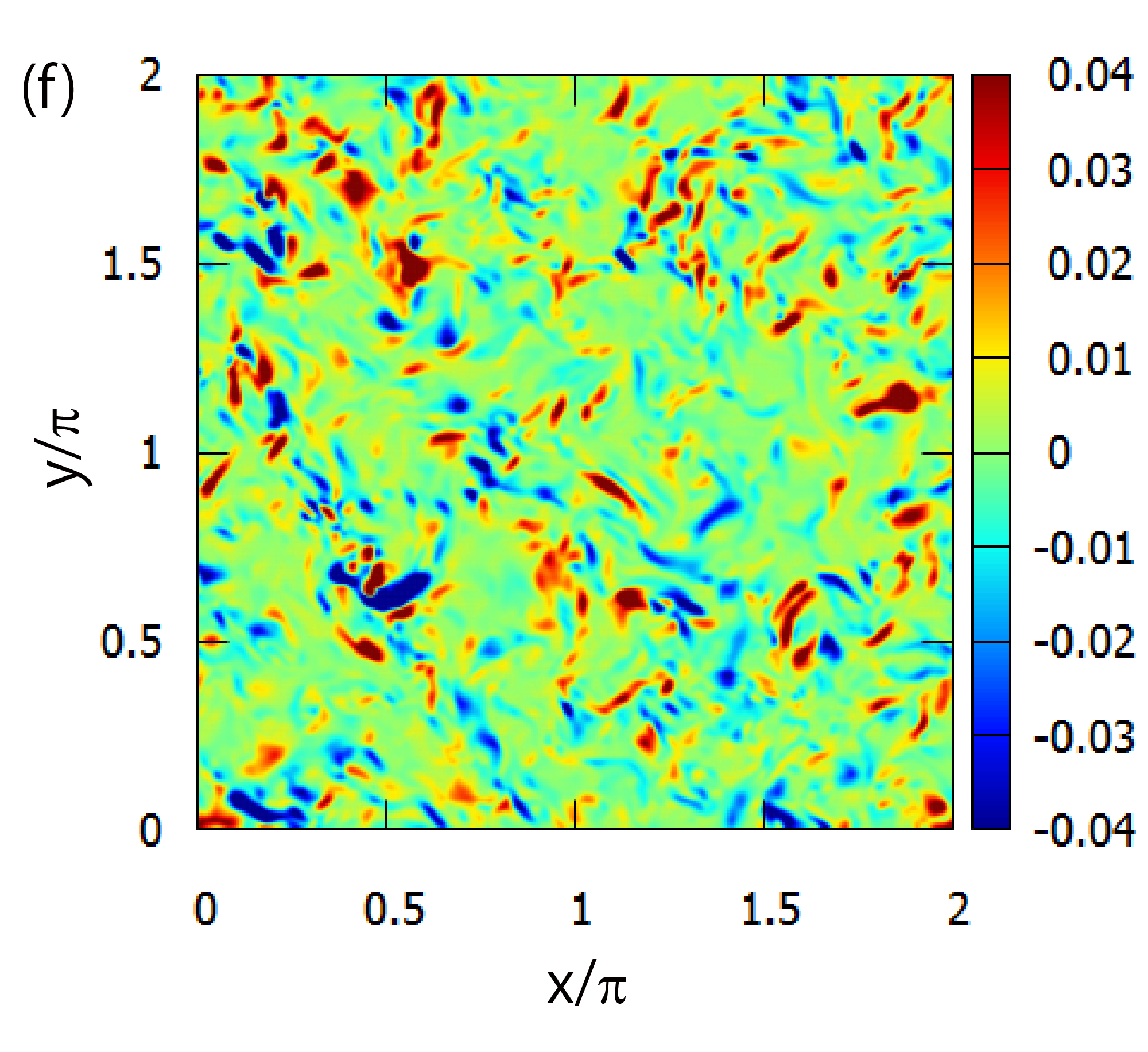}
   \includegraphics[width=55mm]{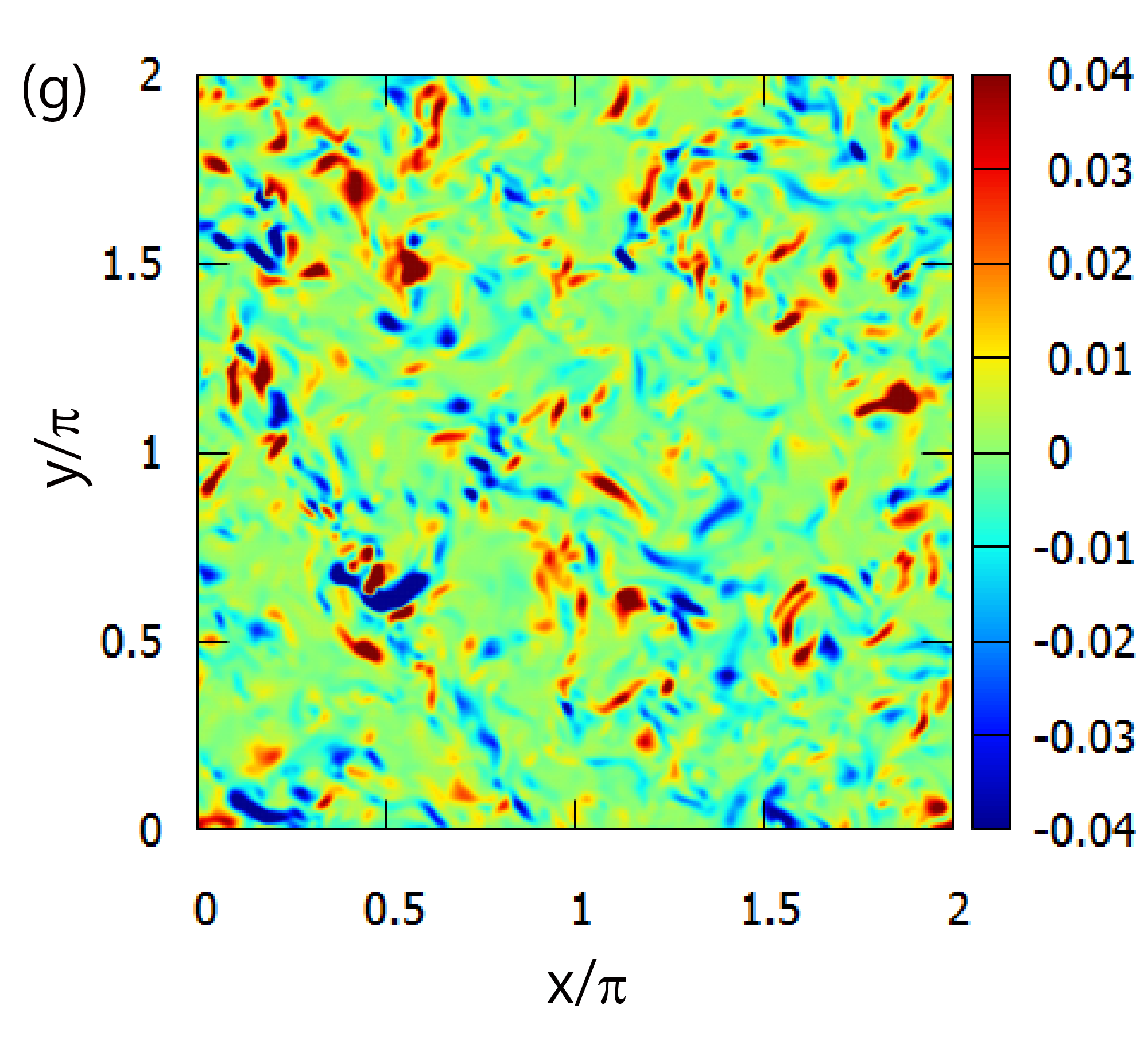}
  \end{center}

 \caption{Spatial distribution of SGS stress $\tau_{12}$ on $z=0$. 
Case 1,\,$\overline{\Delta}=12.2\eta$. 
(a) (Filtered) DNS, (b) NN-D1, 
(c) NN-G1, (d) GM, (e) NN-D2, (f) NN-G2, (g) EGM.}
 \label{fig:apriori512-8n}
\end{figure}

Figures \ref{fig:apriori512-8d} and \ref{fig:apriori512-8n} show 
instantaneous distributions of the SGS stress components $\tau_{11}$ and $\tau_{12}$ on $z=0$. 
The exact SGS stress, 
the prediction by the neural networks (NN-D1, NN-G1, NN-D2, and NN-G2), 
and the gradient and the extended gradient models are compared. 
All distributions are similar to the exact distribution; 
however, the prediction by NN-D1 involves small-scale structures 
which do not exist in the exact distribution, 
while similar structures are also visible for NN-G1. 
The distribution of the gradient model looks quite similar to the exact distribution, 
although the values are smaller than those of the exact one 
as the gradient model underestimates the SGS stress. 
On the other hand, NN-D2 and NN-G2 and the extended gradient model 
give nearly perfect prediction as the distributions are difficult to distinguish 
with each other. 
These results are in accordance with those on the correlation 
and the joint p.d.f.s. 

\subsection{Dependence on filter width}
\label{sec-flsize}

\begin{figure}[htb]
  \begin{center}
   \includegraphics[width=70mm]{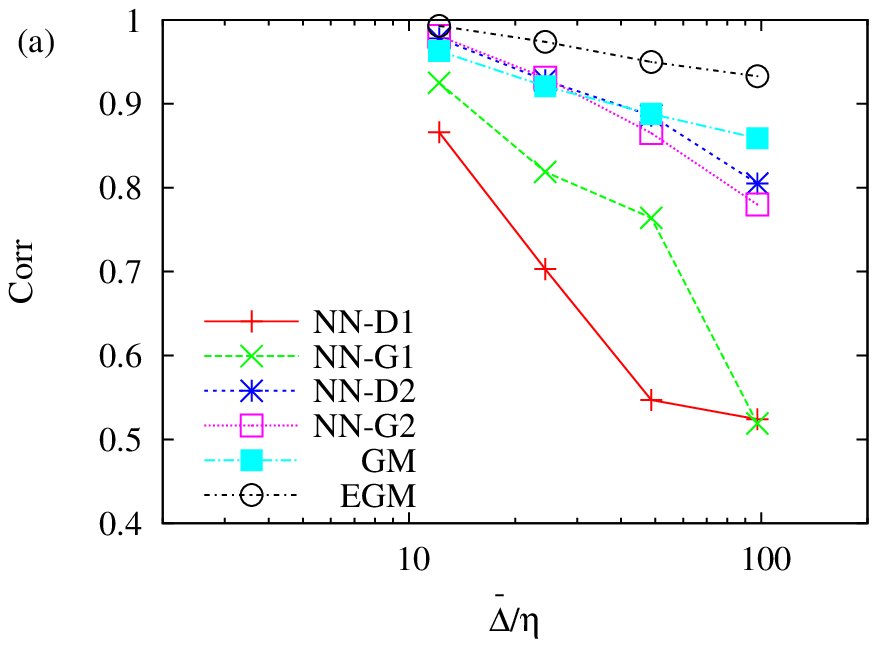}
   \includegraphics[width=70mm]{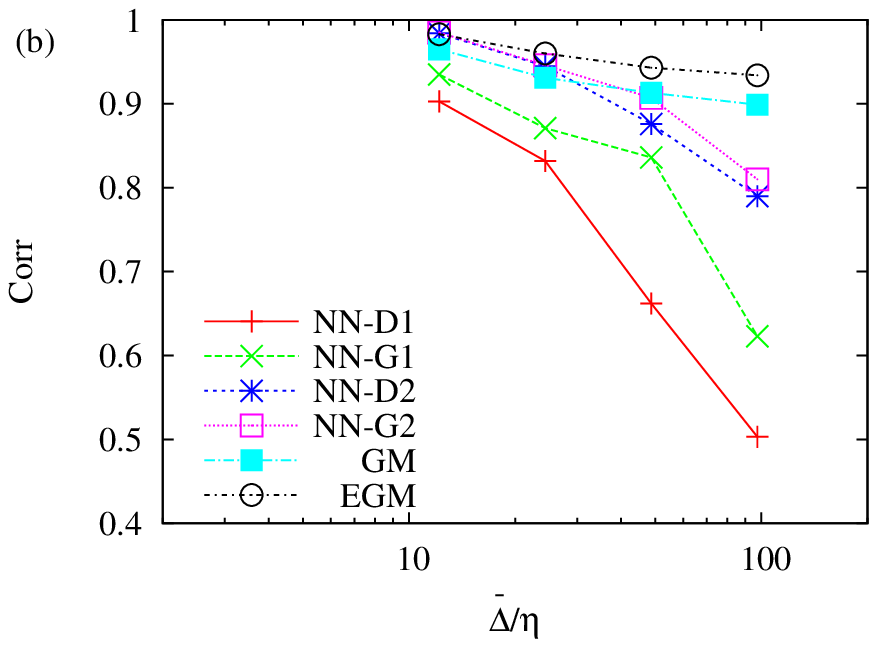}

   \includegraphics[width=70mm]{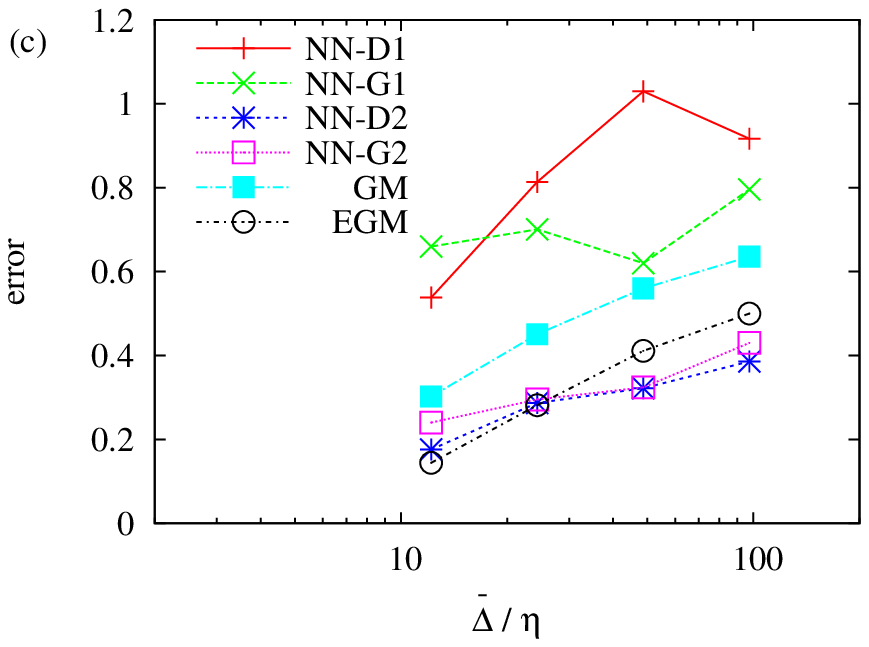}
   \includegraphics[width=70mm]{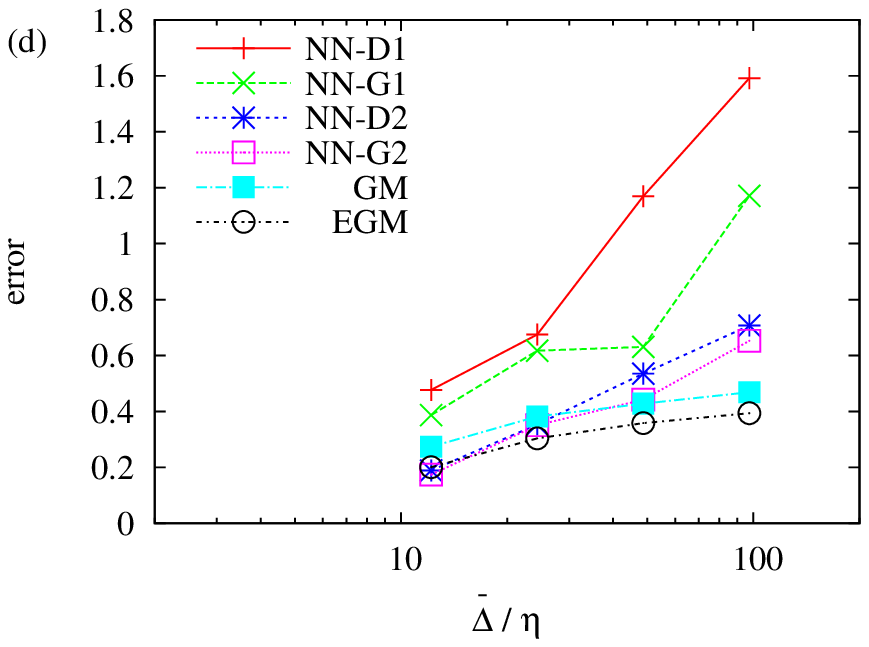}
  \end{center}
 \caption{(a,b) Correlation coefficients 
between exact SGS stress and prediction by neural network
and (c,d) error;  
those between the exact SGS stress and the gradient and the extended 
gradient models are included. 
Dependence on the filter size is shown. 
(a,c) $\tau_{11}$, (b,d) $\tau_{12}$. 
}
 \label{fig:CorrInput}
\end{figure}

Next we investigate how the filter width $\overline{\Delta}$ affects the 
regression ability of the neural network. 
The neural networks of Sets D1, G1, D2, and G2 are considered, 
while the gradient and the extended gradient models are included for comparison. 
The range of the filter width is 
$8\Delta_{\rm{DNS}}=12.2\eta \le \overline{\Delta} \le 64\Delta_{\rm{DNS}}=97.4\eta$. 
In Fig.~\ref{fig:CorrInput} the correlation coefficients 
and the error are plotted against the filter width. 
Naturally correlation becomes weak as the filter width becomes large, 
while the errors increase with the filter width. 
The correlation coefficients decrease rapidly 
for NN-D1 and NN-G1 which use only the first derivative of velocity. 
For the gradient and the extended gradient models, 
correlation is still high for the largest filter width,  
Corr being $0.94$ at $\overline{\Delta}=64\Delta_{\rm{DNS}}$ for the latter. 

On the other hand, 
the error of $\tau_{11}$ is smaller for NN-D2 and NN-G2 
than for the extended gradient model;  
on the whole, the error is not much different between NN-D2, NN-G2, 
the gradient model, and the extended gradient model. 
As observed in previous subsections, 
the gradient and the extended gradient models underestimate 
the magnitude of the SGS stress, 
which explains why the error is not so small 
in spite of the high correlation. 

As we see in the next subsection, 
it is most likely that the model constructed by the neural networks (NN models) 
is close to (but not identical with) the gradient 
and the extended gradient models; 
one of the advantages of the neural networks is 
to correct the trend of underestimation of the gradient models by training. 

\subsection{Comparison between NN model and gradient model}
\label{sec-grad}
 
The results so far suggest that the models constructed by neural networks 
are close to the gradient and the extended gradient models. 
There are three points which support it: 
(i) the prediction by NN-G1 and NN-G2  
is better than NN-D1 and NN-D2, respectively, 
which implies that the prediction is improved by 
removing the input variables which do not appear 
in the gradient or the extended gradient models 
from Set D1 or Set D2;   
(ii) adding second-order derivatives to the input variables of the neural network 
improves prediction significantly, 
which is the case for the extended gradient model 
compared to the gradient model; 
(iii) it is most likely that the trained neural network 
has simple structures since the number of neurons in the hidden layer 
required for high correlation is not so large in spite of the shallow structure of the neural network.  
 
\begin{figure}
  \begin{center}
   \includegraphics[width=55mm]{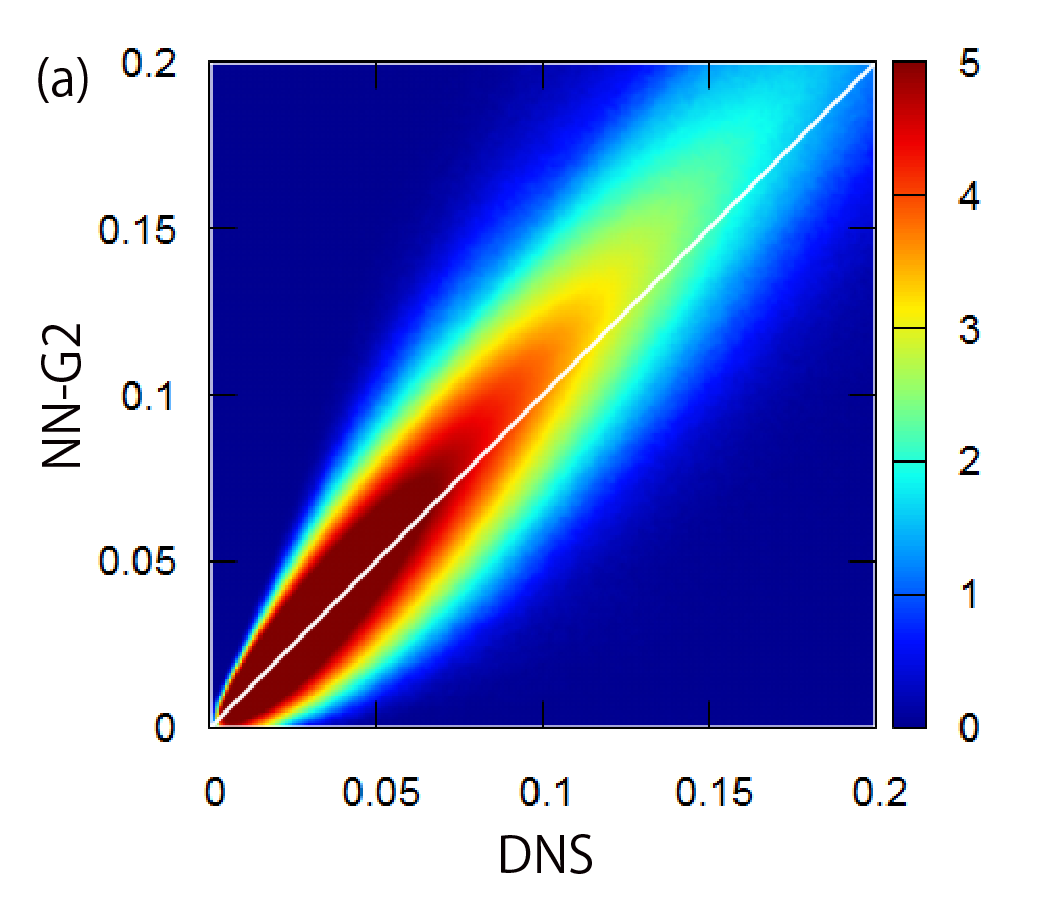}
   \includegraphics[width=55mm]{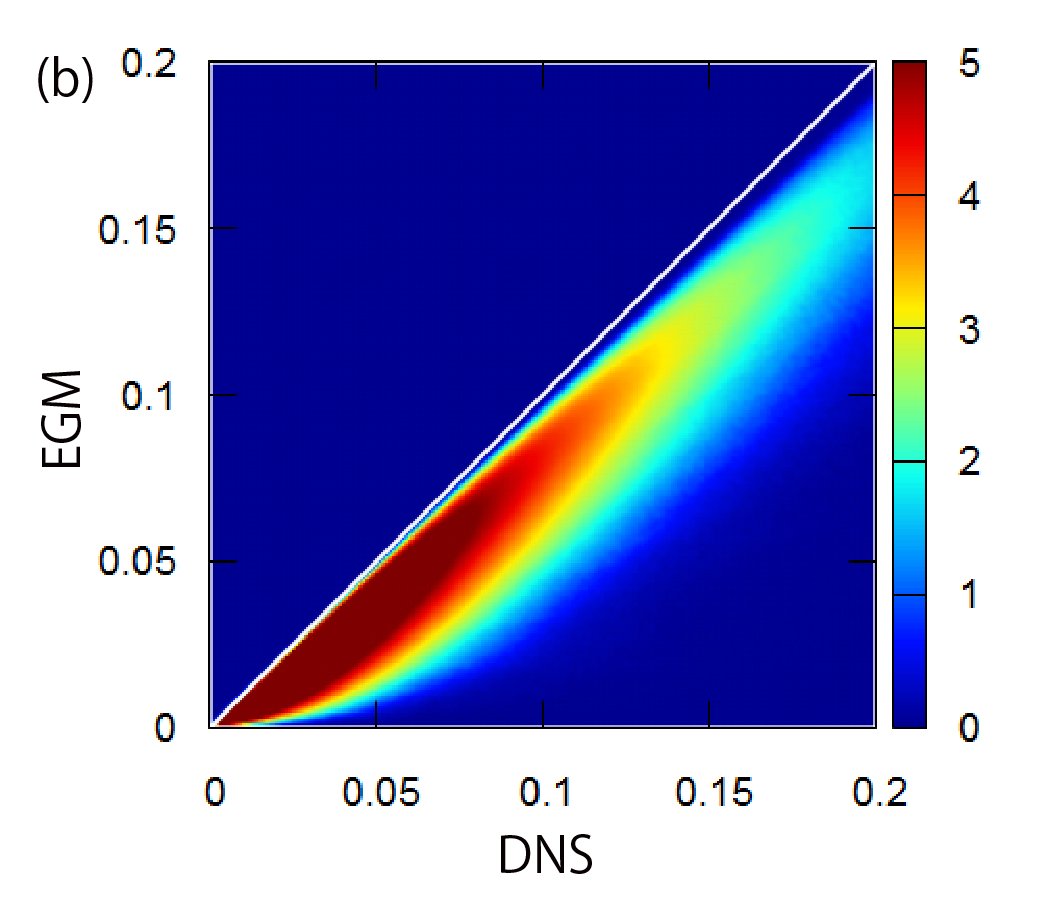}
   \includegraphics[width=55mm]{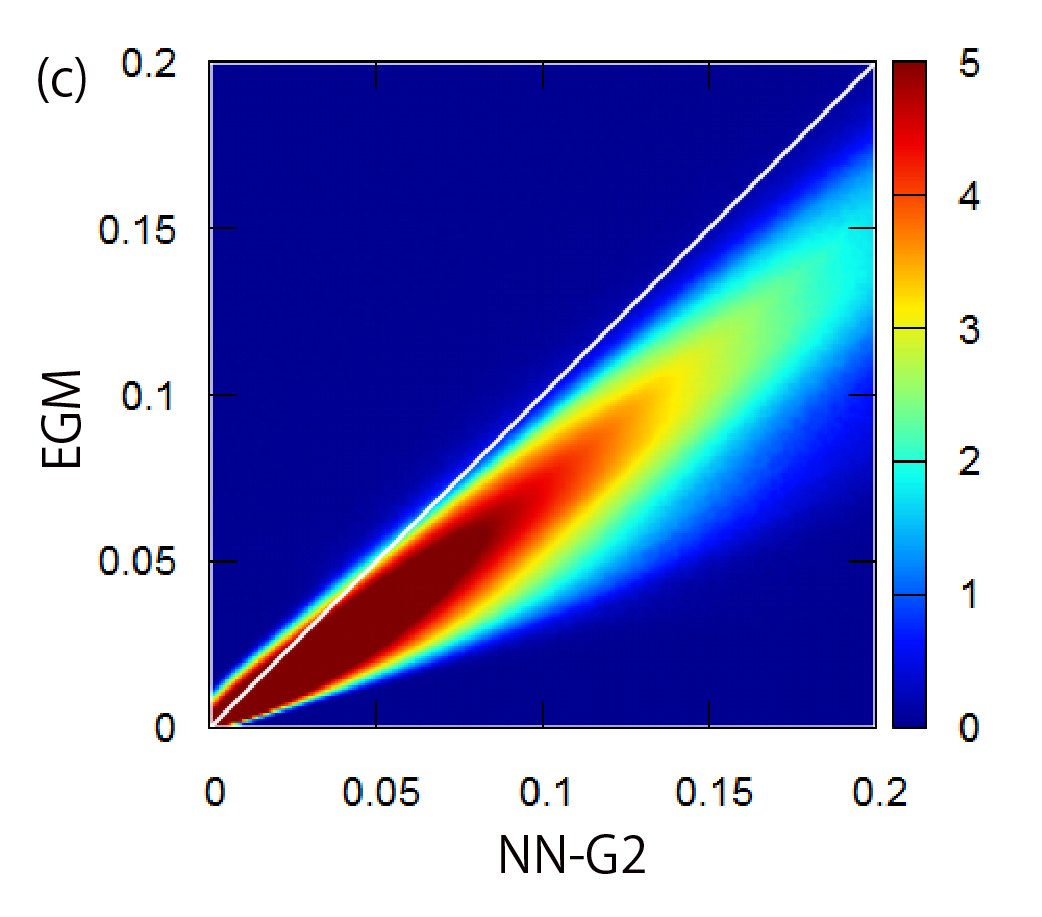}

   \includegraphics[width=55mm]{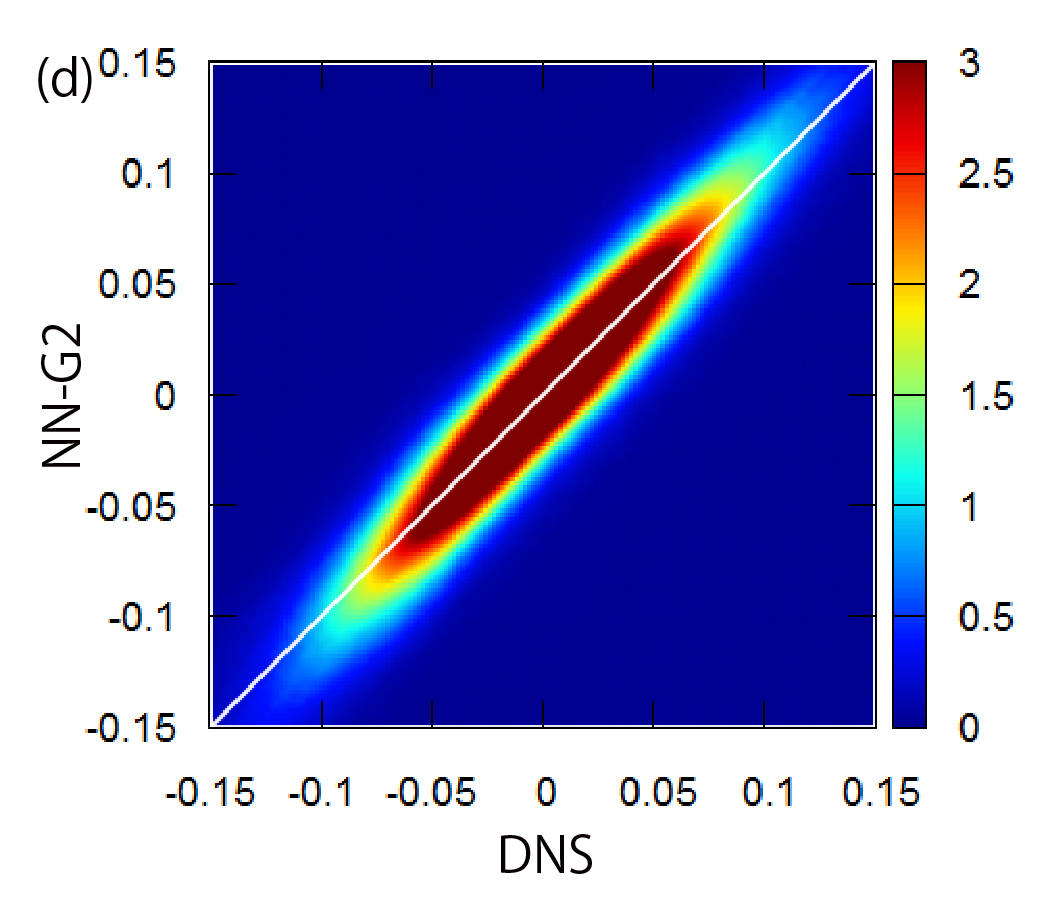}
   \includegraphics[width=55mm]{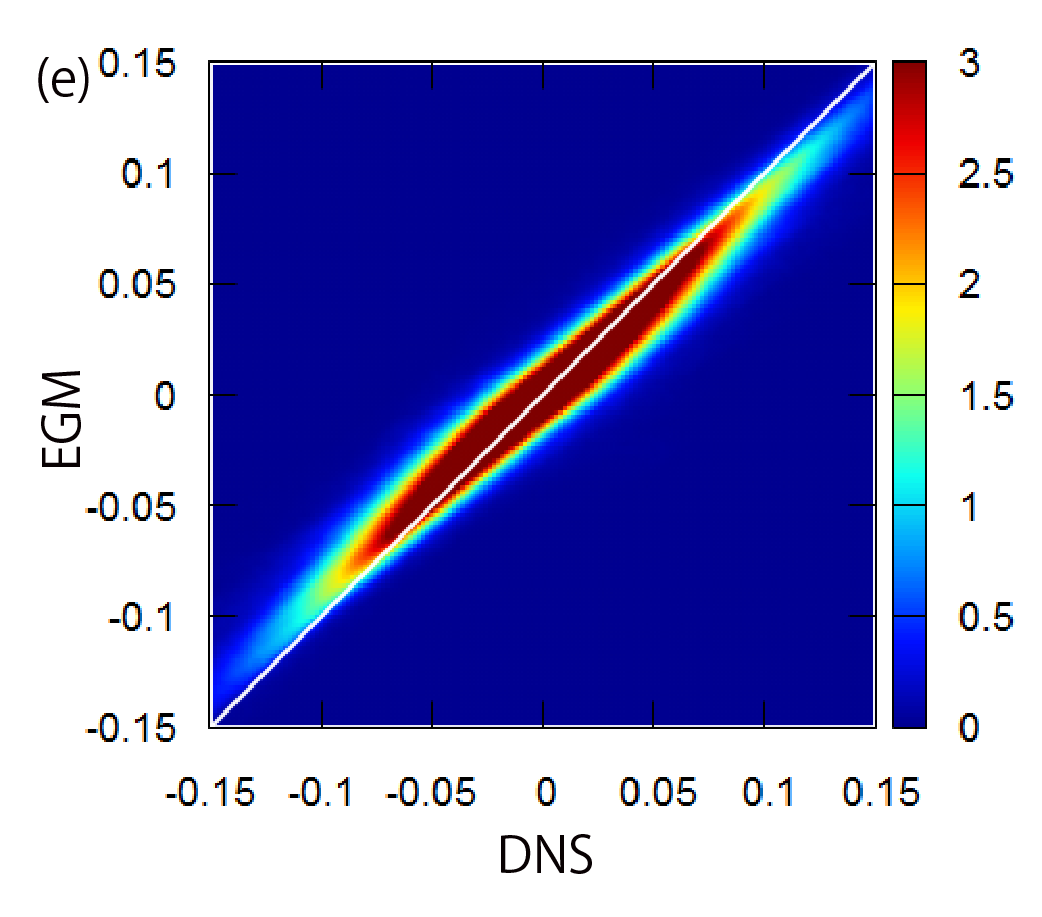}
   \includegraphics[width=55mm]{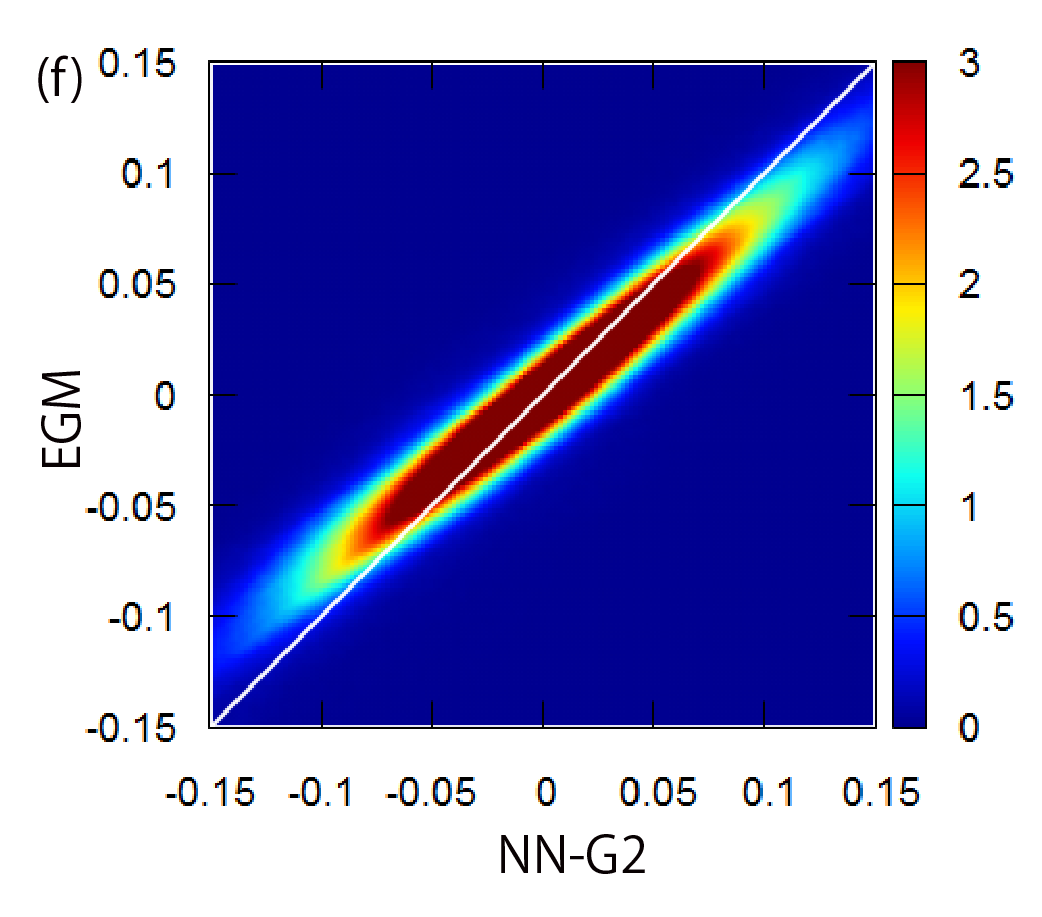}
  \end{center}

 \caption{Joint p.d.f. of SGS stress. 
Case 1,\,$\overline{\Delta}=24.4\eta$. 
(a) $\tau_{11}$, (b) $\tau_{12}$. 
(a,d) DNS vs. NN-G2, (b,e) DNS vs. EGM, (c,f) NN-G2 vs. EGM. }
 \label{fig:jPDF512-16}
\end{figure}

In order to check the relation between the models constructed by neural networks 
and the gradient models, 
cross correlation between them is investigated 
for the filter width larger than that in Sec.~\ref{sec-input-res}: 
$\overline{\Delta}=16\Delta_{\rm{DNS}} = 24.4\eta$ (Table \ref{ta:relation}). 
The correlation between NN-G2 
and the extended gradient model is higher than 
that between NN-G2 and the exact SGS stress, 
supporting that the trained neural network is close to the extended gradient model.  
However, there is difference in the magnitude 
as observed in Fig.~\ref{fig:jPDF512-16}, 
which shows the joint p.d.f.s of the pairs of 
the neural network, the extended gradient model, and the exact SGS stress.  
Therefore, 
the trained neural network is close to but not identical with 
the extended gradient model. 
 
\begin{figure}[h]
  \begin{center}
   \includegraphics[width=55mm]{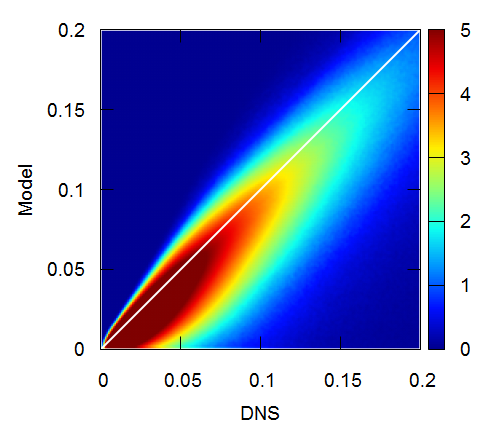}
  \end{center}
 \caption{Joint p.d.f. of exact value of $\tau_{11}$ and prediction by NN-G2 
trained for $\tau_{12}$. 
Case 1,\,$\overline{\Delta}=24.4\eta$. }
 \label{fig:jPDF512-16-cross}
\end{figure}

If the trained neural network is close to the extended gradient model, 
the network trained for $\tau_{12}$ can be also used for $\tau_{11}$ 
by giving appropriate input variables according to eq.~(\ref{eq:EGM}). 
Figure \ref{fig:jPDF512-16-cross} shows the joint p.d.f. of 
the exact value of $\tau_{11}$ and the prediction by NN-G2 
trained for $\tau_{12}$. 
The distribution is slightly wider than Fig.~\ref{fig:jPDF512-16-cross}(c); 
the correlation coefficient is $0.915$, while the error is $0.316$, 
which are smaller and larger, respectively, than those between DNS and NN-G2 in Table \ref{ta:relation}. 
Thus, correlation is slightly weaker.  

\begin{table}[h]
\begin{center}
  \vspace{5mm}
  \caption{Correlation between NN-G2 and extended gradient model (EGM)
and the error. 
Those with the exact SGS stress (DNS) are included for comparison. }
  \begin{tabular}{c|cc|cc|cc}
\hline\hline 
	&\multicolumn{2}{c|}{(DNS,EGM)}&\multicolumn{2}{c|}{(DNS,NN-G2)}&\multicolumn{2}{c}{(EGM,NN-G2)}\\
	& $\tau_{11}$ & $\tau_{12}$ & $\tau_{11}$ & $\tau_{12}$ & $\tau_{11}$ & $\tau_{12}$ \\ \hline
	Corr  &0.975&0.980&0.956&0.947&0.971&0.965\\
	error &0.282&0.221&0.223&0.346&-&- \\
\hline\hline 
  \end{tabular}
  \label{ta:relation}
\end{center}
\end{table}

\subsection{Results on SGS production term}
\label{sec-prod}

\begin{figure}
  \centering
  \includegraphics[width=70mm]{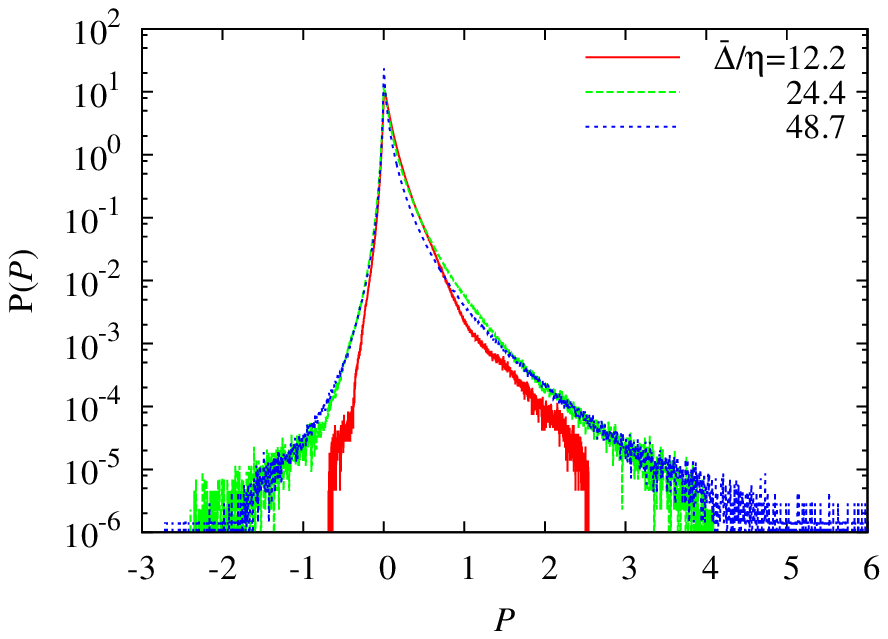}
  \caption{P.d.f. of SGS production term $P$. 
Case 1,\,$\overline{\Delta}=12.2\eta, 24,4\eta, 48.7\eta$.}
  \label{fig:sgsE_DNS}
\end{figure}

In LES the SGS production term $P=-\tau_{ij}\overline{S}_{ij}$, 
which is the energy transfer from the GS component to the SGS component, 
is an important quantity. 
In this subsection we focus on the SGS production term; 
correlation between the exact value of $P$ and that evaluated 
using the NN models and the existing models is investigated.  
In addition, we also train a neural network to predict the SGS production term directly 
and compare the results to those obtained in the preceding subsections. 
The DNS data of Case 1 are used, while the filter width is set to 
$\overline{\Delta}=8\Delta_{\rm{DNS}} = 12.2\eta$. 
Figure \ref{fig:sgsE_DNS} shows the p.d.f. of $P$ calculated by filtering the DNS data. 
It shows that there is backscatter $P<0$, although it is much smaller than $P>0$.

Table \ref{ta:SGSP} shows the correlation of $P$ 
between the exact value calculated by filtering the DNS data 
and that calculated by the NN models and the existing models. 
The ratio of the spatial average of $P$ calculated as 
$\langle P_{\rm model}\rangle/\langle P_{\rm DNS}\rangle$ is also shown;  
it is not shown for the Smagorinsky model since 
the ratio depends on the Smagorinsky coefficient; 
actually, the ratio is unity for $C_s=0.10$ which is smaller than 
the value $C_s=0.164$ based on the Kolmogorov theory. 
In Table \ref{ta:SGSP}, NN-D1p and NN-D2p are 
the neural networks trained to predict $P$ with the input variables of Sets D1 and D2, 
respectively.  
For the NN models, 
the correlation coefficients are slightly smaller than those for $\tau_{ij}$ 
(Table \ref{ta:input}). 
For the gradient and extended gradient models, the values are similar 
in Tables \ref{ta:input} and \ref{ta:SGSP}. 
It is noteworthy that the Smagorinsky model shows high correlation of ${\rm{Corr}}=0.816$, 
although correlation is weak for the SGS stress (Table \ref{ta:input}). 
NN-D1p and NN-D2p 
give higher or similar correlation than NN-D1 and NN-D2, respectively. 
The ratio $\langle P_{\rm model}\rangle/\langle P_{\rm DNS}\rangle$ 
is close to unity for NN-D1p and NN-D2p, 
while it is smaller than unity for NN-D1 and NN-D2. 
This implies that setting the SGS production term as the target function 
can improve LES modeling. 
This point will be further addressed in the future work. 
It is pointed out that the gradient and the extended gradient models also underestimate $P$.

\begin{table}
\begin{center}
  \vspace{5mm}
  \caption{Correlation of SGS production term $P=-\tau_{ij}\overline{S_{ij}}$ 
between exact value 
and prediction by NN models and existing models. 
The ratio of the spatial average of $P$ is also shown 
except for the Smagorinsky model. 
Case 1,\,$\overline{\Delta}=12.2\eta$.}
  \begin{tabular}{c||c|c|c|c||c|c|c||c|c}
\hline\hline 
	Model& NN-D1 & NN-G1 & NN-D2 & NN-G2 & GM & EGM &Smag. & NN-Dp& NN-D2p \\ \hline
	Corr &0.521&0.792&0.890&0.923&0.949&0.992&0.816&0.826&0.902\\
	$\langle P_{\rm model}\rangle/\langle P_{\rm DNS}\rangle$
	&0.577&0.803&0.639&1.028&0.886&0.984&-&0.928&1.085\\
\hline\hline 
  \end{tabular}
  \label{ta:SGSP}
\end{center}
\end{table}

\begin{figure}
  \begin{center}
   \includegraphics[width=55mm]{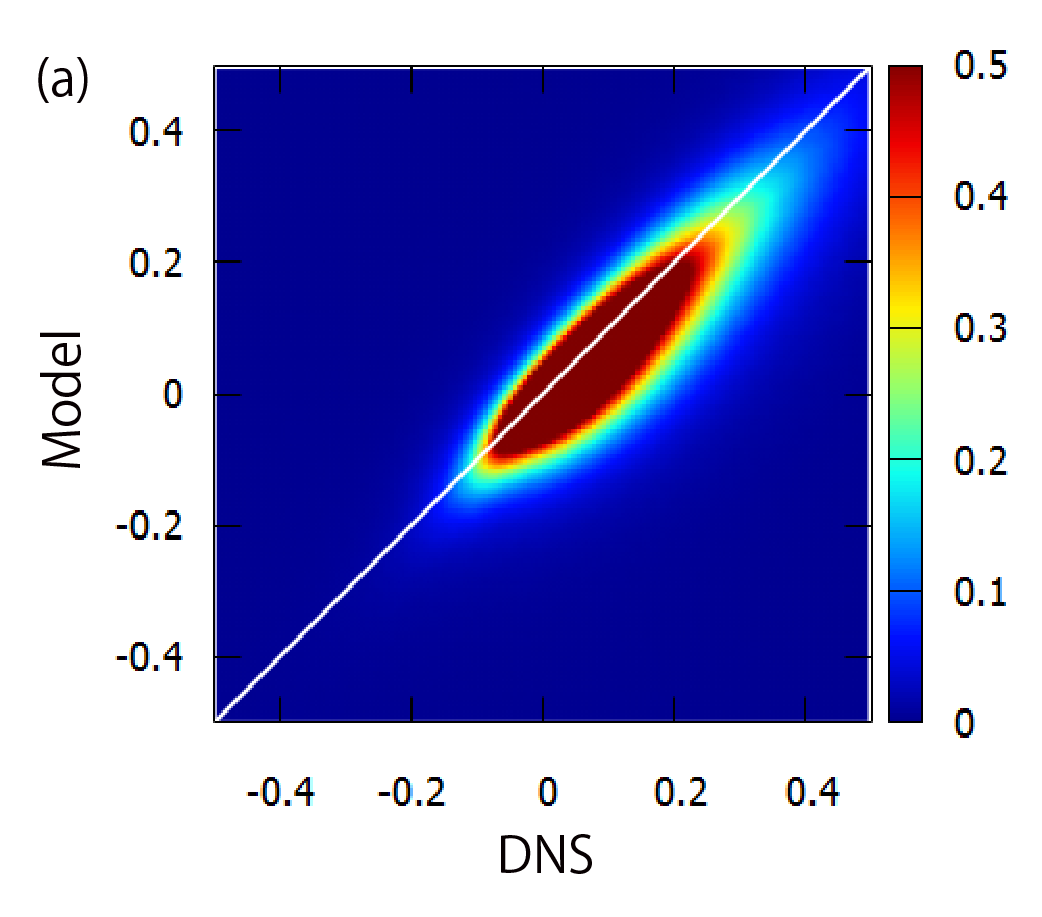}
   \includegraphics[width=55mm]{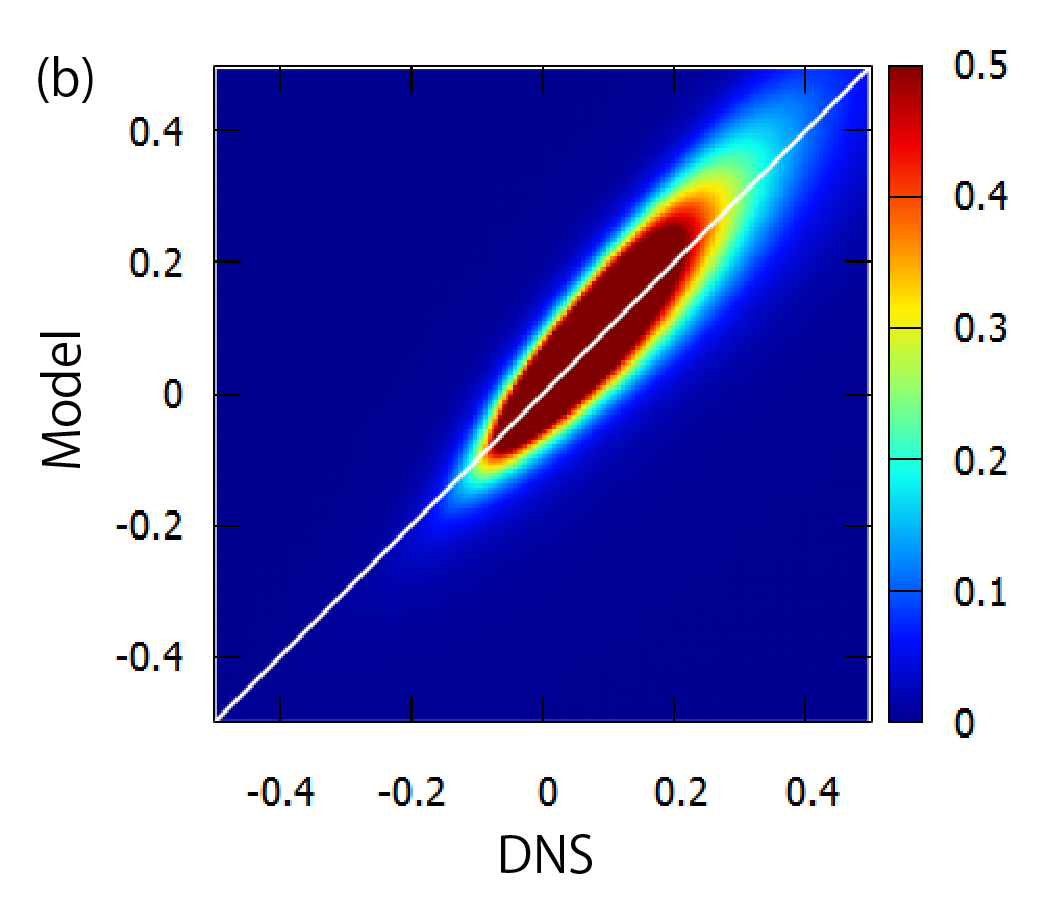}
   \includegraphics[width=55mm]{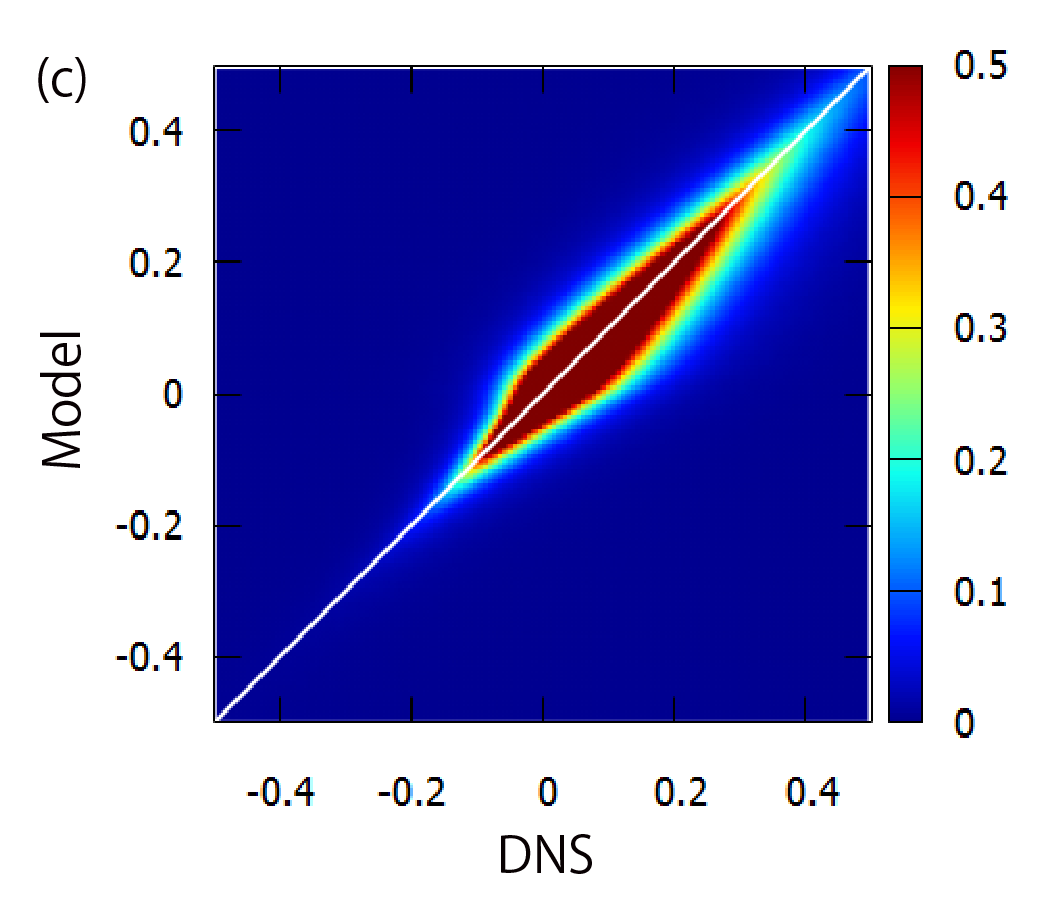}

   \includegraphics[width=55mm]{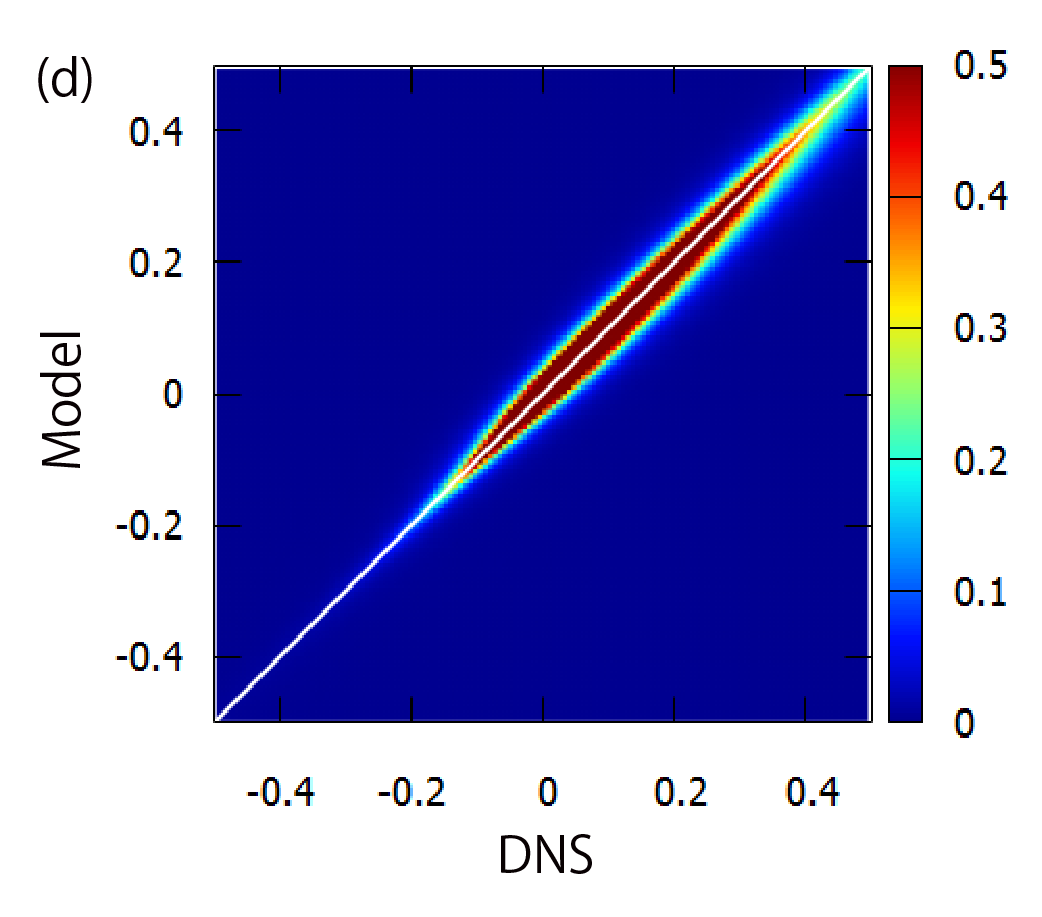}
   \includegraphics[width=55mm]{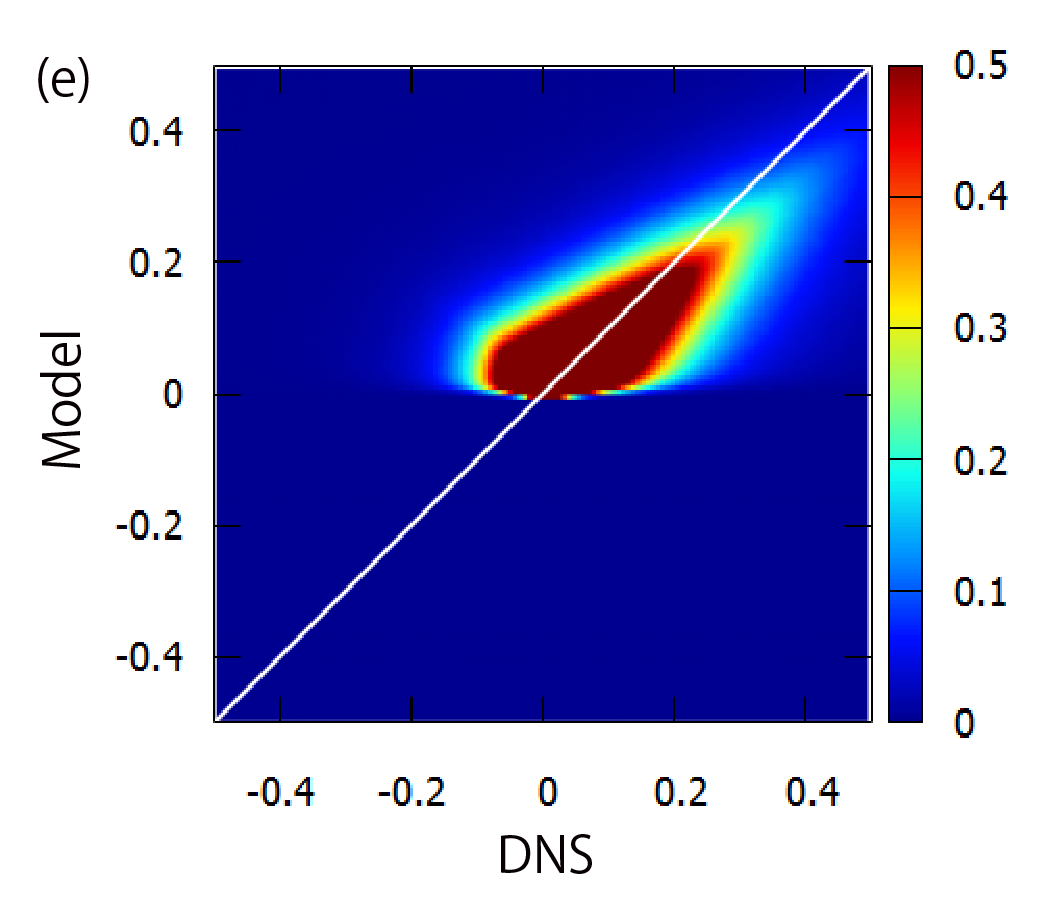}
   \includegraphics[width=55mm]{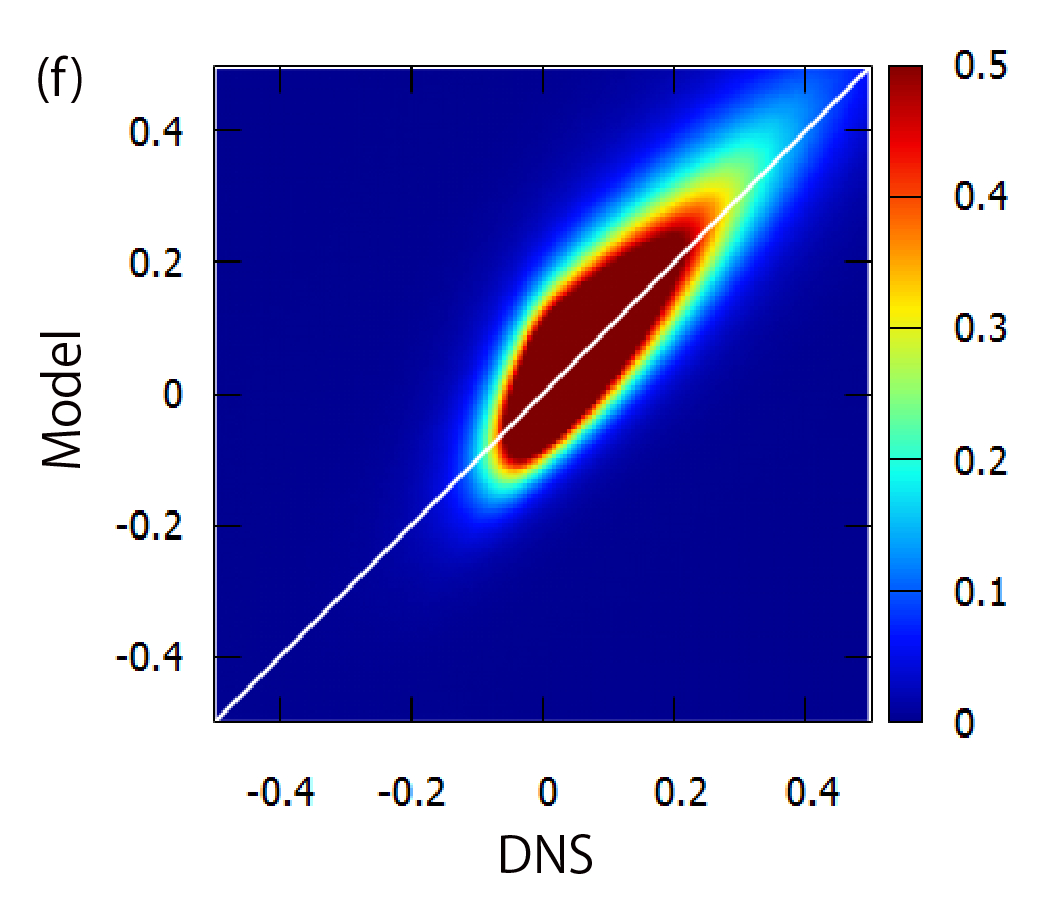}
  \end{center}

 \caption{Joint p.d.f. of SGS production term $P$. 
Case 1, $\overline{\Delta}= 12.2\eta$. 
The horizontal axis is the exact value obtained by filtering the DNS data. 
The vertical axes are predictions by 
(a) NN-D2, (b) NN-G2, 
(c) gradient model, (d) extended gradient model, 
(e) Smagorinsky model, (f) NN-D2p. }
 \label{fig:jPDF512-8P}
\end{figure}

\begin{figure}
  \begin{center}
   \includegraphics[width=55mm]{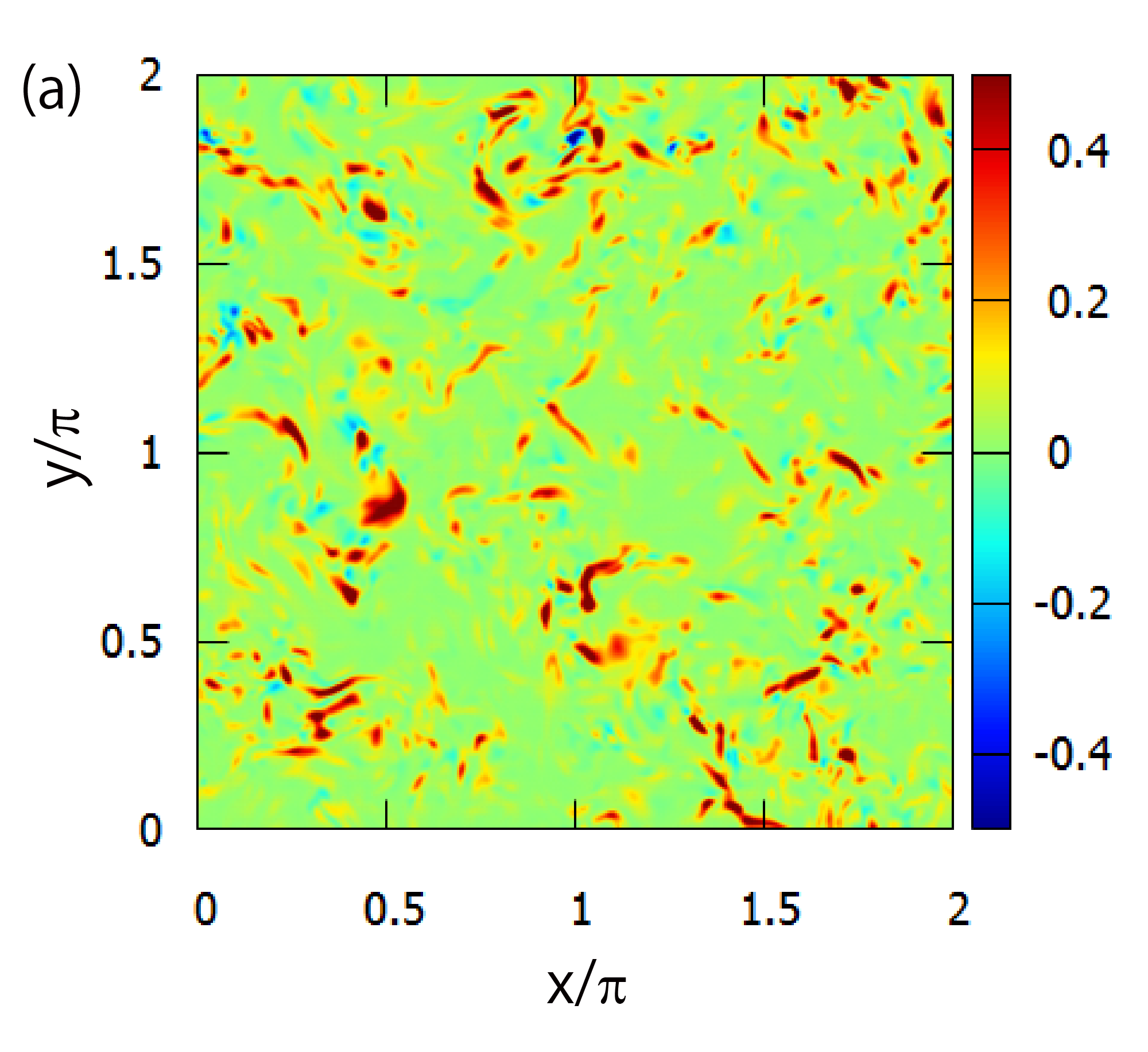}
   \includegraphics[width=55mm]{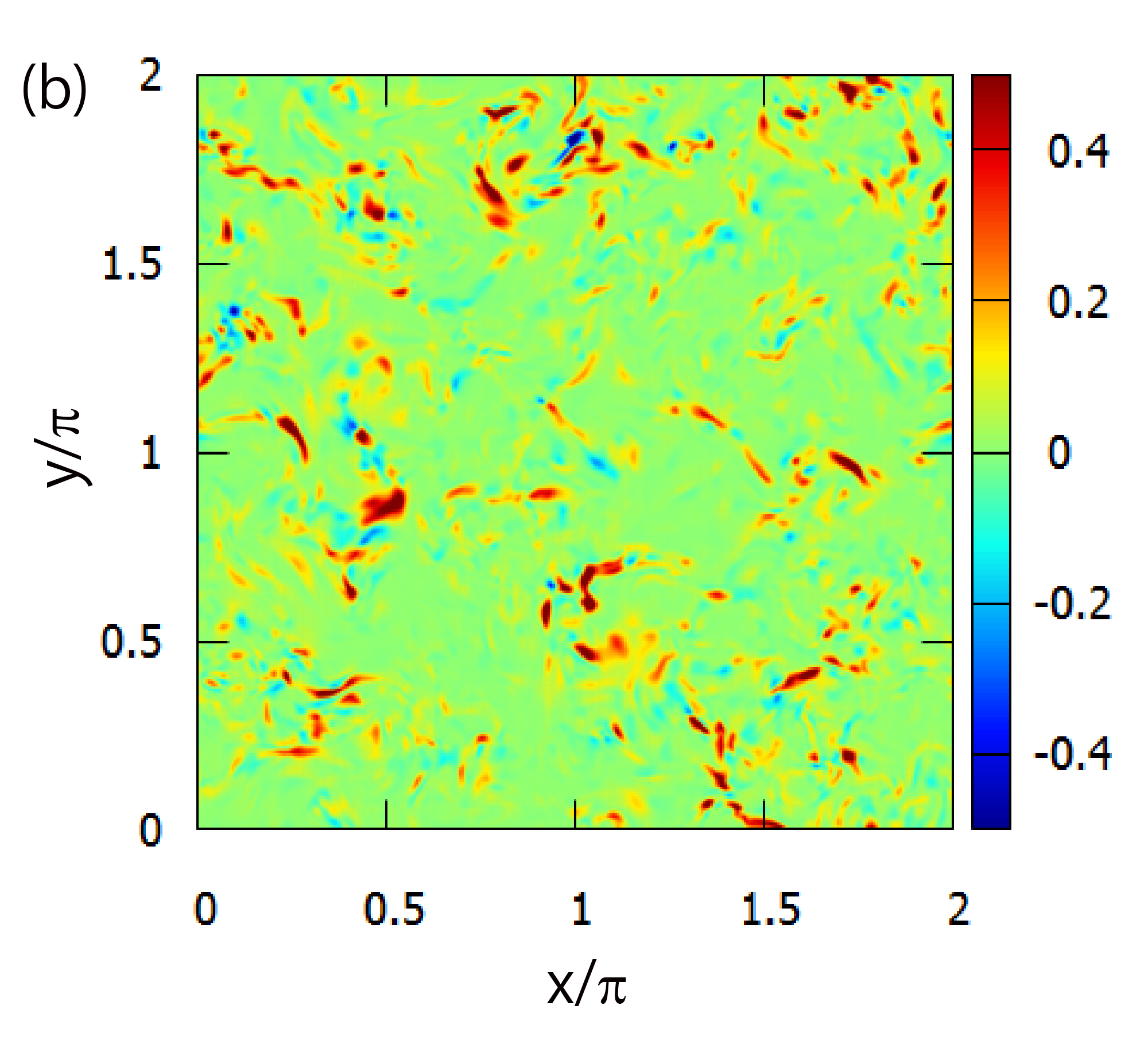}
   \includegraphics[width=55mm]{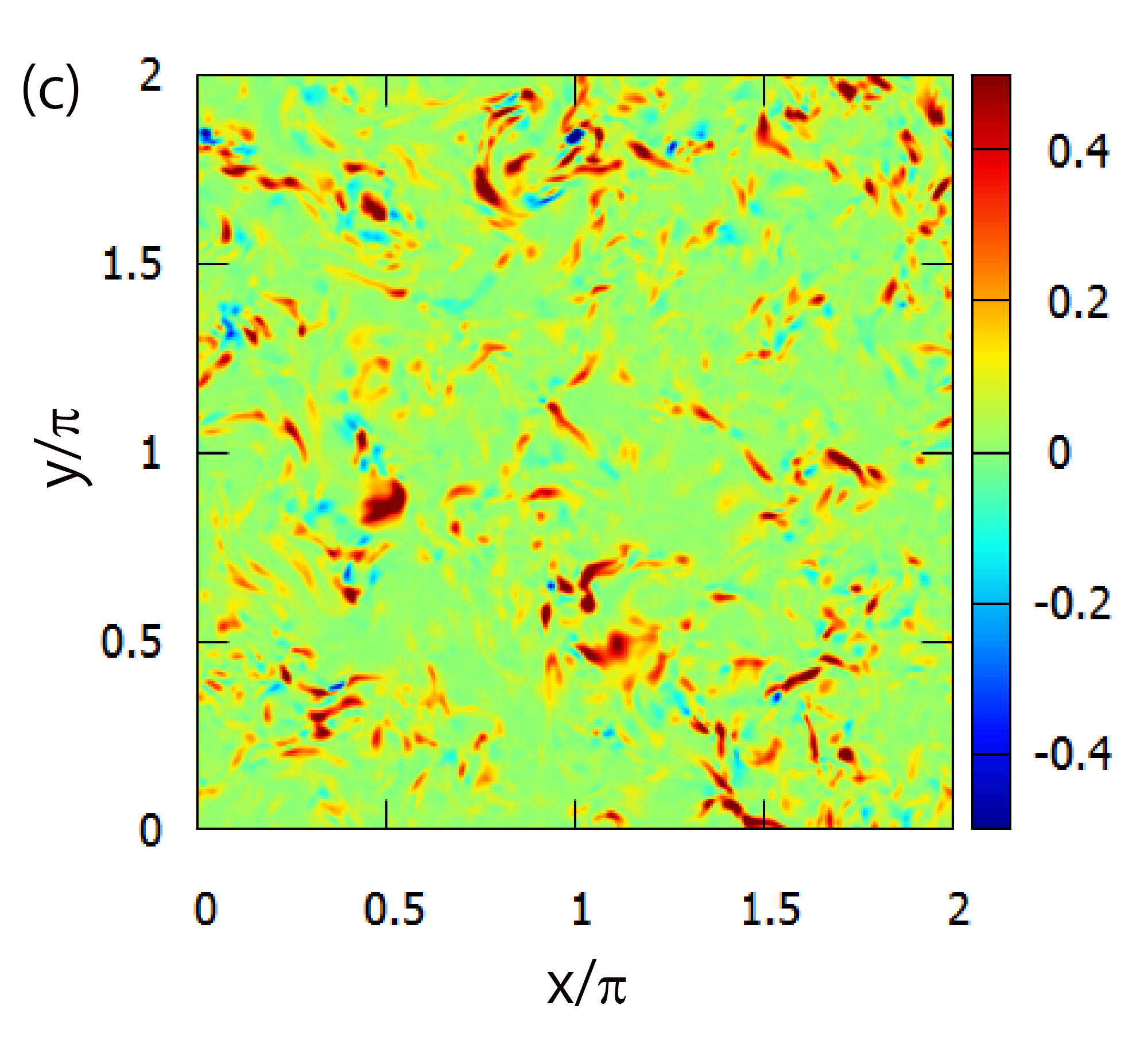}

   \includegraphics[width=55mm]{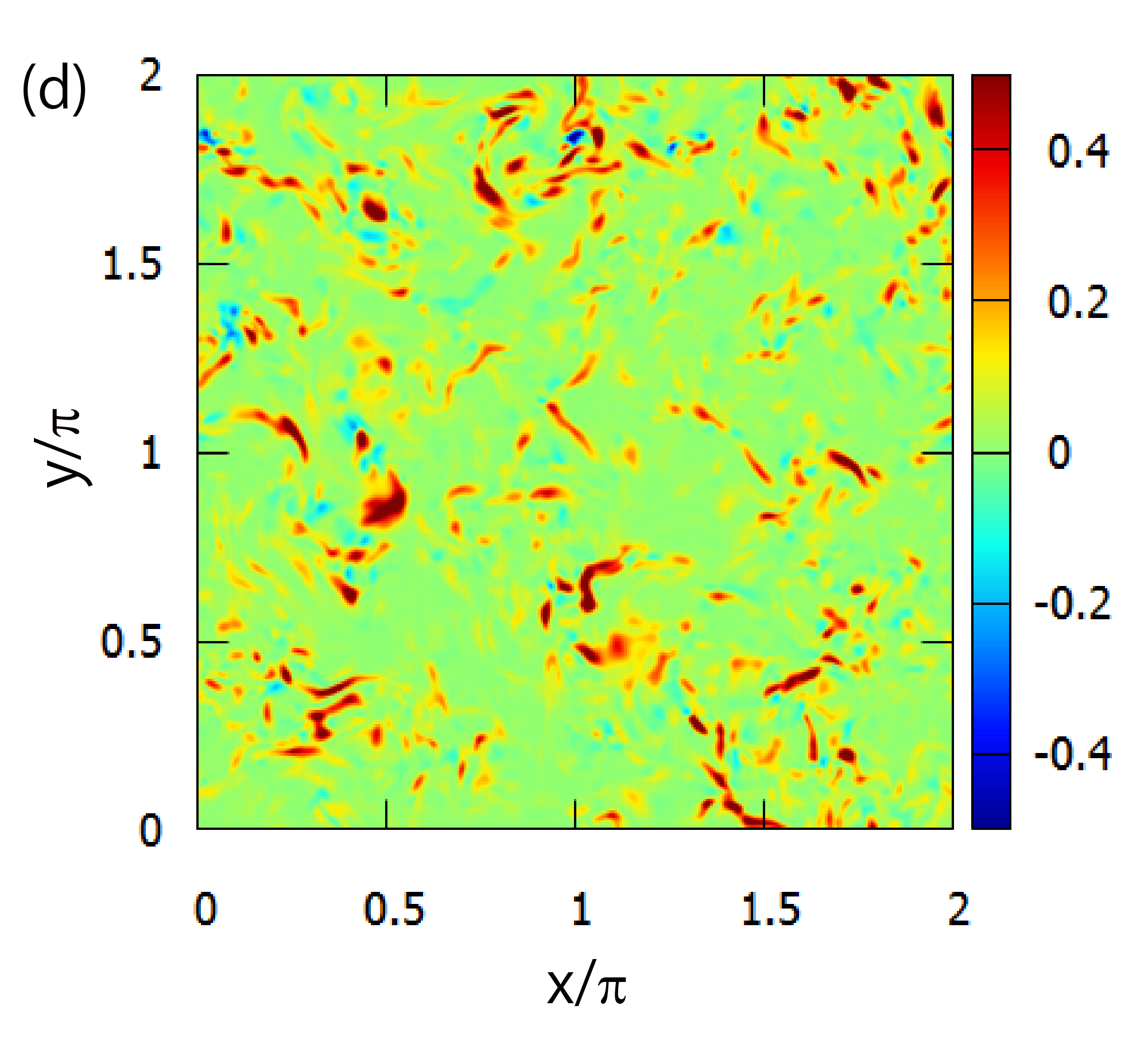}
   \includegraphics[width=55mm]{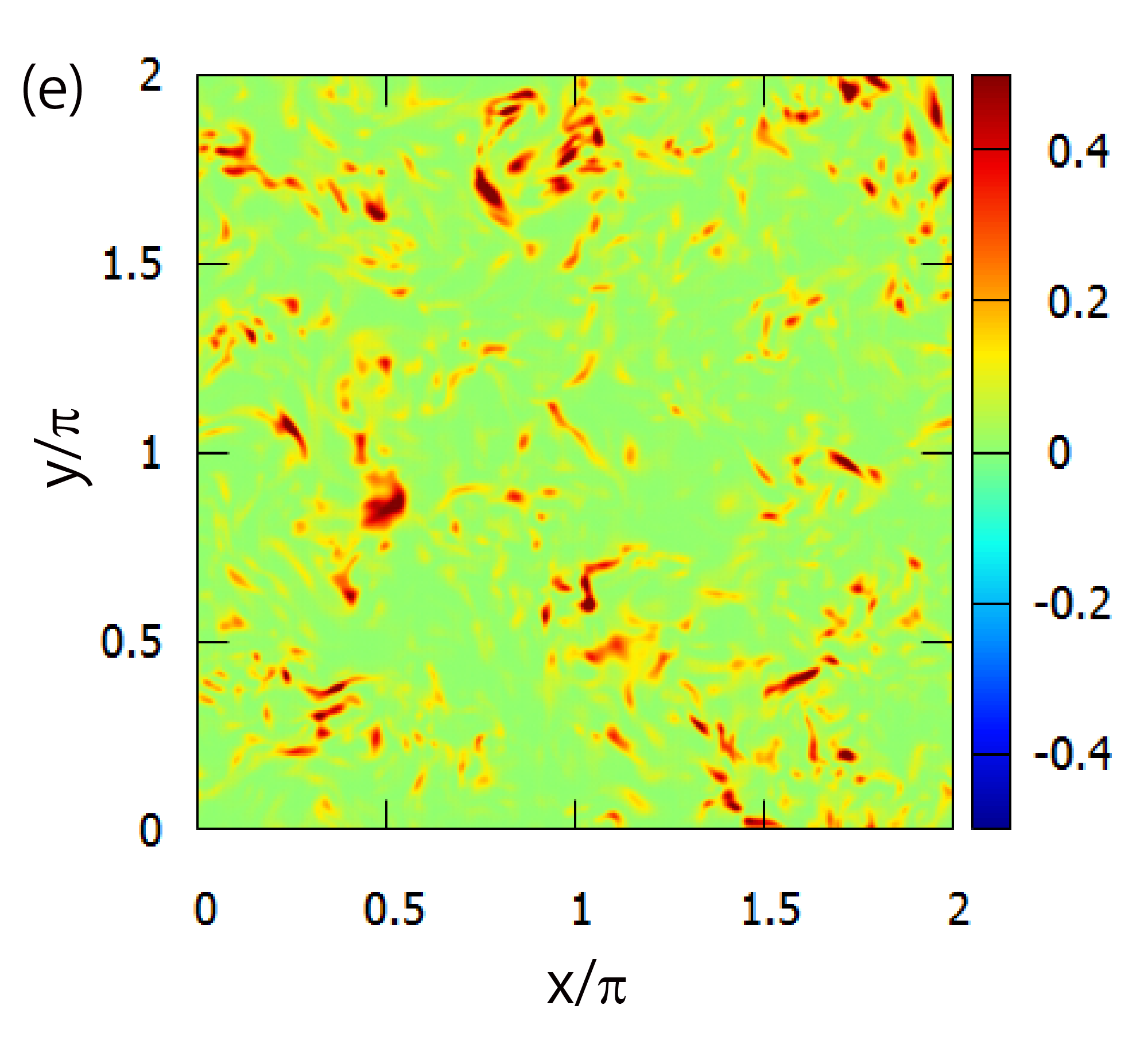}
   \includegraphics[width=55mm]{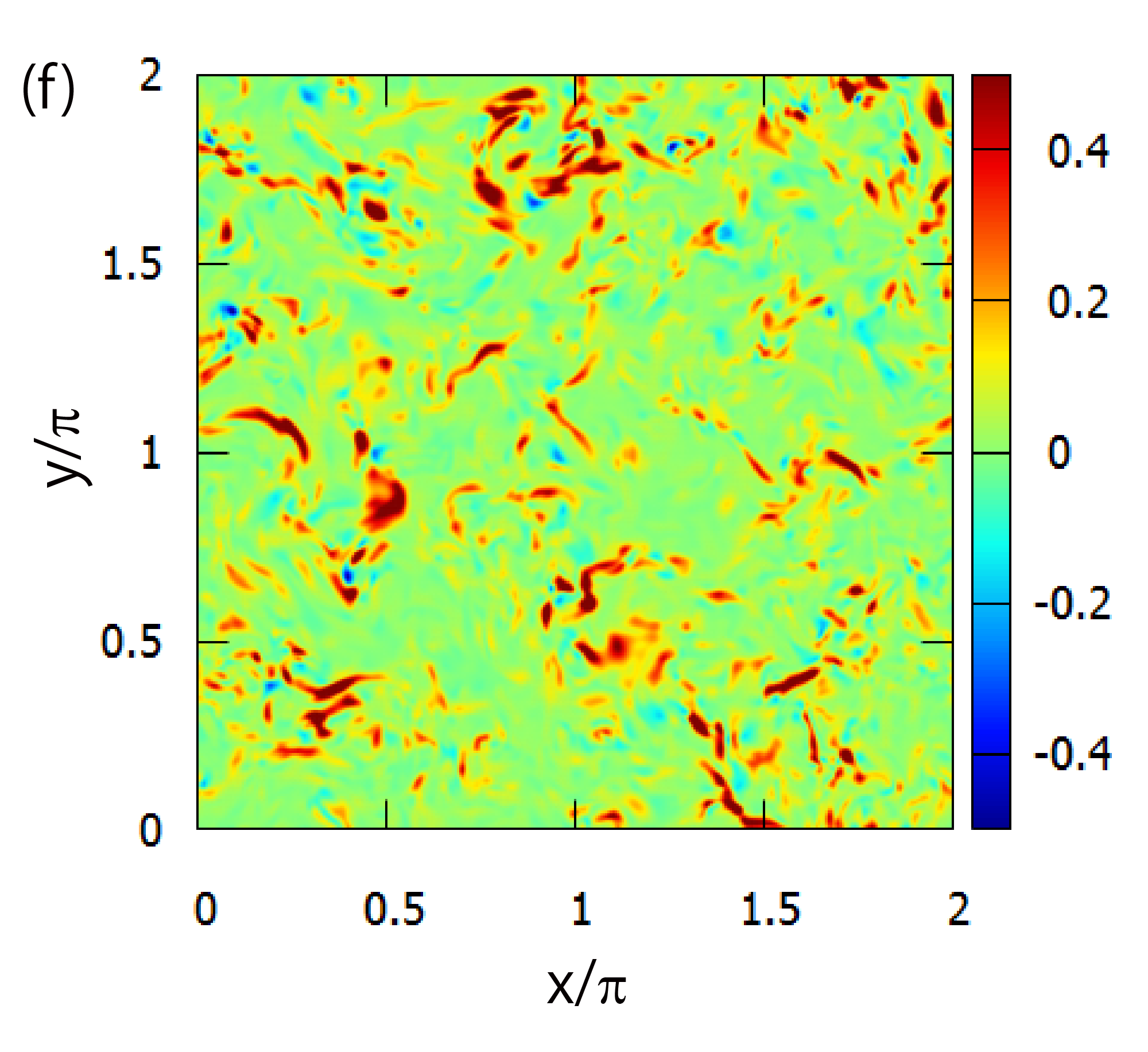}
  \end{center}

 \vspace{5mm}
 \caption{Spatial distribution of SGS production term $P$. 
Case 1, $\overline{\Delta}= 12.2\eta$. 
(a) DNS, (b) NN-D2, (c) NN-G2, (d) extended gradient model, 
(e) Smagorinsky model, (f) NN-D2p. }
 \label{fig:plane512-8P}
\end{figure}

Figure \ref{fig:jPDF512-8P} shows the joint p.d.f.s of 
the exact SGS production term calculated using the DNS data 
and the prediction by the NN models and the existing models. 
In the joint p.d.f. of NN-D2, 
the distribution is concentrated below the diagonal line 
implying that it underestimates $P$ on average. 
This trend is not observed for the other models. 
It is also observed that $P \ge 0$ for the Smagorinsky model, 
which is regarded as one of the drawbacks of the Smagorinsky model. 
Figure \ref{fig:plane512-8P} compares spatial distributions of $P$ on $z=0$. 
All of the NN models and the existing models 
reproduce successfully the exact distribution on the whole, 
although negative values are absent in the Smagorinsky model 
and NN-D2 and NN-D2p 
overestimate backscatter in several regions. 
Based on the results in this subsection, 
we use NN-G2 in {\textit{a posteriori}} test in the next section. 

\section{Results of {\textit{a posteriori}} test}
\label{sec-aposteriori}

\subsection{Stabilization}
\label{sec-stab}

\begin{figure}
  \centering
  \includegraphics[width=7cm]{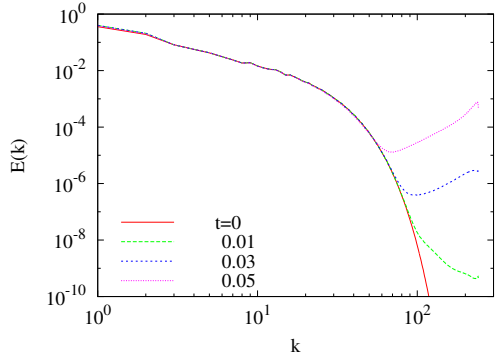}
  \caption{Time evolution of energy spectrum of homogeneous isotropic turbulence. 
LES with the NN model without stabilization. 
}
  \label{fig:stabSpectrum}
\end{figure}

In this section we implement the neural network with the input variables of Set G2 (NN-G2) 
to actual LES ({\textit{a posteriori}} test) 
and investigate the accuracy by comparing the results with those obtained 
with two existing models: the Smagorinsky and Bardina models.  
An important remark here is that the NN model should be stabilized.  
Figure \ref{fig:stabSpectrum} shows the energy spectrum of homogeneous isotropic turbulence 
obtained by LES using the NN model without stabilization. 
The energy at small scales grows rapidly which is unphysical;  
the calculation diverges eventually. 
Similar divergence has been reported by Beck et al.~\cite{BFM-2019}. 
Thus we stabilize the NN model by clipping. 
Namely, the SGS stress of the NN model $\tau_{ij}$ is clipped as 
\begin{eqnarray}
\tau_{ij}^{\ast}&=&\left\{
 \begin{array}{ll}
 \tau_{ij}, & \tau_{ij}\overline{S}_{ij} \le 0 \\
 0, & {\mbox{otherwise,}}
 \end{array}
 \right. 
\label{eq:modeStabilized}
\end{eqnarray}
which is introduced in Lu et al.~\cite{LPA-2010} for stabilizing the gradient model. 
By this clipping procedure the SGS stress is forced to be zero to prevent backscatter. 
The stabilized NN model is successfully used in LES as shown below. 
The same clipping procedure was applied to the extended gradient model. 

\subsection{Homogeneous isotropic turbulence}
\label{sec-hit}

First, we show the results of LES of homogeneous isotropic turbulence. 
Four cases shown in Table \ref{ta:IHTLEScond} are considered. 
The initial conditions were obtained by filtering the 
DNS data of the homogeneous isotropic turbulence (Table \ref{tab:DNS_CASE}). 
In Table \ref{ta:IHTLEScond}, $N_{\rm{LES}}^3$, $R_\lambda$, $\Delta_{\rm{DNS}}$, $\eta$, 
$k_{\rm{max}}$, and $E$ are 
the number of modes used in LES, 
the Reynolds number based on the Taylor microscale, 
the grid size of the corresponding DNS, 
the Kolmogorov scale, 
the magnitude of the largest effective wavevector, 
and the energy of the corresponding (unfiltered) DNS data;  
$\overline{E}$ is the energy after filtering.  
LES with the existing models were also performed for comparison;  
the Smagorinsky coefficient $C_s$ was adjusted by try and error to give better results. 
The filter width is the same for Cases HIT1-1 and HIT2-1, 
while it is the same for Cases HIT1-2 and HIT2-2;  
the ratio of the filter width to the Kolmogorov scale is the same 
for Cases HIT1-2 and HIT2-1. 
The grid size in LES is set to 
$\Delta_{\rm{LES}}=\overline{\Delta}/2$. 
The size of the time step is fixed to $10^{-3}$. 
Forcing was introduced at low wavenumbers to keep the kinetic energy constant as in DNS. 
The simulation was stopped at $t=6$, which is about three times the eddy turnover time 
based on the integral length scale; 
the statistics below were calculated using the data at the final time. 
For the NN model, the neural network was trained using the corresponding DNS data 
with the filter width shown in Table \ref{ta:IHTLEScond}. 

\begin{table}
\begin{center}
  \vspace{5mm}
  \caption{Simulation parameters of LES of isotropic homogeneous turbulence.}
  \begin{tabular}{c|cccccc|c}
\hline\hline 
    Case& $N_{\rm{LES}}^3$ &$R_\lambda$&$\overline{\Delta}/\Delta_{\rm{DNS}}$& $\overline{\Delta}/\eta$ & $k_{\rm max}$& $\overline{E}/E$ & $C_s$ \\ \hline
	HIT1-1&$128^3$  &173& 8 & 12.2  & 60 & 0.942&0.10 \\
	HIT1-2&$64^3$ &173& 16 & 24.4  & 30 & 0.865&0.10 \\
	HIT2-1&$128^3$  &268& 16 & 24.4  & 60 & 0.911&0.12 \\
	HIT2-2&$64^3$ &268& 32 & 48.7 & 30 & 0.831&0.12 \\
\hline\hline 
  \end{tabular}
  \label{ta:IHTLEScond}
\end{center}
\end{table}

The rate of energy dissipation, the integral length scale, and the Taylor microscale 
are compared in Table \ref{ta:IHTLES1-1}; 
also included are the skewness and the flatness of the p.d.f.s of 
$\partial_1 \overline{u_1}$ and $\partial_1 \overline{u_2}$, 
which will be discussed later. 
The values of DNS are included for reference. 
The results obtained by LES should be compared to the values 
obtained for the filtered DNS data, which are denoted by 'Filtered' in the table.  
The rate of energy dissipation in Table \ref{ta:IHTLES1-1} is the viscous dissipation 
of the filtered velocity field; 
it is smaller than the rate of dissipation of the total energy 
since energy is transferred to the subgrid scales through the production term $P$, 
of which ratio increases with the filter width;    
$67\%$ and $93\%$ of the energy dissipation are due to 
the energy transfer to the subgrid scales in Cases HIT1-1 and HIT2-2, respectively. 
 
Table \ref{ta:IHTLES1-1} shows that 
the Bardina model overpredicts the energy dissipation in comparison to the other models;  
the parameters should be carefully tuned by e.g. a dynamic procedure 
to make the Bardina model accurate. 
The integral length scale $\lambda_I$ calculated by 
\begin{eqnarray}
\lambda_I = \frac{3\pi}{4} \frac{\int_0^{k_{\rm{max}}} k^{-1}E(k) dk}{\int_0^{k_{\rm{max}}} E(k) dk}
\end{eqnarray}
does not change very much by filtering:  
$\lambda_I$ of DNS data increases $5 \sim 16\%$ by filtering (Table \ref{ta:IHTLES1-1}). 
The integral length scale obtained with the NN model 
is not far from that of the filtered data, 
although it is slightly larger than those obtained with the other models.  
The longitudinal Taylor microscale $\lambda_{\parallel}$ and 
the transverse Taylor microscale $\lambda_{\perp}$ are calculated by 
\begin{eqnarray}
	\lambda_{\parallel}^2=\frac{\langle u_1^2 \rangle }{ \Bigl\langle \Bigl( \pd{u_1}{x_1} \Bigr)^2\Bigr \rangle}\,,\quad
	\lambda_{\perp}^2=\frac{\langle u_2^2 \rangle }{ \Bigl\langle \Bigl( \pd{u_2}{x_1} \Bigr)^2\Bigr \rangle}\,.\label{eq:lam_der}
\end{eqnarray}
In contrast to the integral length scale 
the Taylor microscales increase by filtering 
since small-scale fluctuations below the filter width are removed:   
$\lambda_\parallel$ of the filtered data is about $1.5 \sim 3.4$ times 
that of DNS (Table \ref{ta:IHTLES1-1}). 
The relation $\lambda_\parallel=\sqrt{2}\lambda_\perp$, 
which is known for the homogeneous isotropic turbulence, is also confirmed in Table \ref{ta:IHTLES1-1}. 
On the whole the Taylor microscales obtained with the NN model 
are in good agreement with those of the filtered data.

\begin{table}
\begin{center}
  \vspace{5mm}
  \caption{Comparison of turbulence characteristics.}
  \begin{tabular}{cc|cccc|ccc}
\hline\hline
Case & Model & $10^2\overline{\varepsilon}$ & $\lambda_I$ & $\lambda_{\parallel}$ & $\sqrt{2}\lambda_{\perp}$ & $S_\parallel$ & $F_\parallel$ & $F_\perp$ \\ 
\hline
HIT1-1 & DNS & 8.123 & 1.248 & 0.208 & 0.207 & -0.53 & 5.81 & 8.77 \\ 
& Filtered & 3.495 & 1.315 & 0.307 & 0.306 & -0.50 & 4.35 & 5.29 \\
& NN & 3.303 & 1.283 & 0.322 & 0.325 & -0.32 & 4.32 & 5.73 \\
& Smagorinsky & 3.601 & 1.239 & 0.311 & 0.311 & -0.36 & 4.11 & 5.91 \\
& Bardina & 4.577 & 1.202 &  0.274 & 0.275 & -0.86 & 5.91 & 7.34 \\
& EGM & 3.286 & 1.288 & 0.323 & 0.325 & -0.41 & 4.26 & 5.50 \\
\hline
HIT1-2 & DNS & 8.123 & 1.248 & 0.208 & 0.207 & -0.53 & 5.81 & 8.77 \\ 
& Filtered & 1.523 & 1.409 & 0.445 & 0.443 & -0.39 & 3.68 & 4.19 \\
& NN & 1.488 & 1.497 & 0.440 & 0.446 & -0.22 & 4.01 & 4.27 \\
& Smagorinsky & 1.593 & 1.266 & 0.432 & 0.430 & -0.31 & 3.64 & 5.91 \\
& Bardina & 2.030 & 1.221 & 0.376 & 0.384 & -0.69 & 4.52 & 7.34 \\
& EGM & 1.469 & 1.448 & 0.447 & 0.451 & -0.30 & 3.80 & 5.50 \\
\hline
HIT2-1 & DNS & 8.270 & 1.096 & 0.130 & 0.130 & -0.56 & 6.96 & 8.76 \\
& Filtered & 1.630 & 1.181 & 0.279 & 0.279 & -0.42 & 3.98 & 5.06 \\
& NN & 1.654 & 1.227 & 0.278 & 0.277 & -0.33 & 4.05 & 4.42 \\
& Smagorinsky & 1.609 & 1.140 & 0.281 & 0.280 & -0.45 & 4.03 & 5.59 \\
& Bardina & 1.726 & 1.212 & 0.270 & 0.272 & -0.73 & 4.88 & 5.80 \\
\hline
HIT2-2 & DNS & 8.270 & 1.096 & 0.130 & 0.130 & -0.56 & 6.96 & 8.76 \\
& Filtered & 0.590 & 1.271 & 0.441 & 0.440 & -0.40 & 3.83 & 4.33 \\
& NN & 0.550 & 1.354 & 0.471 & 0.447 & -0.35 & 3.73 & 3.94 \\
& Smagorinsky & 0.631 & 1.179 & 0.426 & 0.424 & -0.39 & 3.60 & 4.71 \\
& Bardina & 0.686 & 1.275 & 0.408 & 0.406 & -0.68 & 4.38 & 5.67 \\
\hline\hline
  \end{tabular}
  \label{ta:IHTLES1-1}
\end{center}
\end{table}

\begin{figure}
 
  \begin{center}
   \includegraphics[width=70mm]{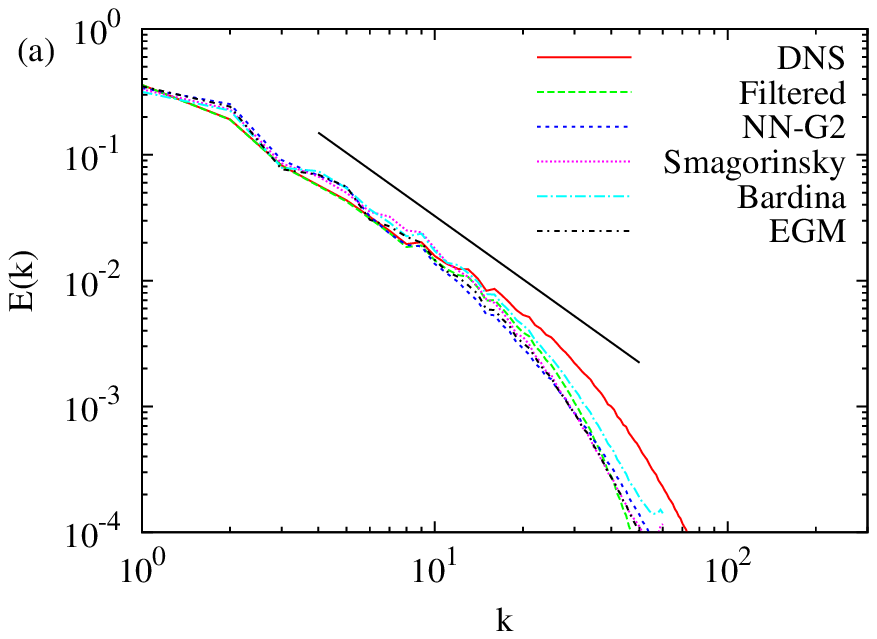}
   \includegraphics[width=70mm]{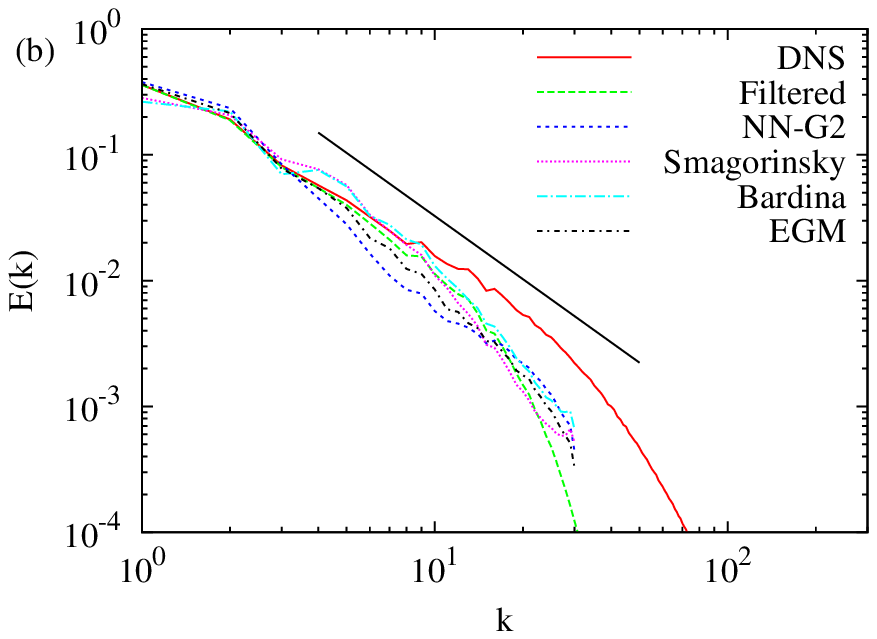}

   \includegraphics[width=70mm]{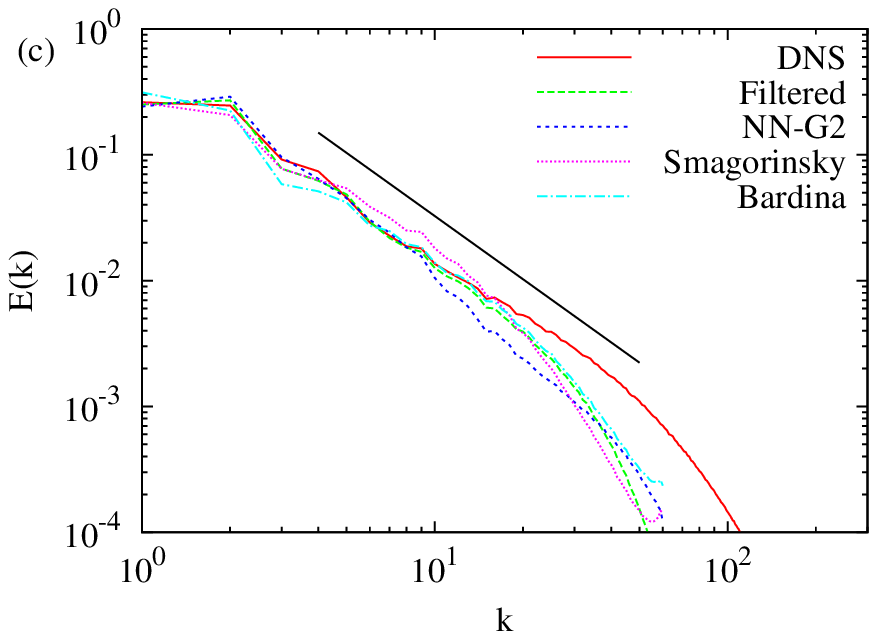}
   \includegraphics[width=70mm]{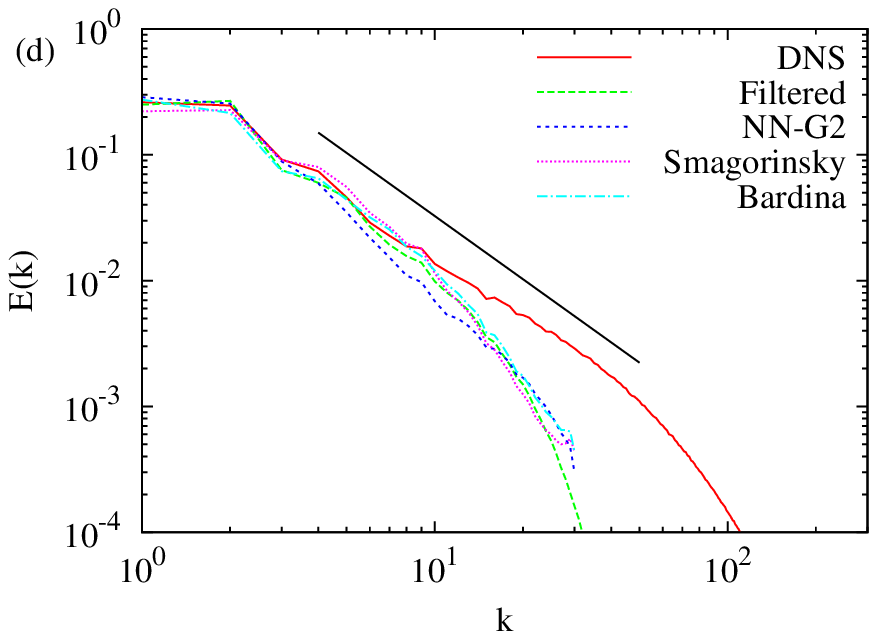}
  \end{center}
 
 \caption{Energy spectrum of homogeneous isotropic turbulence. 
Comparison between DNS, filtered DNS, and LES.   
The solid lines show $E(k) \propto k^{-5/3}$. 
(a) Case HIT1-1, (b) Case HIT1-2, (c) Case HIT2-1, (d) Case HIT2-2. }
 \label{fig:IHTapostarioriSpec}
\end{figure}

Figure \ref{fig:IHTapostarioriSpec} compares the energy spectrum between 
DNS, filtered DNS, and LES.   
The spectrum of the filtered DNS deviates from that of DNS 
at a wavenumber within the inertial subrange except for Case HIT1-1, 
showing that the wavenumber corresponding to the filter width is within the inertial subrange 
for the other three cases. 
All spectra of LES are in reasonable agreement with the filtered DNS, 
although the energy spectrum of the NN model 
is slightly smaller than that of the filtered DNS.

\begin{figure}
 
  \begin{center}
   \includegraphics[width=55mm]{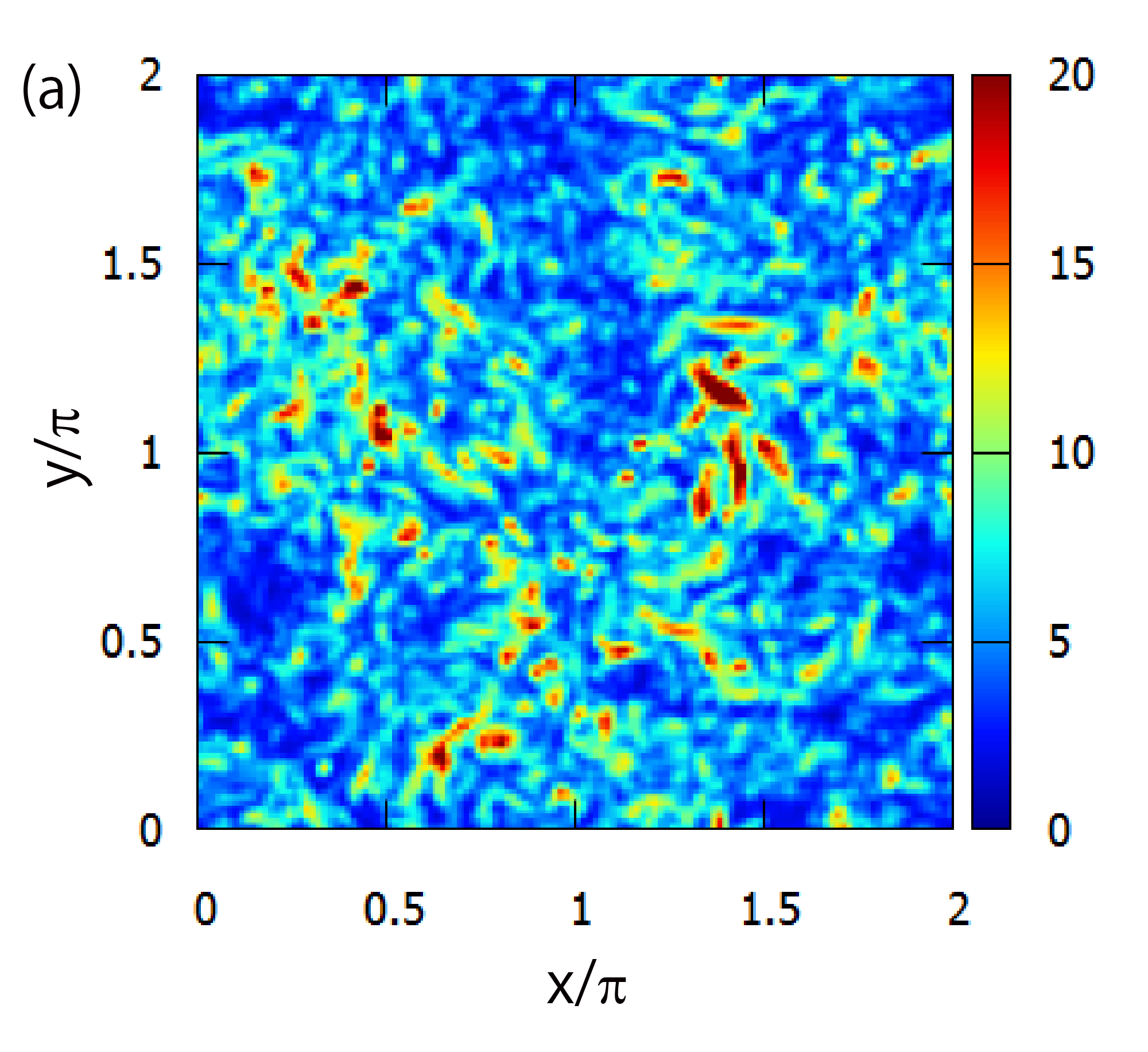}
   \includegraphics[width=55mm]{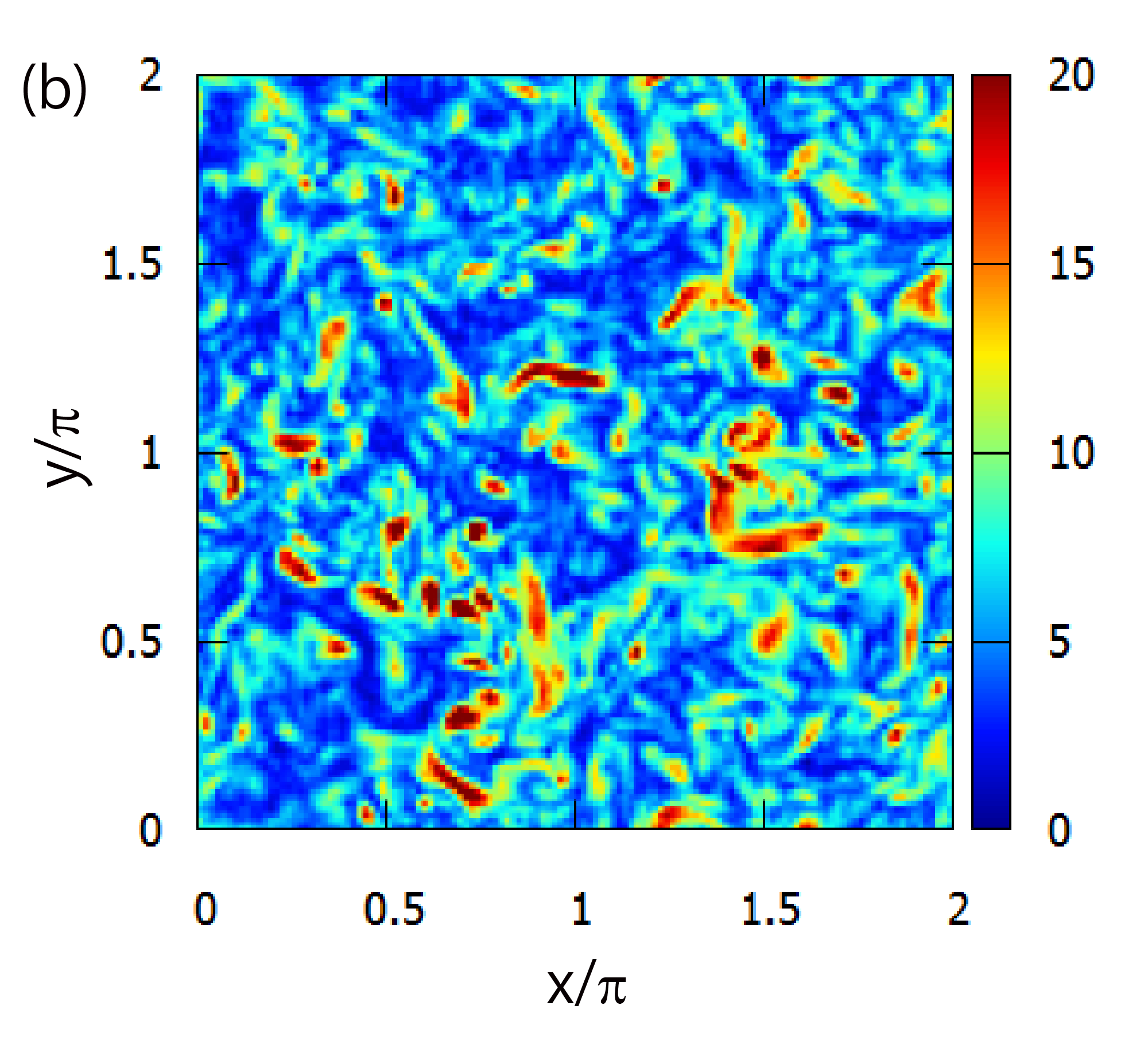}

   \includegraphics[width=55mm]{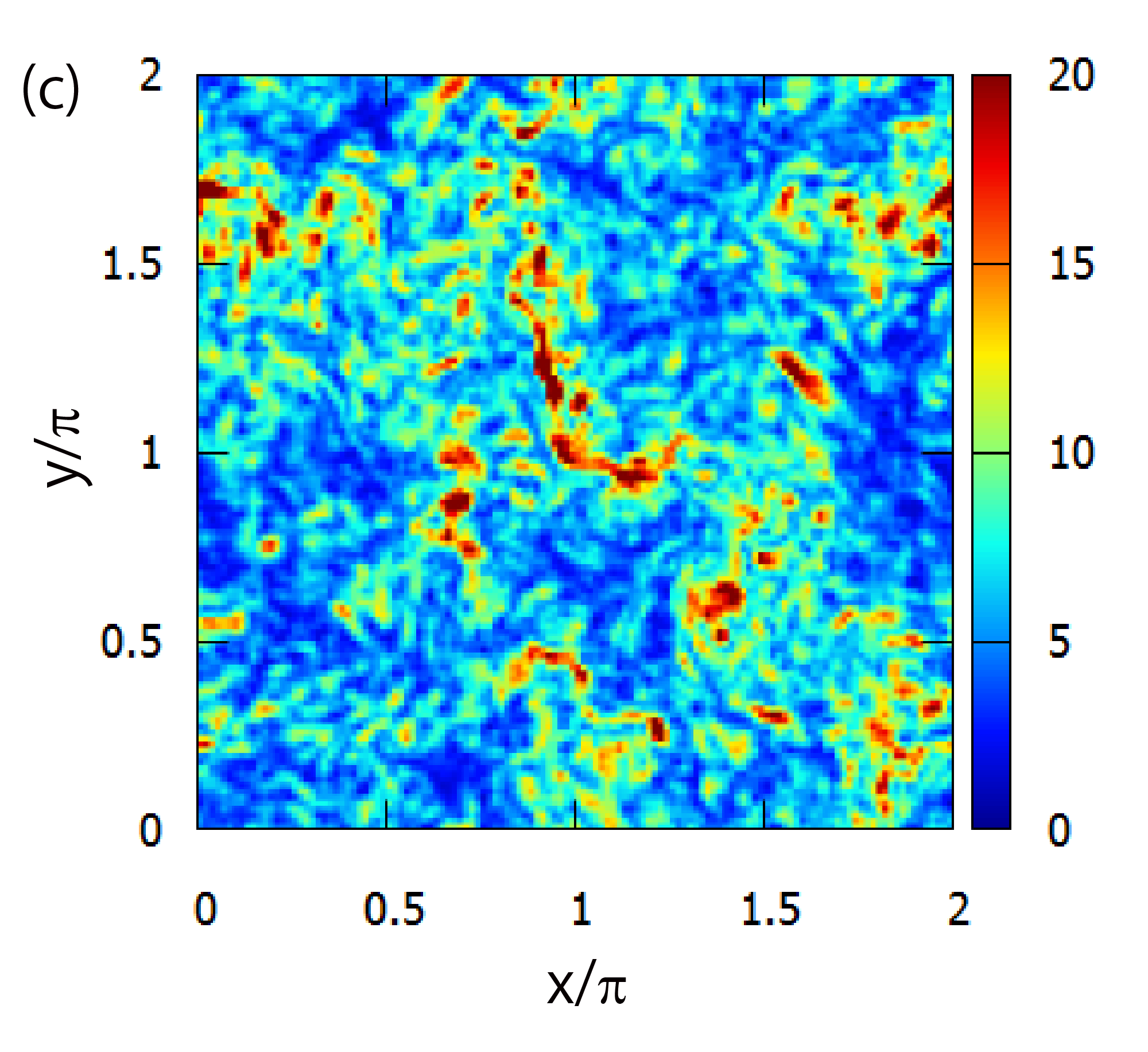}
   \includegraphics[width=55mm]{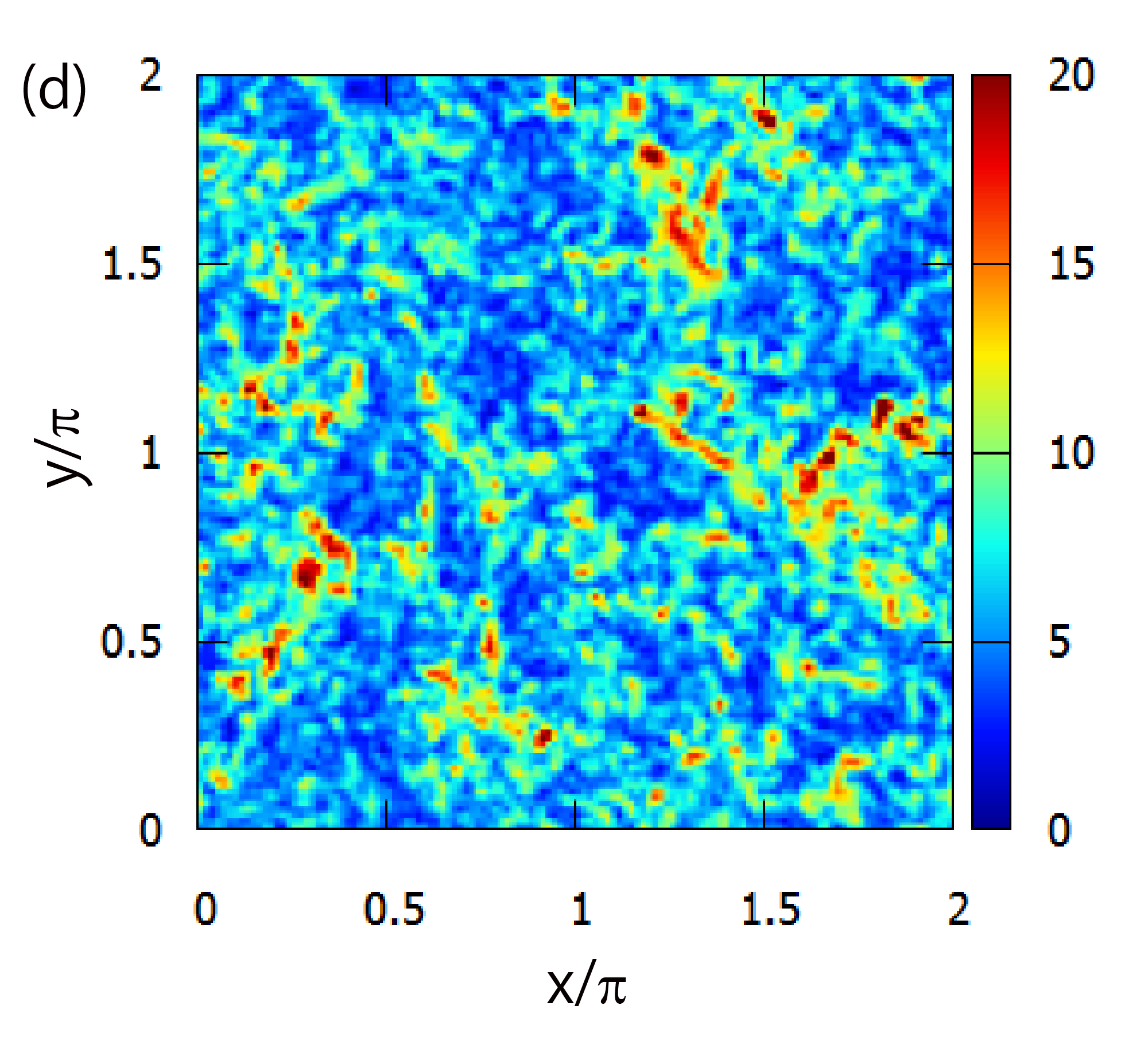}
  \end{center}
 
 \caption{Distributions of magnitude of vorticity on $z=0$. 
Case HIT2-1. 
(a) Filtered DNS, (b) Smagorinsky model, (c) Bardina model, (d) NN-G2.  
}
 \label{fig:IHTapostarioriVor2-1}
\end{figure}

Figure \ref{fig:IHTapostarioriVor2-1} shows the distributions of the magnitude of vorticity 
on $z=0$ for the filtered DNS data and LES of Case HIT2-1. 
We cannot expect agreement between the distributions 
since small differences are amplified exponentially. 
However, some features of the distributions can be compared: 
tube-like structures are observed in all cases;   
the magnitude of vorticity is comparable between all cases. 
In particular, it is confirmed that the NN model successfully 
gives vorticity distributions whose features are close to the filtered DNS. 

\begin{figure}
 
  \begin{center}

   \includegraphics[width=70mm]{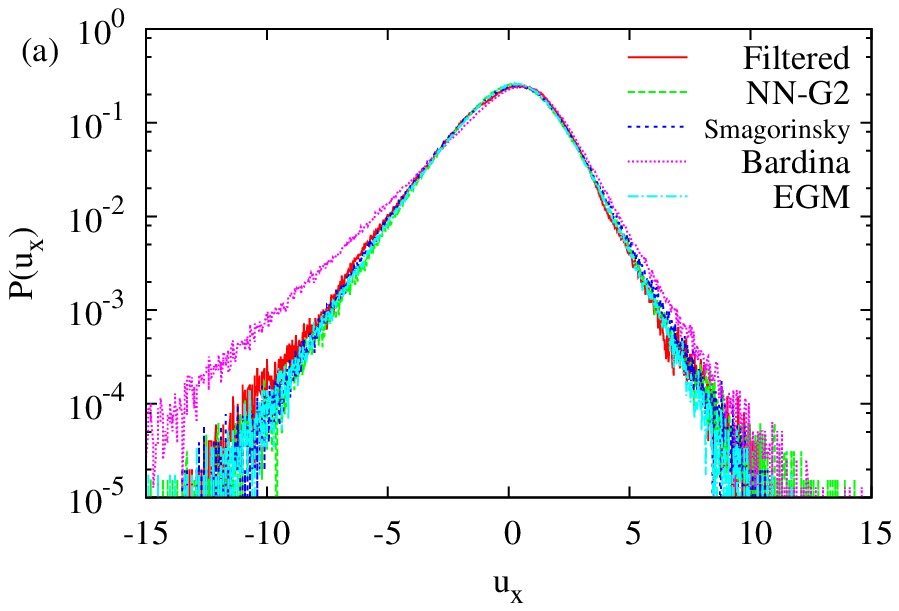}
   \includegraphics[width=70mm]{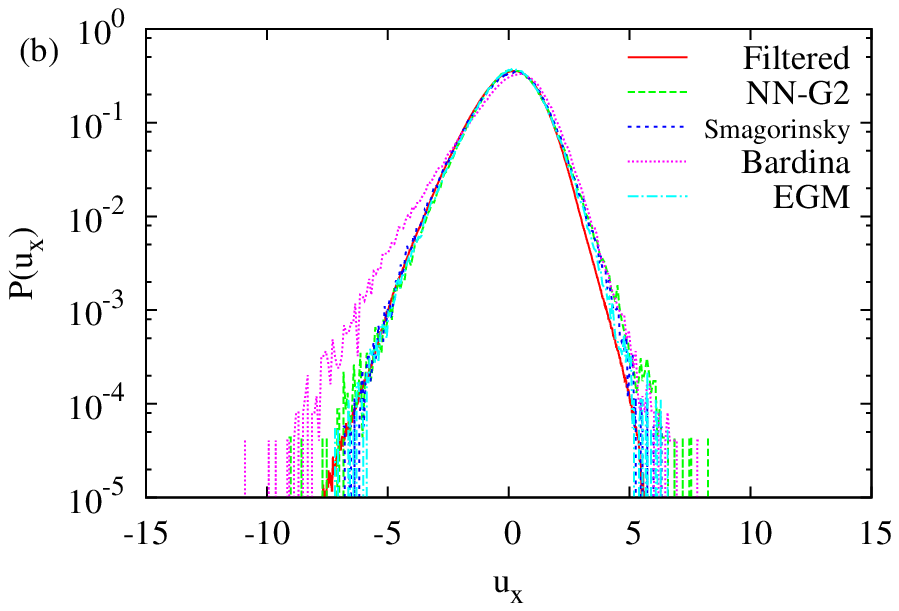}
 
   \includegraphics[width=70mm]{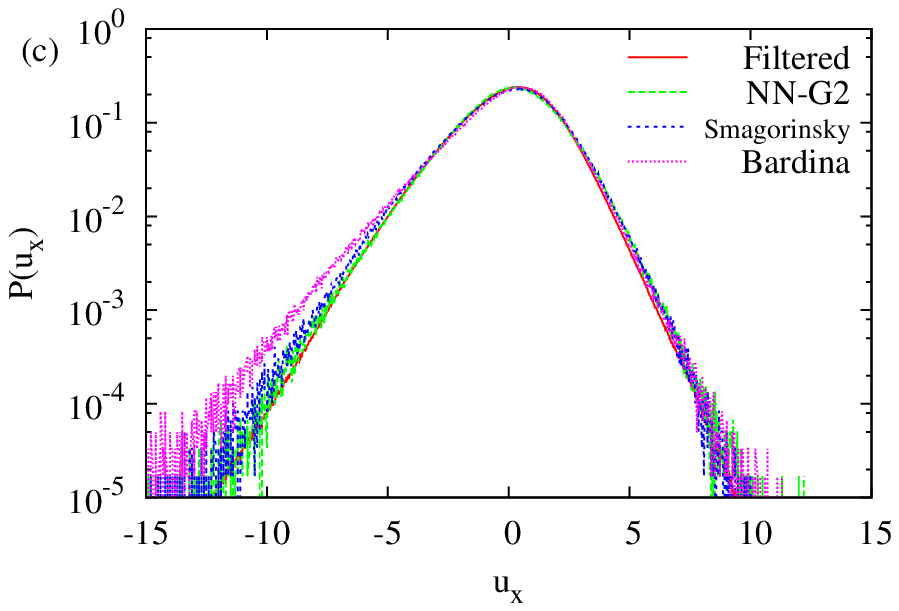}
   \includegraphics[width=70mm]{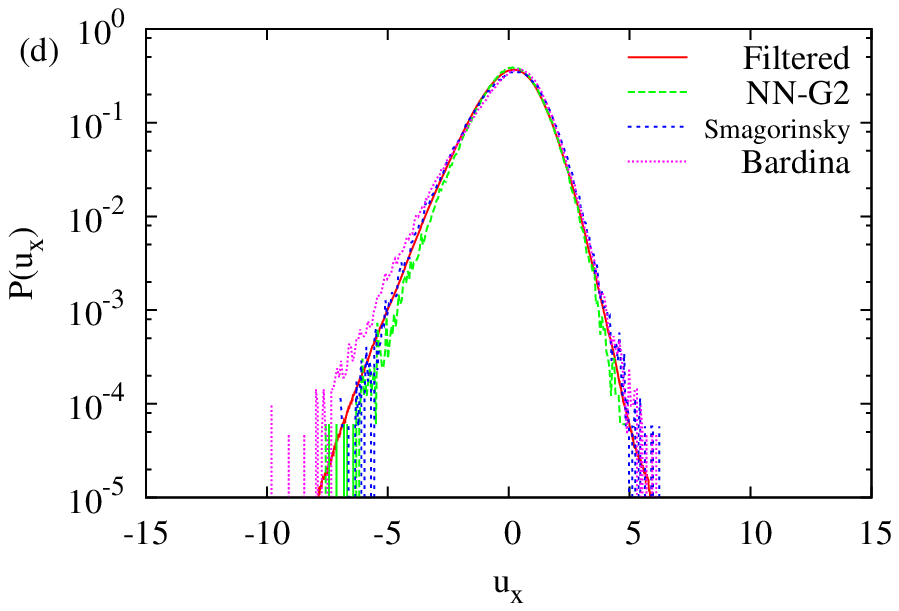}
  \end{center}
 
 \caption{P.d.f.s of longitudinal derivative of velocity $\partial_1 \overline{u}_1$. 
(a) Case HIT1-1, (b) Case HIT1-2, (c) Case HIT2-1, (d) Case HIT2-2. }
 \label{fig:IHTapostariori_u_x}
\end{figure}

Figure \ref{fig:IHTapostariori_u_x} compares the p.d.f.s of the 
longitudinal derivative of velocity $\partial_1 \overline{u_1}$. 
They are not normalized so that the difference in the standard deviation 
can be observed. 
In all cases p.d.f.s are skewed but less intermittent than 
those of DNS since the small scales are removed. 
All p.d.f.s except those of the Bardina model nearly collapse. 
The skewness and flatness factors are compared in Table \ref{ta:IHTLES1-1}; 
the flatness factors of the p.d.f.s of the transversal derivative $\partial_2 \overline{u_2}$ 
are also included. 
The values of the NN model and the Smagorinsky model are in reasonable agreement with 
those of the filtered DNS data.

\subsection{Initial-value problem of Taylor-Green vortices}
\label{sec-TG}

\begin{figure}
  \begin{center}
   \includegraphics[width=70mm]{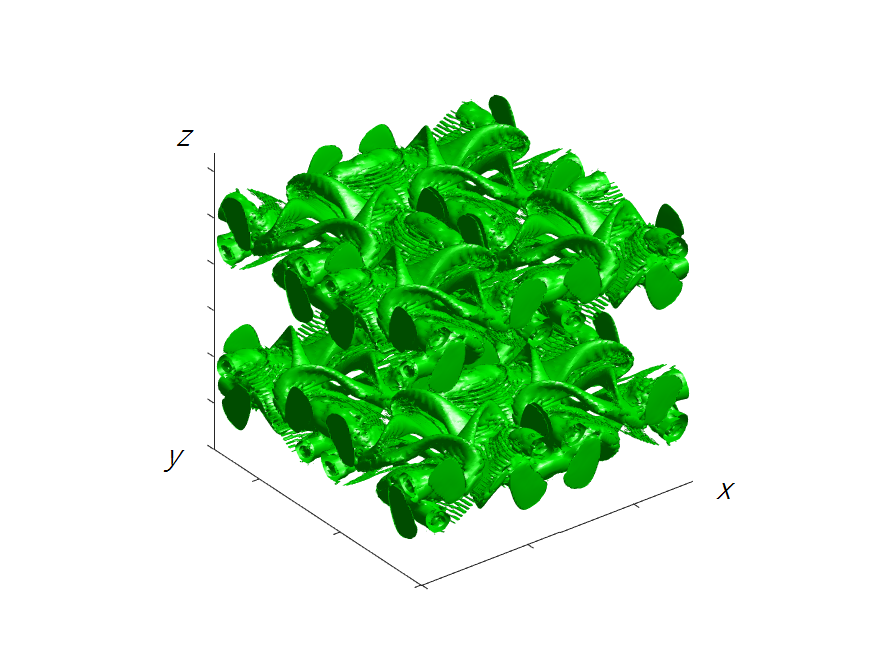}
   \includegraphics[width=70mm]{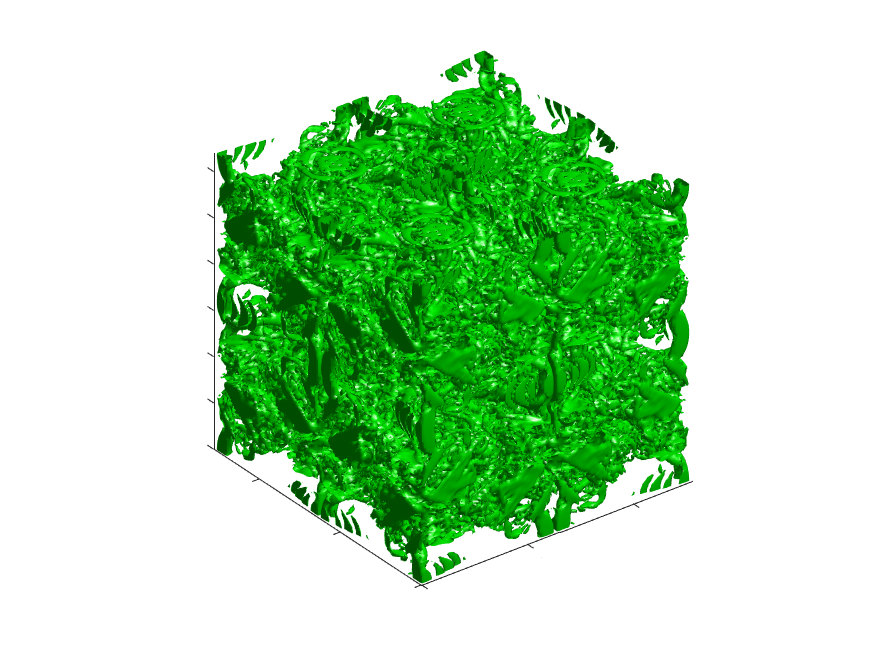}
  \end{center}
 \caption{Isosurface of magnitude of vorticity. DNS. Case TGV-1. 
(Left) $t=6$, (right) $t=12$. }
 \label{fig:3DTGVisosurface}
\end{figure}

The initial-value problem of the three-dimensional Taylor-Green vortices 
is often used to check the accuracy of a numerical scheme in computational fluid dynamics. 
In this section we consider this problem in which the initial velocity field is given by 
\begin{eqnarray}
u(0, \mathbd{x})&=&u_0\,\sin(x)\cos(y)\cos(z)\,,\label{eq:3DTGVx}\\
v(0, \mathbd{x})&=&-u_0\,\cos(x)\sin(y)\cos(z)\,,\label{eq:3DTGVy}\\
w(0, \mathbd{x})&=&0\,.\label{eq:3DTGVz}
\end{eqnarray}
The initial coherent large-scale vortices develop into turbulence 
which contains fine-scale structures 
as shown in Fig.~\ref{fig:3DTGVisosurface}. 
The four cases listed in Table \ref{ta:3DTGVcase} are considered. 
For each case LES is performed 
with two values of the filter width: $\overline{\Delta}=8\Delta_{\rm{DNS}}$ and $32\Delta_{\rm{DNS}}$. 
The NN models trained in Case HIT-1 ($\overline{\Delta}=8\Delta_{\rm{DNS}}$) 
is used for LES with grid points $N_{\rm{LES}}^3=128^3$,  
while the NN model trained in Case HIT-2 ($\overline{\Delta}=32\Delta_{\rm{DNS}}$) 
is used for LES with grid points $N_{\rm{LES}}^3=64^3$. 
It is pointed out that the filter width in the physical space is the same for training and LES. 
LES with the Smagorinsky model and the Bardina model is also performed. 
The time step in LES is fixed to $\Delta t=1.0\times10^{-3}$. 

\begin{table}[h]
\begin{center}
  \vspace{5mm}
  \caption{Simulation parameters of DNS of initial-value problem of three-dimensional 
Taylor-Green vortices. }
  \begin{tabular}{c|rrrrrr}
\hline \hline 
    Case& $N^3$ &$10^4\nu$& $k_{\rm max}$& $10^3 \Delta t$ \\ \hline
TGV-1 &$512^3$&7.0&248&1.0\\
TGV-2 &$512^3$&4.0&248&1.0\\
TGV-3 &$1024^3$&2.8&483&0.625\\
TGV-4 &$1024^3$&1.1&483&0.625\\
\hline \hline 
  \end{tabular}
  \label{ta:3DTGVcase}
\end{center}
\end{table}

\begin{figure}[h]
 
  \begin{center}
   \includegraphics[width=70mm]{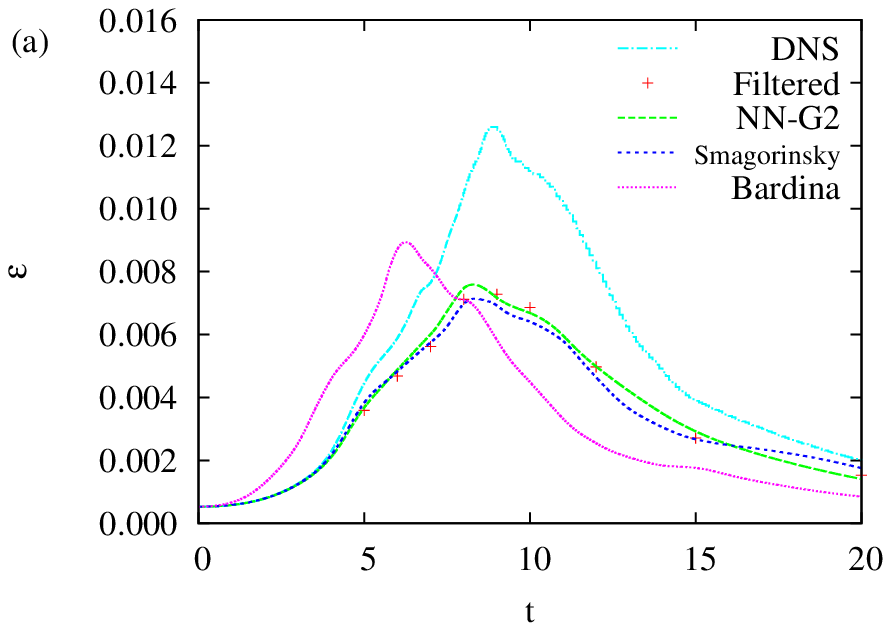}
   \includegraphics[width=70mm]{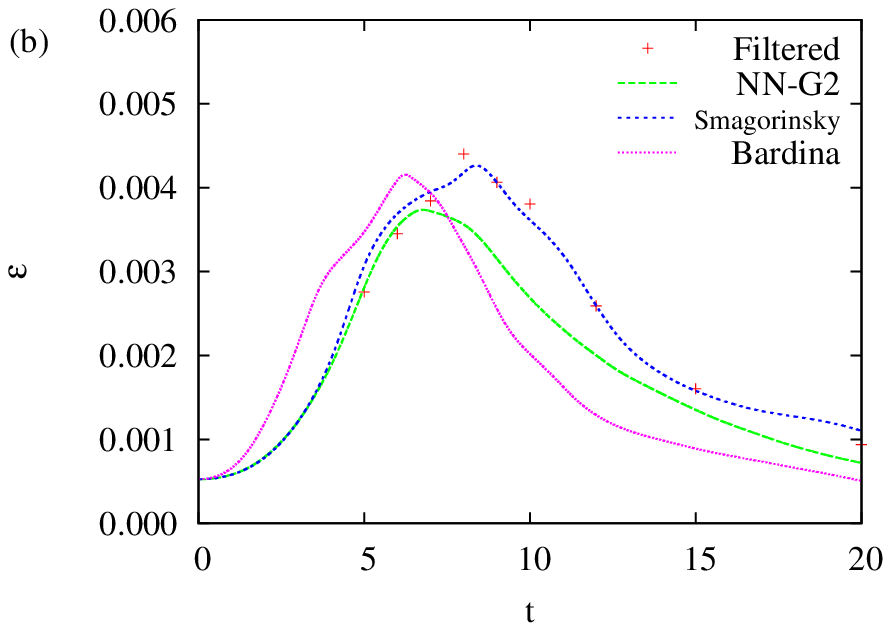}

   \includegraphics[width=70mm]{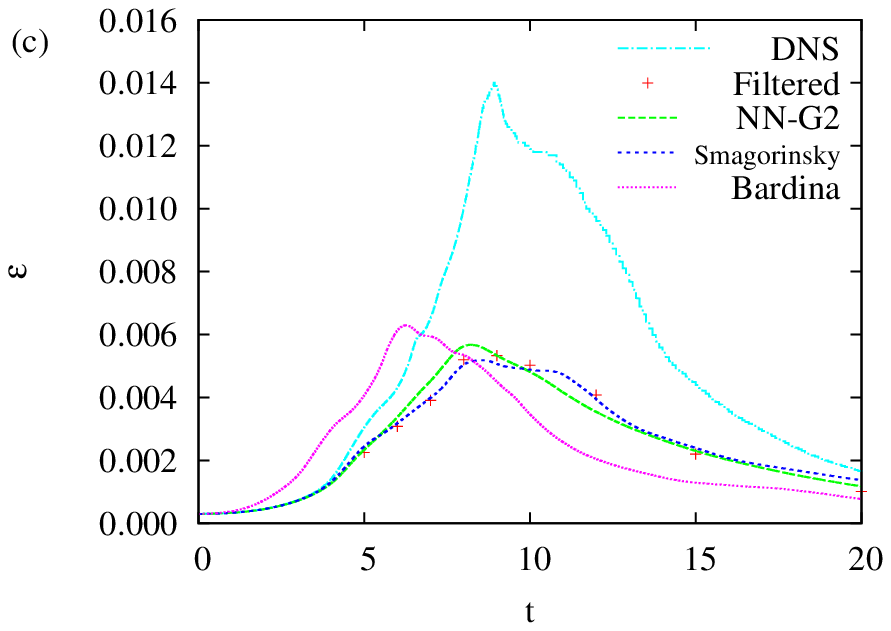}
   \includegraphics[width=70mm]{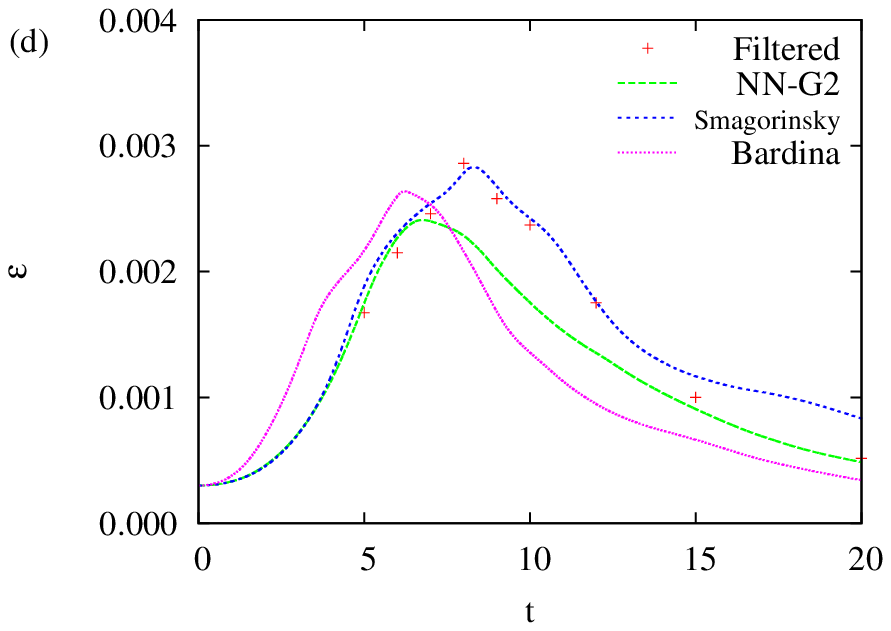}
 
   \includegraphics[width=70mm]{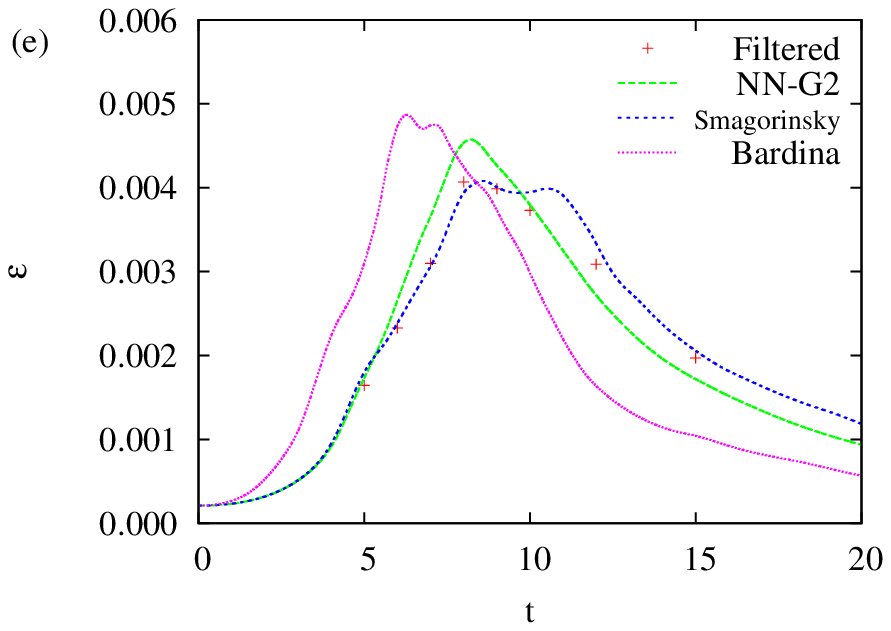}
   \includegraphics[width=70mm]{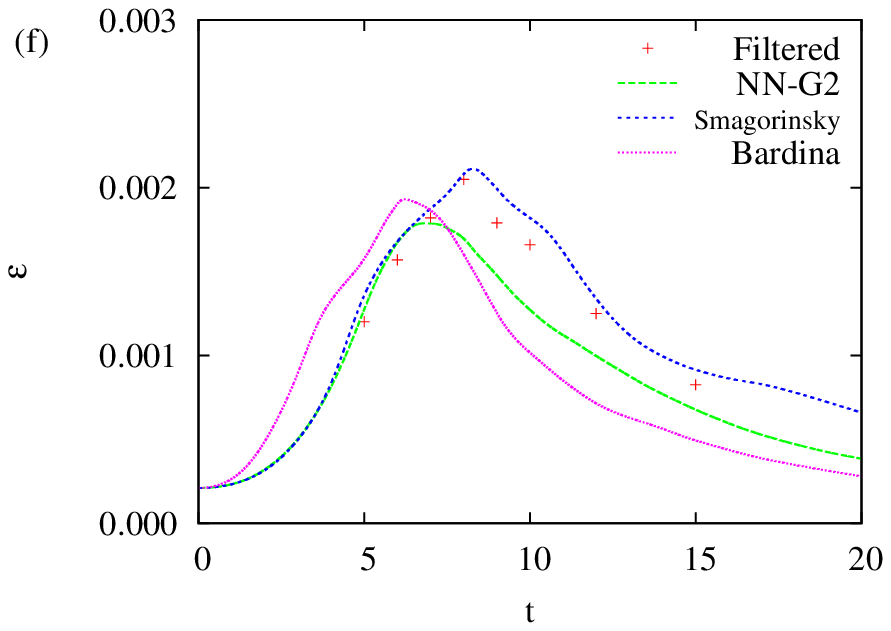}

   \includegraphics[width=70mm]{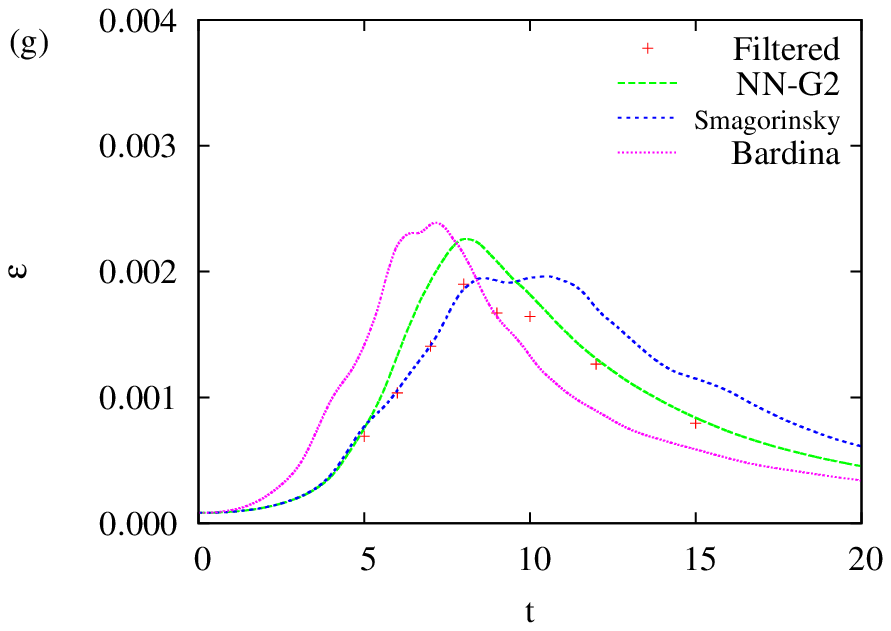}
   \includegraphics[width=70mm]{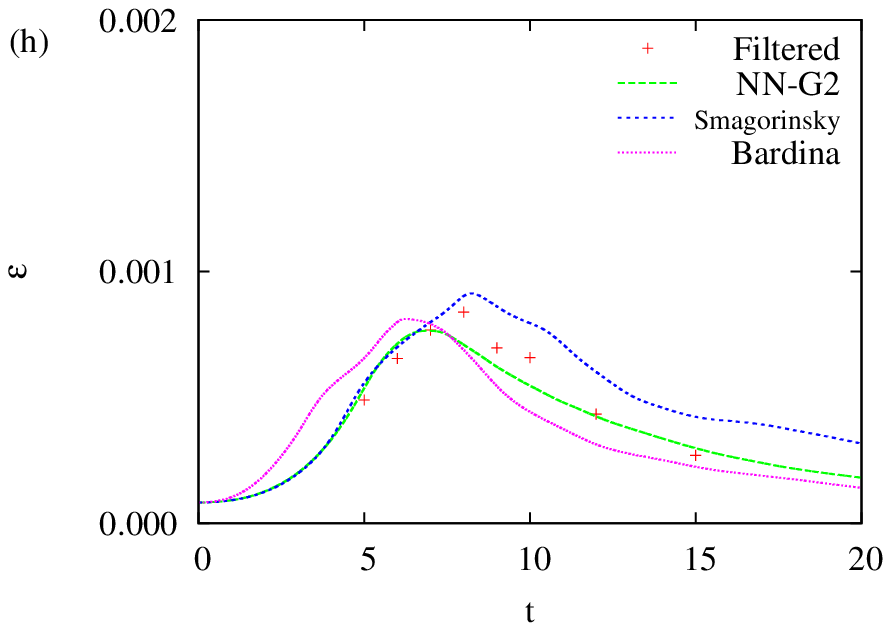}

  \end{center}
 
 \caption{Time evolution of rate of energy dissipation. 
Comparison between DNS, filtered DNS, LES. 
(a) Case TGV-1, $N_{\rm{LES}}^3=128^3$, 
(b) Case TGV-1, $N_{\rm{LES}}^3=64^3$, 
(c) Case TGV-2, $N_{\rm{LES}}^3=128^3$, 
(d) Case TGV-2, $N_{\rm{LES}}^3=64^3$, 
(e) Case TGV-3, $N_{\rm{LES}}^3=128^3$, 
(f) Case TGV-3, $N_{\rm{LES}}^3=64^3$, 
(g) Case TGV-4, $N_{\rm{LES}}^3=128^3$, 
(h) Case TGV-4, $N_{\rm{LES}}^3=64^3$.  
}
 \label{fig:TGVens}
\end{figure}

Figure \ref{fig:TGVens} shows time evolution of the rate of energy dissipation. 
The DNS results are included for reference only in Fig.~\ref{fig:TGVens}(a) and (c) 
since the values are much larger than the values of filtered DNS and LES in the other cases. 
The rate of energy dissipation increases until $t \approx 9$ and then decreases. 
The three models (the NN, Smagorinsky, and Bardina models) reproduce this behavior, 
while the maximum of the rate of the energy dissipation is earlier for the Bardina model 
than the filtered DNS and the other models; 
this is because the added eddy viscosity is too large so that 
the energy dissipation is overpredicted.  
The Smagorinsky model gives good results for larger filter width $\overline{\Delta}=32\Delta_{\rm{DNS}}$ 
except for Case TGV-4, 
while the performance is comparable to the NN model 
for $\overline{\Delta}=8\Delta_{\rm{DNS}}$. 
The results show that the neural network trained for forced homogeneous isotropic turbulence 
can be used for decaying turbulence. 

\begin{figure}[h]
 
  \begin{center}
   \includegraphics[width=55mm]{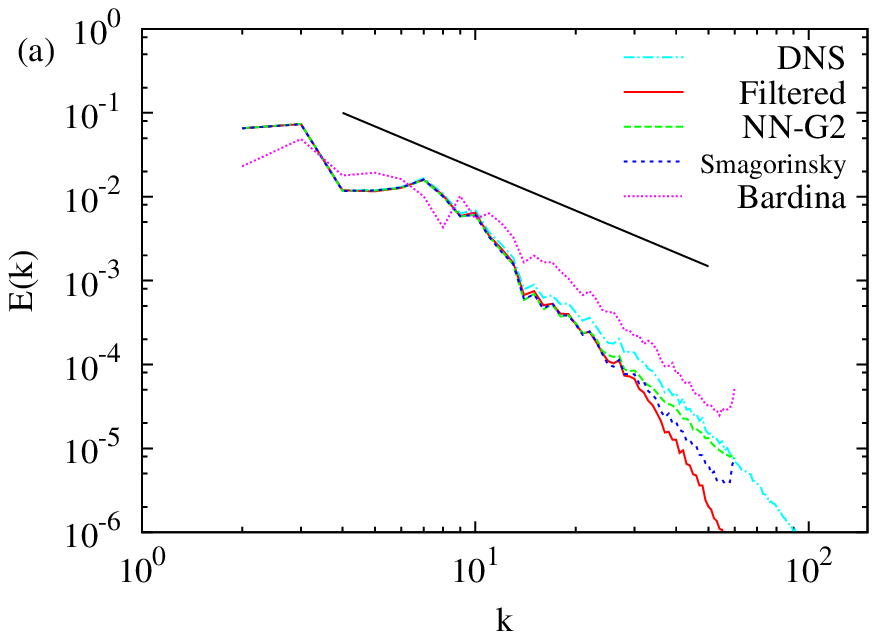}
   \includegraphics[width=55mm]{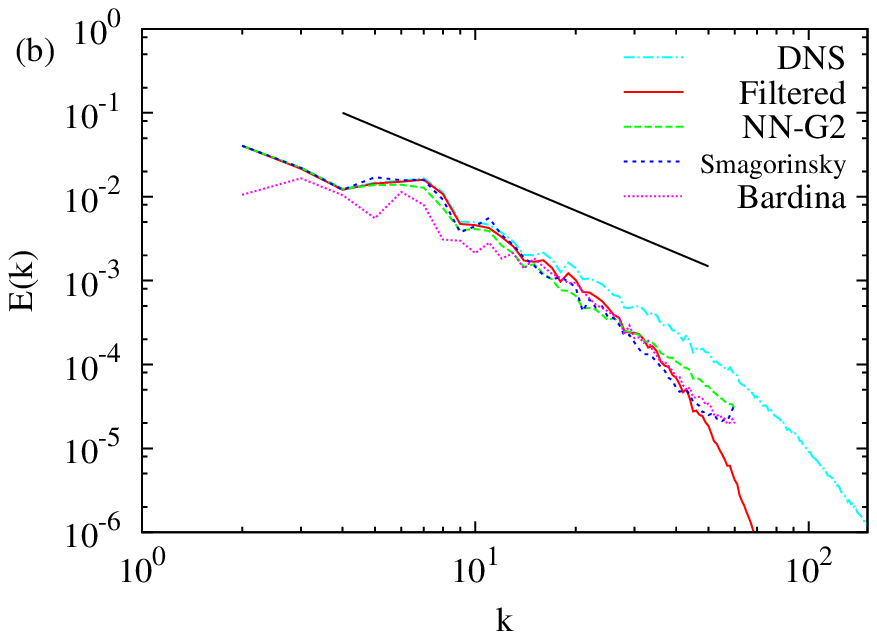}
   \includegraphics[width=55mm]{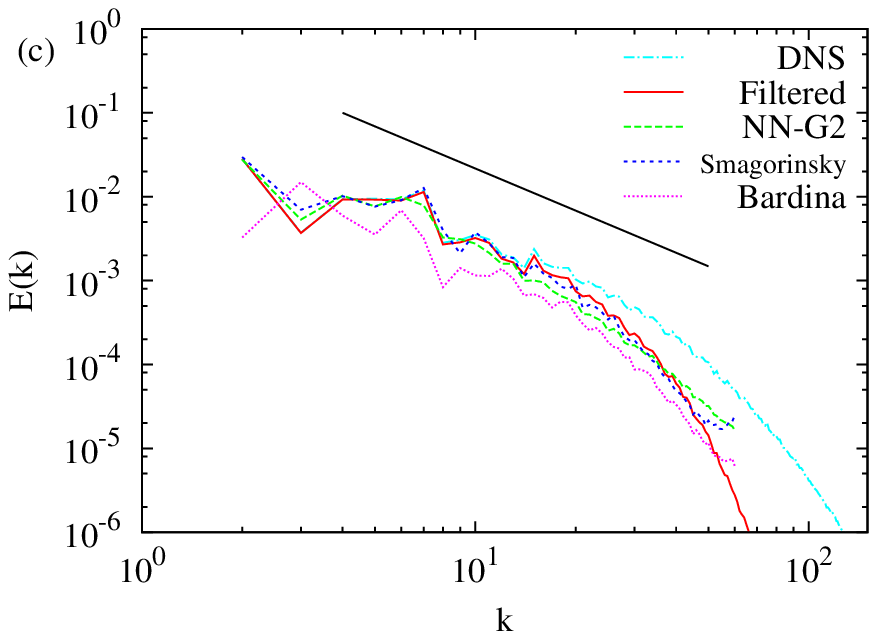}

   \includegraphics[width=55mm]{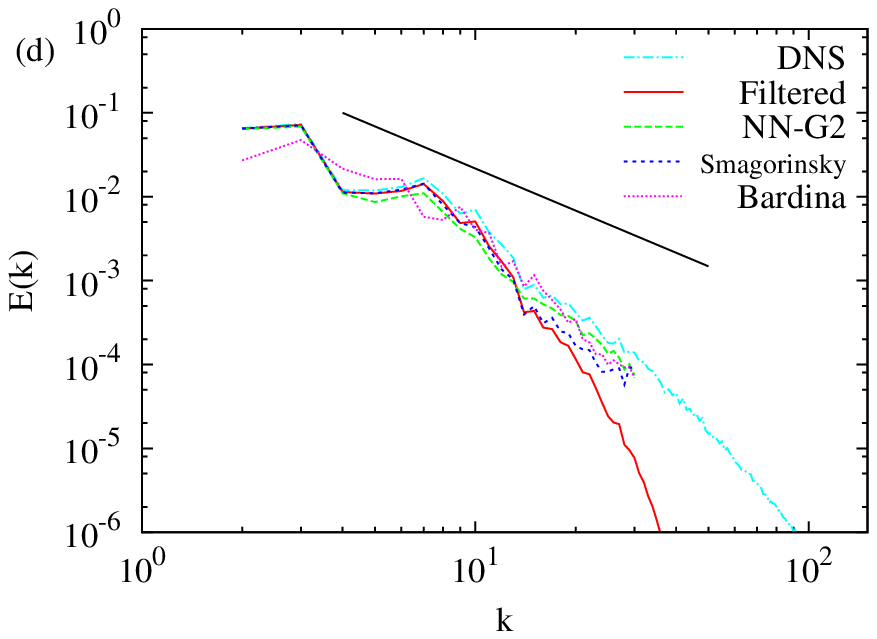}
   \includegraphics[width=55mm]{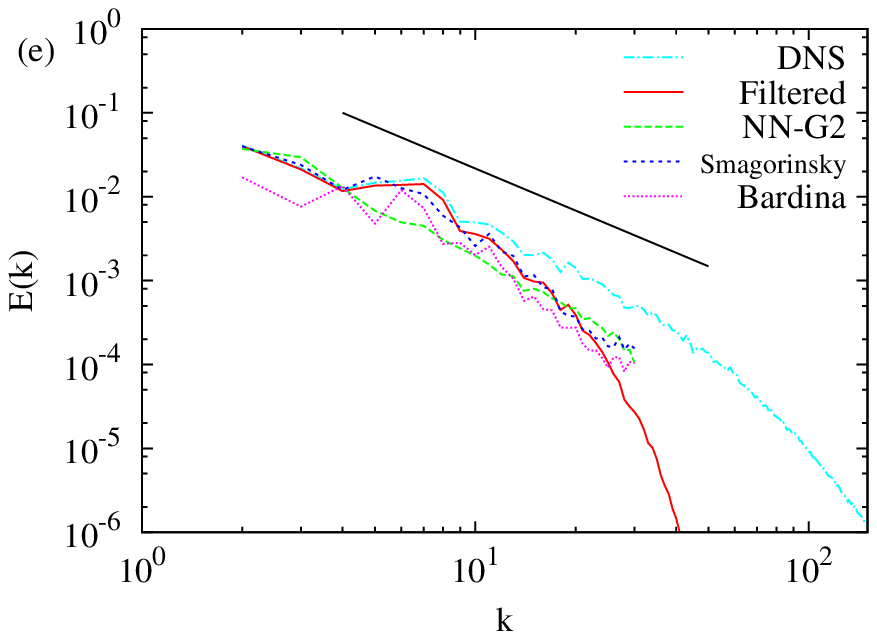}
   \includegraphics[width=55mm]{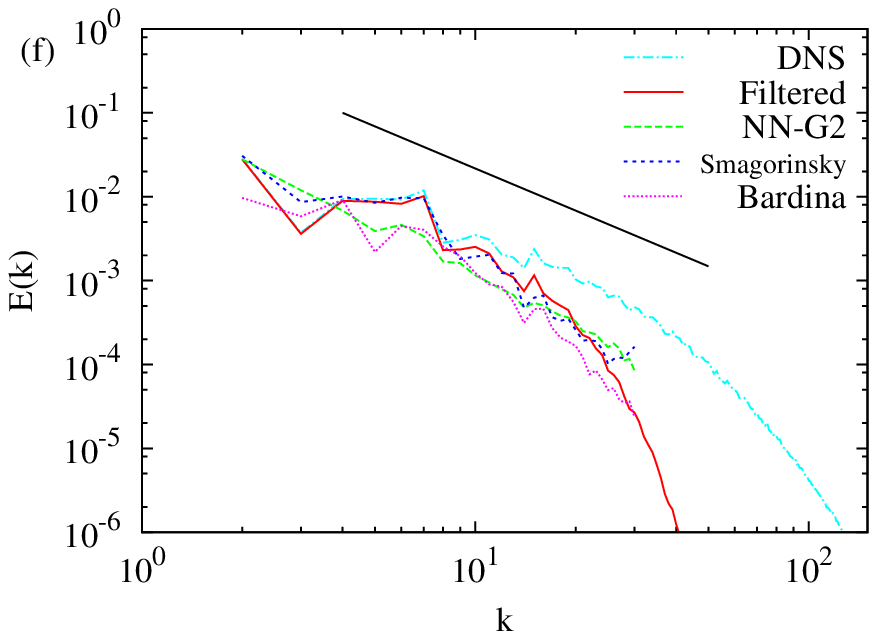}
  \end{center}

 \caption{Comparison of energy spectrum between DNS, filtered DNS, and LES. 
Case TGV-2. 
The solid lines show $E(k) \propto k^{-5/3}$. 
(a--c) $N_{\rm{LES}}^3=128^3$, (d--f) $N_{\rm{LES}}^3=64^3$. 
(a,d) $t=6$, (b,e) $t=9$, (c,f) $t=12$.}
 \label{fig:TGVapostariori2}
\end{figure}

Figure \ref{fig:TGVapostariori2} shows the energy spectrum 
for Case TGV-2 at three instants: 
(i) $t=6$ at which the the turbulent structures are being created, 
(ii) $t=9$ at which the rate of energy dissipation is maximum, 
and (iii) $t=12$ at which turbulence is decaying. 
The inertial subrange in which $E(k) \propto k^{-5/3}$ is established at $t=9$ and $12$. 
For $N_{\rm{LES}}^3=128^3$ the NN model and the Smagorinsky model 
reproduce the spectrum at $t=6$ which retains the features of the initial Taylor-Green vortices 
successfully.  
At $t=12$, however, 
the energy spectrum of the NN model deviates from the filtered DNS at low wavenumbers, 
while they are in reasonable agreement at large wavenumbers. 

\begin{figure}[h]
 
  \begin{center}
   \includegraphics[width=55mm]{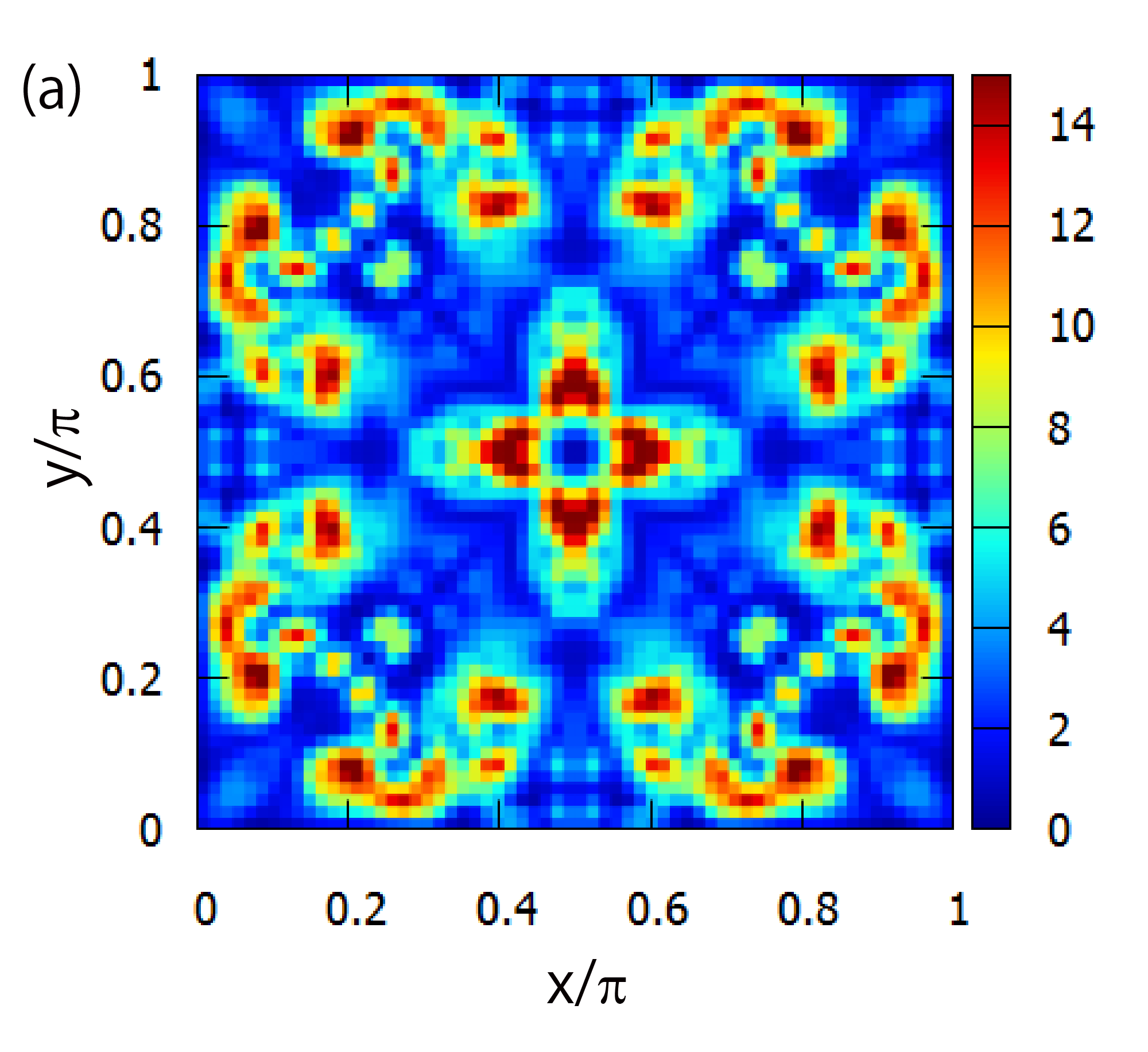}
   \includegraphics[width=55mm]{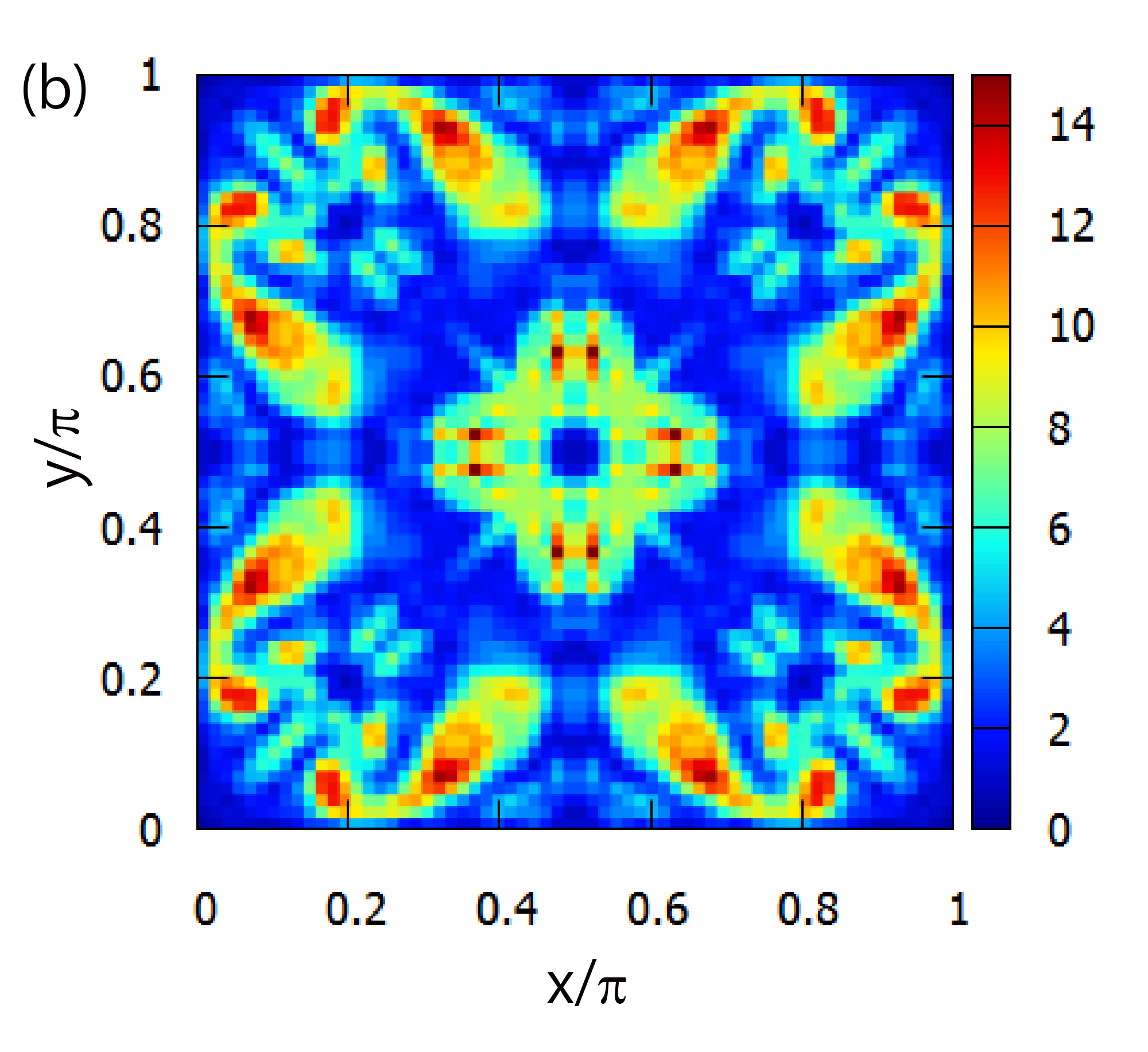}
   \includegraphics[width=55mm]{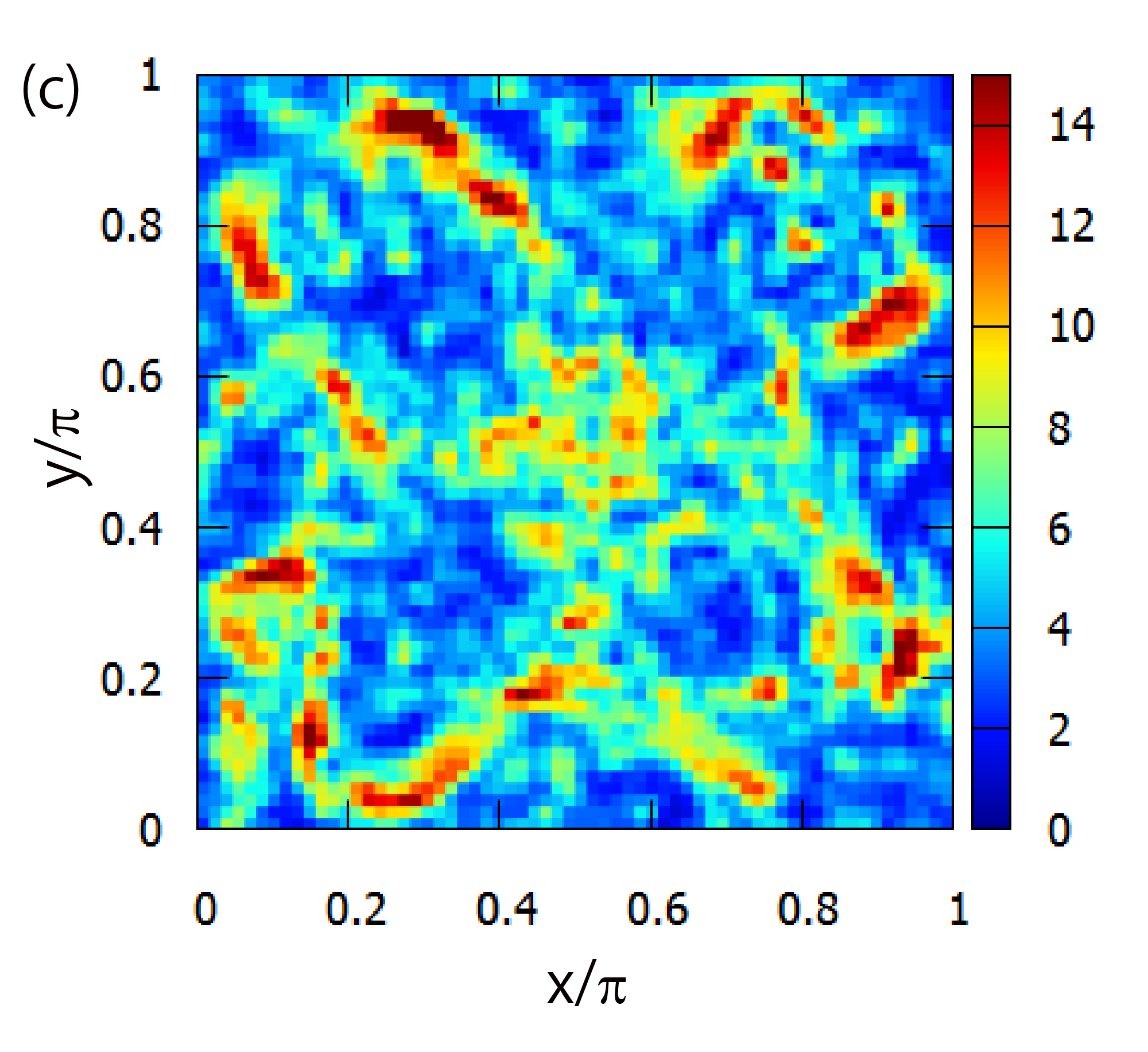}
  \end{center}

 \caption{Distributions of $\omega_z$ on $z=\pi/2$ at $t=9$. Case TGV-2, $N_{\rm{LES}}^3=128^3$. 
(a) Filtered DNS, (b) Smagorinsky model, (c) NN model. 
}
 \label{fig:TGVapostarioriPl}
\end{figure}

\begin{figure}[h]
 
  \begin{center}
   \includegraphics[width=55mm]{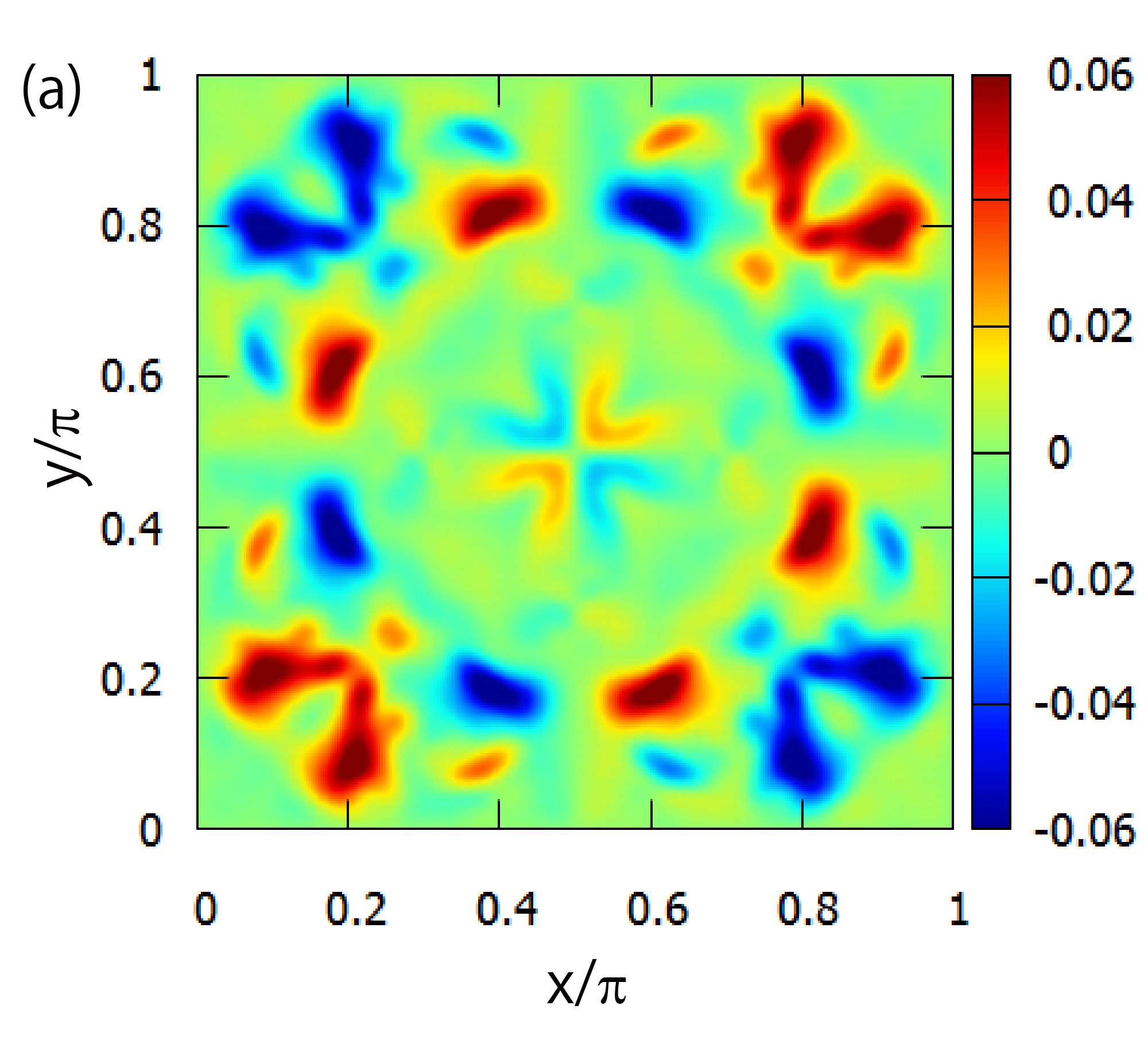}
   \includegraphics[width=55mm]{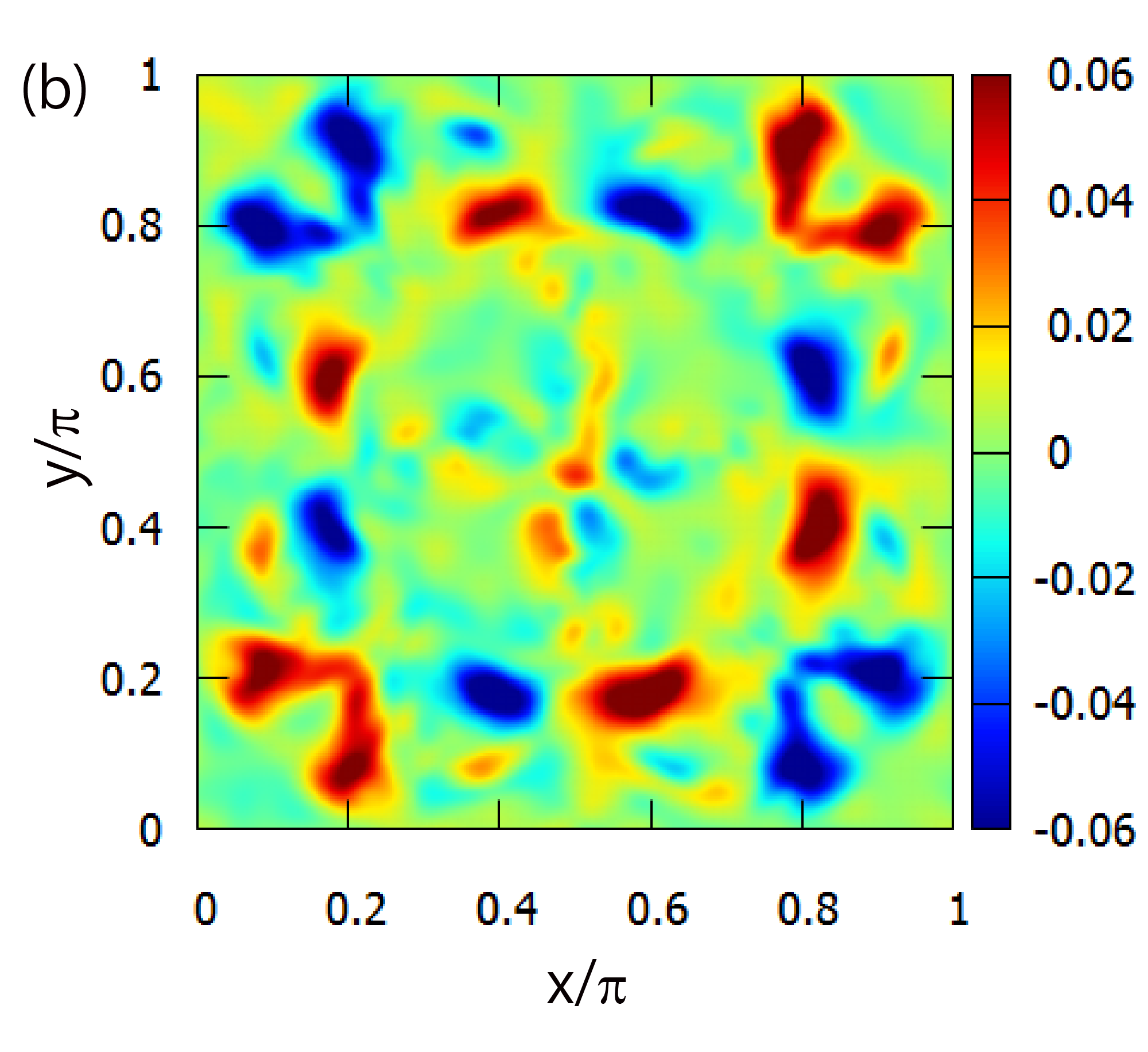}
  \end{center}

 \caption{Distribution of $\tau_{12}$ on $z=\pi/2$ at $t=9$
predicted by NN model ({\textit{a priori}} test).
Case TGV-2, $N_{\rm{LES}}^3=128^3$. 
(a) Filtered DNS, (b) NN model. 
}
 \label{fig:TGVapriori_diag}
\end{figure}

The reason for the difference between the NN model and the filtered DNS 
can be understood by 
Fig.~\ref{fig:TGVapostarioriPl}, 
which shows the vorticity distribution on $z=\pi/2$ at $t=9$. 
Symmetry with respect to $x=\pi/2$, $y=\pi/2$, and $x=\pm y$ 
satisfied by the initial velocity field 
is preserved for the filtered DNS and the Smagorinsky model, 
while it is broken for the NN model. 
This is because the symmetry under the orthogonal transformations 
is not incorporated in the present NN model, 
although it respects the symmetry under a limited number of transformations 
such as $(x,\,y,\,z) \to (y,\,z,\,x)$ by its construction. 
Figure \ref{fig:TGVapriori_diag} showing the results of {\textit{a priori}} test at $t=9$ 
confirms that the symmetry is slightly broken by the NN model. 
The small asymmetry in the SGS stress grows in a chaotic manner 
leading to the broken symmetry observed in Fig.~\ref{fig:TGVapostarioriPl}.  
It may be possible to keep the symmetry using e.g. the approach by Ling et al.~\cite{LKT-2016}.  
However, the number of tensors and scalars which express 
a general form of the SGS stress would be large 
when the second-order derivatives of the velocity are included in the input variables; 
it is not evident that training is successful in this case. 

\section{Concluding Remarks}
\label{sec-summary}

The turbulence models for the SGS stress in LES were developed by neural networks 
which consist of three layers. 
Two methods were shown to be effective for improvement of regression accuracy 
of the neural network: 
one is to introduce weight into data sampling so that the SGS stress of large magnitude 
contributes to training; 
the other is to include the second-order derivatives of velocity 
in the input variables.  
As a result strong correlation of the SGS stress between the exact values 
and those predicted by the NN models was observed; 
the correlation coefficient is about $0.9$ and $0.8$ 
for large filter widths $\overline{\Delta}=48.8\eta$ and $97.4\eta$, respectively,  
although correlation is slightly lower than the gradient and extended gradient models. 
It was also shown that the NN models are close to but not identical with 
the gradient and extended gradient models. 
The NN model was used in LES of the homogeneous isotropic turbulence and 
the initial-value problem of the Taylor-Green vortices. 
The NN model should be stabilized to assure numerical stability. 
The results obtained with the stabilized NN model 
were in good agreement with those of the filtered DNS and LES with the Smagorinsky model. 
However, the NN model could not keep the symmetry of the flow 
in the initial-value problem of the Taylor-Green vortices. 

We emphasize that neural networks of simple structure 
can predict the SGS stress accurately using methods for improvement. 
The NN model developed in the present study would allow us to 
infer a formula which is established by neural networks 
and has physical interpretation. 
Similar approach in Zhou et al.~\cite{ZHWJ-2019} 
using only the velocity gradient tensor (and the filter width)   
gave correlation coefficients of about $0.9$ for $\overline{\Delta} \approx 70\eta$ 
(estimated by their simulation parameters); 
their results may be further improved by using our methods. 
In Xie et al.~\cite{XWE-2020} high correlation with ${\rm{Corr}} \approx 0.99$ 
has been achieved for $\overline{\Delta} \approx 33\eta$ (also estimated) 
by using higher-order derivatives of velocity at the points in the neighborhood;  
although their approach seems effective for accurate regression of the SGS stress, 
it would be difficult to be implemented into e.g. a body-fitted coordinate system. 

One important task which should be done before pursuing an explicit form of the SGS stress 
is to establish a numerically stable NN model; 
our NN model should be stabilized in {\textit{a posteriori}} test, 
which is also the case in other recent works \cite{ZHWJ-2019, BFM-2019, XWE-2020}. 
It would be worth trying to seek a method different from addition of eddy viscosity 
and clipping.



%
%

%

\begin{acknowledgments}
Numerical calculations were performed on the Altix UV1000, UV2000, 
and the Supercomputer system ``AFI-NITY'' at the 
Advanced Fluid Information Research Center, 
Institute of Fluid Science, Tohoku University.
\end{acknowledgments}


\end{document}